%% file: MHD_rectangular_2phase.tex
\documentclass[pdflatex,sn-mathphys-num]{sn-jnl}

\usepackage[]{graphicx,color}%
\usepackage{multirow}%
\usepackage{amsmath,amssymb,amsfonts}%
\usepackage{amsthm}%
\usepackage{mathrsfs}%
\usepackage[title]{appendix}%
\usepackage{xcolor}%
\usepackage{textcomp}%
\usepackage{manyfoot}%
\usepackage{booktabs}%

\usepackage{bm}							
\usepackage[position=t,caption=false,justification=centering]{subfig}
\usepackage{hyperref}
\usepackage{mathtools}
\usepackage[makeroom]{cancel}
\usepackage{stmaryrd} 
\usepackage{tabularx}
\usepackage{multirow}
\usepackage{float}
\renewcommand{\vec}[1]{{\mbox{\boldmath $ #1 $}}}
\allowdisplaybreaks
\newcommand{\Rey}{Re} 
\newcommand{\Ha}{Ha} 
\newcommand{\Pran}{Pr} 
\newcommand{\Po}{Po} 
\graphicspath{{./figures/}}

\begin{document}

\title[Two-phase stratified MHD flows in rectangular ducts...]{Two-phase stratified MHD flows in rectangular ducts}


\author[1]{\fnm{Subham} \sur{Pal}}\email{subhampal@tauex.tau.ac.il}

\author*[1,2]{\fnm{Ilya} \sur{Barmak}}\email{ilyab@tauex.tau.ac.il}

\author[1]{\fnm{Arseniy} \sur{Parfenov}}\email{ae2parfenov@gmail.com}

\author[1]{\fnm{Alexander} \sur{Gelfgat}}\email{gelfgat@tauex.tau.ac.il}

\author[1]{\fnm{Neima} \sur{Brauner}}\email{brauner@tauex.tau.ac.il}

\affil[1]{\orgdiv{School of Mechanical Engineering}, \orgname{Tel Aviv University}, \orgaddress{\city{Tel Aviv}, \postcode{6997801}, \country{Israel}}}

\affil[2]{\orgname{Soreq NRC}, \orgaddress{\city{Yavne}, \postcode{8180000}, \country{Israel}}}


\abstract{The characteristics of two-phase stratified magnetohydrodynamic (MHD) flow in horizontal rectangular ducts are investigated for a system consisting of a conductive liquid and a non-conductive gas. Numerical and analytical solutions of the governing equations for the velocity and induced magnetic field intensity of fully developed laminar MHD flow are obtained for various combinations of bottom- and side-wall conductivities and for different orientations of an externally applied transverse magnetic field. The relevant set of dimensionless parameters governing the problem is identified.	Unlike in single-phase MHD flows, the presence of a non-conductive gas layer breaks the flow symmetry, leading to a significantly different dependence of the flow characteristics on duct aspect ratio, wall-conductivity configuration, and the strength and orientation of the applied magnetic field. Using mercury--air flow as a representative test case, the solutions are employed to quantify the influence of the gas phase on the in-situ liquid holdup, velocity field, pressure gradient, flow lubrication, and pumping-power requirements. It is shown that, regardless of the magnetic Reynolds number, these characteristics are strongly affected by the wall-conductivity configuration and by the orientation of the external magnetic field.
}


\maketitle

\section{Introduction} \label{Sec: Introduction}

Magnetohydrodynamic (MHD) flow of electrically conducting fluids in rectangular ducts under a transverse magnetic field arises in various applications (e.g., \cite{Branover78,Moreau90,Kirillov95,Muller01,Davidson01,Molokov07,Shanmugadas25}).  These include MHD power generators and MHD pumps, magnetic flow meters and accelerators. They are utilized in both large-scale systems, such as liquid metal cooling blankets in fusion reactors and continuous steel casting in the metallurgical industry and small-scale systems, including medical devices, micromixers, and lab-on-a-chip technologies. The electrical conductivity of the duct walls is a key factor in system design for each specific application.  

In magnetic pumps, the interaction between an applied magnetic field and a supplied current generates a Lorentz force that drives the conductive fluid flow. In other configurations, where the fluid is mechanically driven by a pump, the motion of the conductive fluid through a magnetic field induces electric currents perpendicular to the field. These currents, in turn, produce a Lorentz force opposing the flow direction, thereby increasing the required pressure gradient. The relative magnitude of the Lorentz force to the viscous force is characterized by the Hartmann number, $\displaystyle\Ha=B_0 H \sqrt{\sigma_e/\eta}$, where $H$ is the channel height, $B_0$ is the external transverse magnetic field strength, and $\eta,\sigma_e$ are the fluid dynamic viscosity and electric conductivity, respectively. At high Hartmann numbers, electromagnetic pressure losses, combined with viscous friction losses, can become substantial. Ongoing research focuses on methods to mitigate these losses, such as using flow channel inserts (FCIs) or optimizing the geometry of the flow channels  with the aim of reducing pumping power requirements and mechanical stresses on duct walls (e.g., \cite{Buhler20,Nishio25}).

Analysis of a fully developed two-dimensional laminar MHD single-phase flow of a conductive fluid in ducts have been subject to many theoretical and numerical studies, which enable insights on the impact of the duct's walls electrical conductivity on the MHD flow characteristics and the pressure losses. The duct walls perpendicular to the magnetic field are referred to as the Hartmann walls \cite{Hartman37a}, and those parallel to it are the Shercliff walls \cite{Shercliff53}. Analytical solutions are available for single phase MHD flows in rectangular ducts with perfectly conducting walls \cite{Chang61,Uflyand61}, with insulating walls \cite{Shercliff53}, arbitrary conductivity of the Hartmann walls and insulated Shercliff walls \cite{Hunt65,Sloan66}, and for perfectly conductive Hartmann walls and arbitrary conductivity of the Shercliff walls \cite{Hunt65,Butler69}.  Cases with unsymmetrical wall conductivities were also studied \cite{Tao15}. The analytical solutions are obtained in the form of infinite series of hyperbolic functions, which limit their practical applicability to small Hartmann numbers. Asymptotic solutions for high Ha, indicate that thickness of the Hartmann and Shercliff boundary layers scale with $\Ha^{-1}$ and $\Ha^{-1/2}$, respectively (e.g., \cite{Muller01,Knaepen08}). Numerical solutions, enable obtaining the velocity profiles and pressure-gradient/flow rate relation at high Hartmann numbers. However, high spatial resolution near the duct walls is essential for accurately resolving the velocity gradients within the thin boundary layer and the associated shear stresses (e.g., \cite{Krasnov10,Boeck14}).

Many practical challenges involve two-phase flow systems, particularly within the nuclear and petroleum industries, geophysics and MHD power generation. Examples include two-phase liquid metal magnetohydrodynamics generators \cite{Wang22}, magnetic field-driven micropumps \cite{Yi02,Weston10}, microchannel networks of lab-on-a-chip devices \cite{Bau03,Hussameddine08}, where the presence of a second non-conductive fluid may enhance the mobility of the conducting fluid. 

Although single-phase MHD flow has been examined in depth, much less attention has been paid to two-phase MHD flow. Because the two phases differ in density, stratified flow is frequently observed in horizontal and inclined conduits. A second, non-conducting phase (e.g., gas) flowing concurrently to the conductive fluid may radically change its velocity profile and the accompanying pressure gradient. The available studies on stratified two-phase MHD flows are confined to the idealized case of two infinite parallel plates with a magnetic field perpendicular to the walls (e.g.,\cite{Shail73,Lohrasbi88,Malashetty97,Shah22,Parfenov24}). Moreover, none of these investigations accounts for the effect of the channel walls' electrical conductivity on the induced magnetic field, velocity distribution, and the two-phase flow behavior characteristics.

Using analytical solutions that account for the induced magnetic field, we recently showed that - in sharp contrast to single-phase conductive fluid flow - the velocity profile in two-phase flow depends on whether the bottom wall is electrically conducting or insulating \cite{Barmak25}. Independently of magnetic Reynolds number, an insulating bottom wall amplifies the gas lubrication effect and delivers substantially larger potential savings in pumping power. Complementary numerical simulations examined the role of side walls in wide channels. For high aspect ratios, all cases converge to predictions from the classical two-plate (TP) model with matching bottom wall conductivity. Yet, when the bottom wall is conducting, insulating side walls still exert a pronounced influence. Unexpectedly, the modified induced magnetic field created by those side walls dramatically reshapes the velocity profile and lowers the pressure gradient relative to TP-model forecasts.  However, to the best of our knowledge, two-phase MHD flow in square and rectangular ducts remains virtually unexplored in the literature.

In this study, we refer to MHD gas-liquid stratified flow in rectangular ducts  of  various aspect ratios. The conductive liquid is subject to external magnetic field. Numerical solutions of the two-dimensional problem for the velocity and induced magnetic field profiles were obtained for various combinations of bottom and side wall conductivities. Since the existence of an upper layer breaks the symmetry existing in single phase flow, cases where  the external magnetic field  is  parallel to the fluid--fluid interface cannot  be deduced from those obtained  with the field perpendicular to the interface, and should be examined separately.  For the same reason, ducts of the width-to-height aspect ratios $AR<1$ should be explored as the MHD flow cannot be deduced from solutions obtained for   ducts with $AR>1$.   Analytical solutions obtained for stratified   two-phase MHD flow with several combinations of wall conductivities  are presented and are used  to verify the corresponding numerical solutions.  The verified numerical tool is then used to explore the effects of the  duct geometry, conductivity of the Hartmann and Shercliff walls and the direction of the applied magnetic field on  two-phase flow characteristics.  

\section{Problem Formulation} \label{Sec: Formulation}

\begin{figure}[h!]
	\hspace*{-5mm}
	\subfloat[]
	{\def\svgwidth{0.63\textwidth}
		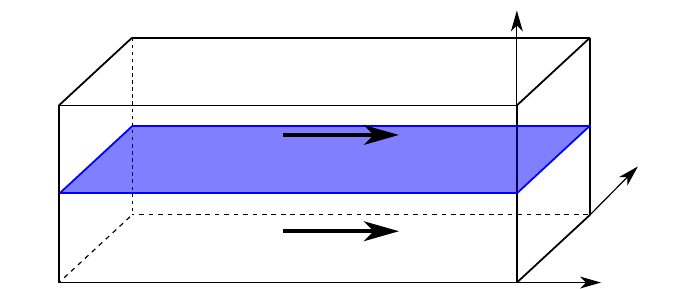}
	\hspace*{-2mm}
	\subfloat[]
	{\def\svgwidth{0.5\textwidth}
		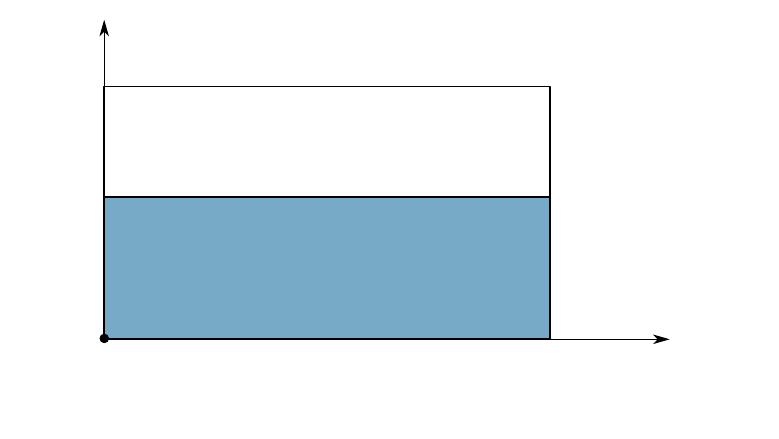}
	\caption{Stratified two-phase flow in a rectangular duct under external magnetic field. (a) Flow schematics and (b) cross-sectional view.}
	\label{Fig: MHD_geometry}
\end{figure}

We address a two-phase stratified MHD flow in a horizontal rectangular duct of aspect ratio $AR = W/H$, with height $H$ and width $W$ (Fig.\ \ref{Fig: MHD_geometry}). The flow is subject to a constant external magnetic field $\vec{B_0}=\bigl(B_{0|x},B_{0|y},0\bigr)$. The superficial velocities of the heavy and light fluids are $U_{1S}$ and $U_{2S}$, respectively, where $U_{1S,2S}= Q_{1,2}/ (H W)$ and $Q_{1,2}$ are the corresponding flow rates. The interface between two fluids is assumed to be plane, and the height of the lower (heavy) phase layer is denoted as $h_1$. Then the holdup is defined as $\displaystyle h=h_1/H$. The heavy phase is electrically conductive (e.g., liquid metal), whereas phase 2 is a lighter fluid assumed to be electrically non-conductive (e.g., gas). The physical properties of the conductive liquid are $\eta_1$ - dynamic viscosity; $\rho_1$ - density; $\sigma_{e1}$ - electric conductivity. 
In this study, the flow is assumed to be steady, fully developed unidirectional in the $z$-direction and laminar. Thereby, the dimensionless velocity (scaled by $U_{1S}$) is $\vec{U} = \bigl(0,0,U(x,y)\bigr)$ and all the flow and electromagnetic variables are independent of $z$. The dimensionless form of the axial components (the only non-zero) of the induction and momentum equations for the conductive phase (\cite{Muller01}):
\begin{align} 
	\frac{\partial^2 b}{\partial x^2} 
	+ \frac{\partial^2 b}{\partial y^2}
	+ \Ha
	\biggl(
		\frac{B_{0|x}}{B_0}
		\frac{\partial U_1}{\partial x}
	+ \frac{B_{0|y}}{B_0}
	\frac{\partial U_1}{\partial y}
	\biggr)
	= 0, 
	\label{Eq: Induction_b}
\\  
	\frac{\partial^2 U_1}{\partial x^2} 
	+ \frac{\partial^2 U_1}{\partial y^2}
	+ \Ha 
	\biggl(	
	\frac{B_{0|x}}{B_0}
	\frac{\partial b}{\partial x} 
	+ \frac{B_{0|y}}{B_0}
	\frac{\partial b}{\partial y}  
	\biggr)		
	= G, \label{Eq: Momentum_b}
\end{align}
where $b$ denotes the (dimensionless) induced magnetic field acting in the flow direction and $G$ is the dimensionless pressure gradient. These equations were rendered dimensionless by scaling the lengths by $H$, the magnetic field by $B_0 \Rey_m / \Ha$ ($B_0$=$\sqrt{B_{0|x}^2 +B_{0|y}^2}$, $\Rey_m=\mu_0\sigma_{e1} U_{1S} H$ is the magnetic Reynolds number, and $\displaystyle \Ha = B_0 H \sqrt{\sigma_{e1}/\eta_1}$), and the pressure gradient by $\displaystyle\eta_1 U_{1S} / H^2$. A derivation of Eqs.\ \ref{Eq: Induction_b} and \ref{Eq: Momentum_b} from the general MHD equations can be found in \cite{Barmak25} similar to the single-phase flow, see \cite{Muller01}.

The dimensionless momentum equation for the non-conductive lighter fluid layer reads:
\begin{equation} \label{Eq: Momentum_2_dim}
	\frac{\partial^2 U_2}{\partial x^2} 
	+ \frac{\partial^2 U_2}{\partial y^2}
	= \eta_{1 2} G,
\end{equation}
where the velocity is scaled by the superficial velocity of conductive liquid $\displaystyle U_2=\hat{U}_2 / U_{1S}$ and $\displaystyle\eta_{1 2} = \eta_1/\eta_2$ is the viscosity ratio of fluids. Note that in a unidirectional flow the pressure gradient, $G$, is uniform in the flow cross-section and same in both phases.

Eqs.\ \ref{Eq: Induction_b}, \ref{Eq: Momentum_b}, and \ref{Eq: Momentum_2_dim} should be solved with the appropriate boundary conditions to find the fully developed velocity profiles in both phases and the induced magnetic field. For the velocity, these are no-slip condition on the duct walls and continuity of the velocity and shear stress across the fluid--fluid interface \cite{Muller01}:
\begin{subequations} \label{Eq: BC_U_rectangular}
	\begin{align}
		U_1(x; y=0) &= 0,\qquad U_1(x=0,AR; y) = 0
		\\
		U_2(x; y=1) &= 0,\qquad U_2(x=0,AR; y) = 0 
		\\
		U_1(x; y=h) &= U_2(x; y=h) = U_i
		\\
		\eta_{1 2} \frac{\partial U_1}{\partial y}\Biggr\rvert_{x; y=h} &=
		\frac{\partial U_2}{\partial y}\Biggr\rvert_{x; y=h}
	\end{align}
\end{subequations}

The boundary conditions for the induced magnetic field depend on the conductivity of the duct walls:
\begin{subequations} \label{Eq: BC_induced_B_rectangular}
	\begin{align}
		&-\frac{\partial b}{\partial y}\Biggr\rvert_{x; y=0}
		+ \frac{1}{c_{bw}} b(x; y=0) 
		= 0 
		\\
		&-\frac{\partial b}{\partial x}\Biggr\rvert_{x=0,AR; y}
		+ \frac{1}{c_{sw}} b(x=0,AR; y) 
		= 0 
	\end{align}
\end{subequations}
where $c_{bw}$ and $c_{sw}$ are the dimensionless electric conductivities of the bottom and side walls, respectively. In this study, we consider only two limiting cases of the walls being either fully insulating $\displaystyle \bigl(c=0\bigr)$ or perfectly conducting $\displaystyle \bigl(c\to\infty\bigr)$ surfaces. In the former case, $b=0$, in the latter -- the normal derivative of $b$ becomes zero at the wall.

The interfacial boundary conditions for the induced magnetic field is then the same as for a fully insulating wall:
\begin{equation} \label{Eq: BC_induced_B_interface}
	b(x;y=h)=0,
\end{equation}
since the upper phase is non-conductive.

Conductive fluids in MHD flows are also characterized by the magnetic Prandtl number, defined as the ratio of momentum diffusivity (i.e., kinematic viscosity) to the magnetic diffusivity, or as the ratio of the magnetic to bulk Reynolds numbers ($\displaystyle\Rey=\rho_1 U_{1S} H / \eta_1$): 
\begin{equation}
\Pran_m = \frac{\Rey_m}{\Rey}=\frac{\mu_0 \sigma_{e1}}{\rho_1/\eta_1} 
\end{equation}
The magnetic Prandtl number is small for liquid metals (e.g., for mercury $\Pran_m=1.383\times 10^{-7}$). Therefore, for laminar flows of moderate $\Rey$, $\Rey_m$ is negligibly small. Moreover, the fully-developed steady flow is independent of $\Rey_m$ (see Eqs.\ \ref{Eq: Induction_b} and \ref{Eq: Momentum_b} above). 

\section{Problem Solution} \label{Sec: Numerics}

For a particular liquid metal--gas system (i.e., known viscosity ratio), the governing equations and boundary conditions of the considered problem (Eqs.\ \ref{Eq: Induction_b}-\ref{Eq: BC_induced_B_interface}) can be solved for velocity and induced magnetic field, if the location of the interface (i.e., holdup), the dimensionless pressure gradient, the Hartmann number and the conductivities of the bottom and side walls are provided. The flow rates (or superficial velocities) of each phase can then be calculated by integration of the velocity profile over the respective cross sections occupied by each fluid. In practice, however, the feed flow rates are prescribed, while the holdup and pressure gradient are unknown. Therefore, an iterative solution procedure is necessary to satisfy the constraints introduced by the specified flow rates. 

For a rectangular duct of a finite aspect ratio, analytical solutions can be obtained for certain combinations of bottom- and side-wall conductivities. Analytical solutions are provided in Appendix\ \ref{Sec: appendix_analyt_vertical} for a problem with the external magnetic field being vertical, i.e., $\vec{B_0}=\vec{B_{0|y}}$, when the duct is fully insulating or with perfectly conducting bottom and insulating side walls. For a problem with $\vec{B_0}=\vec{B_{0|x}}$, analytical solutions are provided in Appendix\ \ref{Sec: appendix_analyt_horizontal} for a perfectly conducting duct and for a duct with an insulating bottom and perfectly conducting side walls. These solutions show that the holdup, $h$, and the (dimensionless) pressure gradient, $G$, are determined by six dimensionless parameters, of which only two, $Q_{21}$ and $\Ha$, define the operational conditions, while four others are set by choosing a particular two-phase system and rectangular duct ($\eta_{1 2}, AR, c_{bw}, c_{sw} $).

Calculation of the analytical solution involves evaluation of series of hyperbolic functions. Performing these calculations with sufficient precision requires taking into account 50 or more terms, when each term (hyperbolic function) should be evaluated with high floating-point precision (i.e., 128 or even 256 decimal places). Therefore, even for cases where analytical solution can be obtained, the numerical solution is advantageous for investigation of a wide range of the parameter values. In the present work, both the "$B_{0|y}$-problem" (Sec.\ \ref{Sec: y-problem}) and "$B_{0|x}$-problem" (Sec.\ \ref{Sec: x-problem}) are solved numerically using the finite-volume method described in \cite{Barmak25}, where more details can be found on the discretization and solution procedure.

{\renewcommand{\arraystretch}{1.5}
	\begin{table}[h!]
		\caption{\label{Tab: Grid convergence_y_problem} Grid convergence for the numerical solution of "$B_{0|y}$-problem" (vertical external magnetic field). Mercury--air stratified flow in a square duct ($AR=1$) with conducting bottom and insulating side walls.}
		\begin{tabular*}{\textwidth}{@{\extracolsep{\fill}}l l c c c c@{\extracolsep{\fill}}}
			\hline%
			\hline
			\multirow{3}{*}{$N_x$} & 
			\multirow{3}{*}{$N_y$} &
			\multicolumn{4}{c}{$Q_{2 1}$} 
			\\
			\cmidrule{3-6}
			& & 
			\multicolumn{2}{c}{$h=0.7$} &
			\multicolumn{2}{c}{$h=0.9$} 
			\\
			\cmidrule{3-6}
			& & $\Ha=5.181$ & $\Ha=103.625$ & $\Ha=5.181$ & $\Ha=103.625$
			\\
			\hline
			400  & 400  & 6.1060 & 428.5151 & 0.2652 & 11.9021
			\\
			& 600       & 6.1057 & 431.1265 & 0.2652 & 11.9085
			\\
			& 800       & 6.1056 & 432.0512 & 0.2652 & 11.9805
			\\
			800 & 800   & 6.1054 & 432.2892 & 0.2652 & 11.9873
			\\
			400 & 1000  & 6.1052 & 433.3095 & 0.2652 & 11.9952 
			\\
			1000 &       & 6.1049 & 433.5092 & 0.2653 & 12.9995					
			\\
			\hline
			\multicolumn{2}{@{}c@{}}{\textbf{Analytical}} &
			\textbf{6.1108}	& \textbf{436.1472} & \textbf{0.2654}	& \textbf{12.0323} 
			\\
			\hline%
			\hline
		\end{tabular*}
	\end{table}
}
{\renewcommand{\arraystretch}{1.5}
	\begin{table}[h!]
		\caption{\label{Tab: Grid convergence_x_problem} Grid convergence for the numerical solution of "$B_{0|x}$-problem" (horizontal external magnetic field). Mercury--air stratified flow in a conducting square duct ($AR=1$).}
		\begin{tabular*}{\textwidth}{@{\extracolsep{\fill}}l l c c c c@{\extracolsep{\fill}}}
			\hline%
			\hline
			\multirow{3}{*}{$N_x$} & 
			\multirow{3}{*}{$N_y$} &
			\multicolumn{4}{c}{$Q_{2 1}$} 
			\\
			\cmidrule{3-6}
			& & 
			\multicolumn{2}{c}{$h=0.7$} &
			\multicolumn{2}{c}{$h=0.9$} 
			\\
			\cmidrule{3-6}
			& & $\Ha=5.181$ & $\Ha=103.625$ & $\Ha=5.181$ & $\Ha=103.625$
			\\
			\hline
			400  & 400  & 6.2949 & 245.9506 & 0.3345 & 16.4651
			\\
			600  &      & 6.2952 & 246.5515 & 0.3345 & 16.5482
			\\
			800  &      & 6.2955 & 246.6815 & 0.3347 & 16.5519
			\\
			800 & 800   & 6.2957 & 246.7115 & 0.3348 & 16.5731	
			\\
			1000 & 400  & 6.2968 & 246.8212 & 0.3349 & 16.6143
			\\
			& 1000      & 6.2999 & 246.9127 & 0.3349 & 16.5921					
			\\
			\hline
			\multicolumn{2}{@{}c@{}}{\textbf{Analytical}} &
			\textbf{6.3017}	& \textbf{247.3914} & \textbf{0.3351}	& \textbf{16.639} 
			\\
			\hline%
			\hline
		\end{tabular*}
	\end{table}
}

Table\ \ref{Tab: Grid convergence_y_problem} and Table\ \ref{Tab: Grid convergence_x_problem} demonstrate that for both "$B_{0|y}$-problem" and "$B_{0|x}$-problem", respectively, with an increase in the number of computational cells, the numerical solution for a square duct (i.e., $AR=1$), in particular flow rate ratio for a given holdup, converges to the analytical solution obtained for particular combinations of bottom and side wall conductivities (see Appendix \ref{Sec: appendix_analyt_vertical} and Appendix \ref{Sec: appendix_analyt_horizontal}). Except for large Hartmann numbers (i.e., $\Ha > 100$), a computational grid with $\displaystyle N_x=N_y=400$ is found to be sufficient to predict holdup and pressure gradient up to three decimal digits compared to the analytical solution (resulting in less than $0.1\%$ deviation in most cases). For high $\Ha$, a finer computational grid with $N_x=N_y=1000$ is adopted.

\section{Results and discussion}\label{Sec: Results} 

The model equations presented in Sec.\ \ref{Sec: Formulation} are used to investigate how the characteristics of two-phase stratified flow of a gas and a conductive liquid in horizontal rectangular ducts of various aspect ratios and different wall conductivity configurations are affected by the intensity and orientation of an external magnetic field. Since the upper (gas) phase is non-conductive, only the electrical conductivity of the bottom and the side-walls, which are in contact with the conductive phase, affect the flow characteristics. To explore the effect of conductivities of the walls, we consider various combinations of the limiting cases of insulating and perfectly conducting bottom and side walls. As a representative two-fluid system, mercury and air are chosen as the conductive liquid and gas, respectively.  Their properties are summarized in Table\ \ref{Tab: Properties}. The range of intensities of external magnetic fields considered in this work is 0.01 to 0.2 Tesla. This range represents the magnetic fields that can be obtained in standard laboratory conditions, whereby the corresponding Hartmann numbers are $\Ha \in [5.181, 103.62]$ in ducts of $H=0.02$m height. However, the results obtained are dimensionless and independent on the duct height.  
{\renewcommand{\arraystretch}{2}
	\begin{table}[h!]
		\caption{\label{Tab: Properties} Properties of mercury (Hg) and air at standard conditions.}	
		\begin{tabular*}{\textwidth}{@{\extracolsep{\fill}}ccccccc@{\extracolsep{\fill}}}
			\hline%
			\hline				
			& $\rho$ & $\eta$ & $\sigma_e$ & $\Pran_m$ & $\eta_{1 2}=\dfrac{\eta_1}{\eta_2}$
			\\				
			& $\bigg[\dfrac{kg}{m^3}\bigg]$ & $\bigg[\dfrac{kg}{m \cdot s}\bigg]$ & $\bigg[\dfrac{1}{\Omega \cdot m}\bigg]$ & &
			\\
			\hline
			Mercury (1) & 13,534.0 & 0.00149 & 1.0$\times 10^{6}$ &
			1.383$\times 10^{-7}$  &
			\multirow{2}{*}{82.778}					
			\\
			Air (2) & 1.0 & 1.818$\times 10^{-5}$ & 0 & 0 
			\\
			\hline%
			\hline
		\end{tabular*}
	\end{table}
}

\subsection{Vertical external magnetic field} \label{Sec: y-problem} 

\begin{figure}[h!]
	\centering
	\subfloat[]{\includegraphics[width=0.49\textwidth,clip]{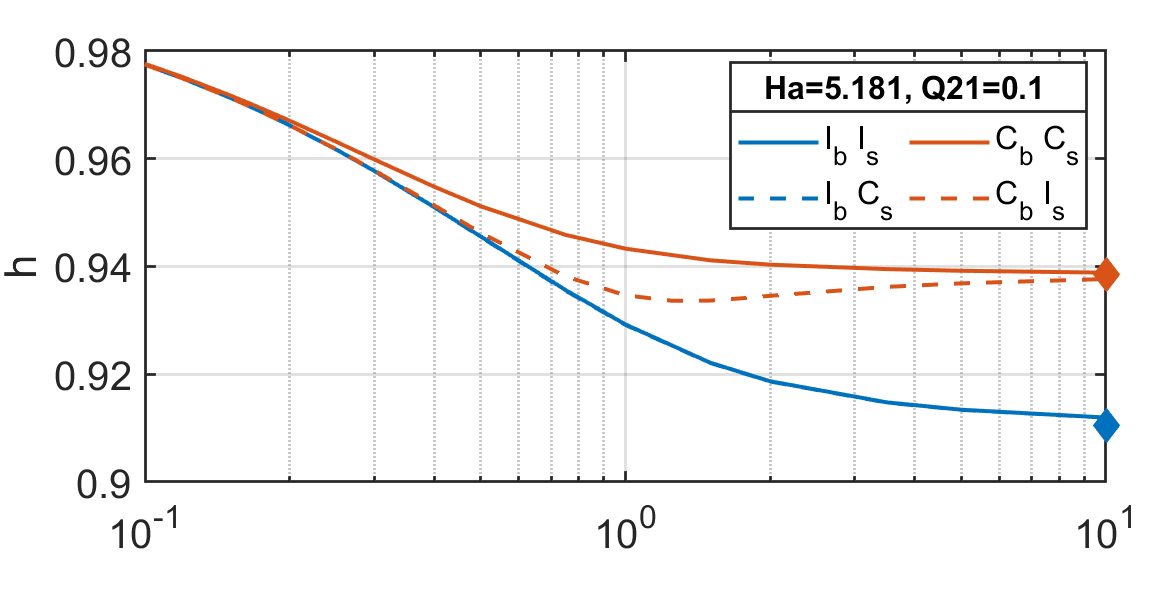}}
	\subfloat[]{\includegraphics[width=0.49\textwidth,clip]{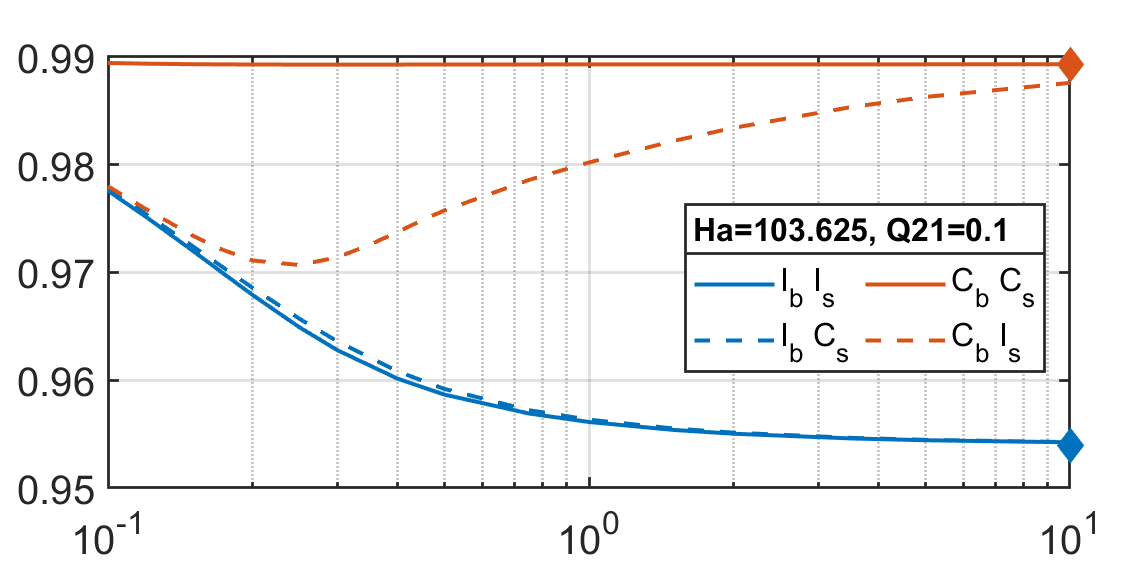}}
	\\
	\subfloat[]{\includegraphics[width=0.49\textwidth,clip]{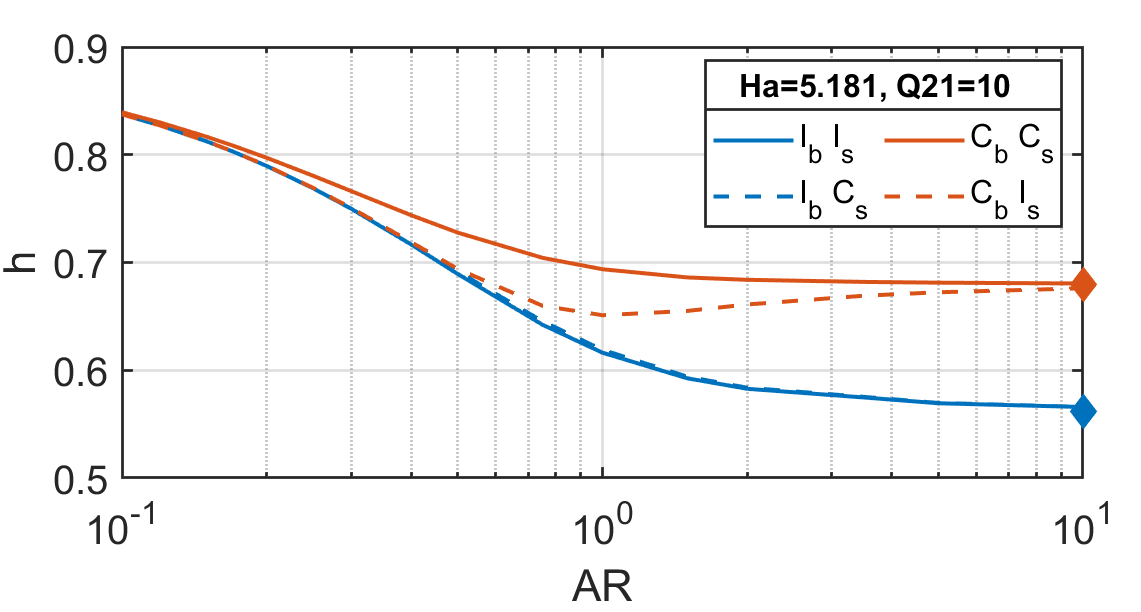}}
	\subfloat[]{\includegraphics[width=0.49\textwidth,clip]{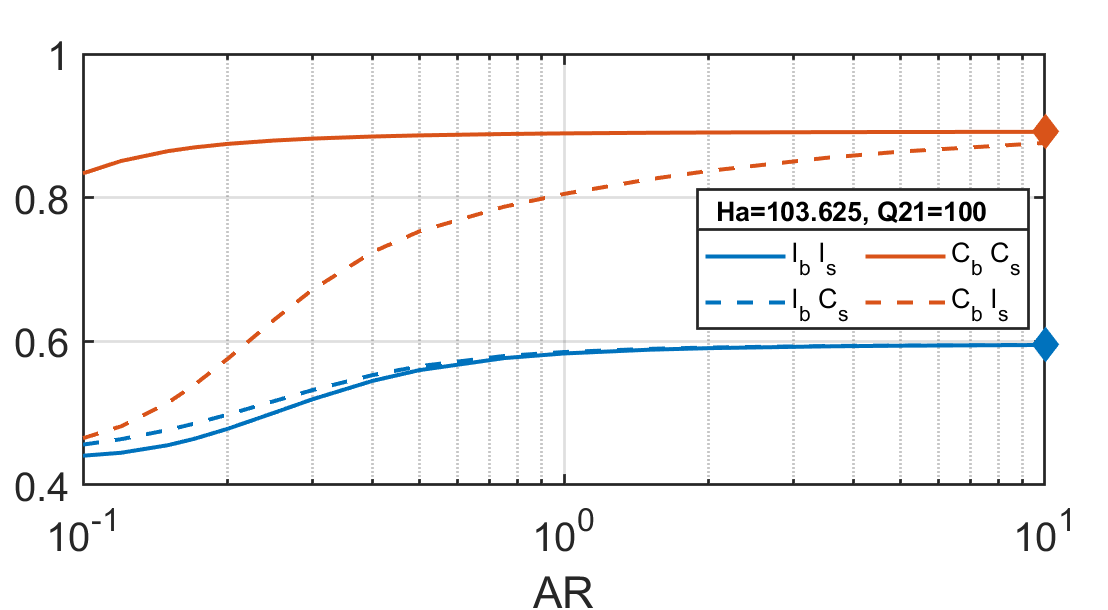}}
	\caption{\label{Fig: 4_1_h_effect_AR}Effect of the aspect ratio, $AR$, on the holdup in ducts with different wall conductivities. (a) $\Ha=5.181$; (b) $\Ha=103.625$. The red (blue) diamond represent the two-plate model value of the holdup in case of conductive (insulating) bottom wall \cite{Barmak25}.}
\end{figure}

An important characteristic of the two-phase flow is the holdup of the liquid, which in turn affects all the other flow characteristics. Figure\ \ref{Fig: 4_1_h_effect_AR} is obtained for a vertical external magnetic field ($B_{0|x}=0$) and demonstrates the effect of the duct aspect ratio on the conductive fluid holdup for four different configurations of the wall conductivities, which include: insulating bottom wall with either insulating or perfectly conducting side walls (denoted in Fig.\ \ref{Fig: 4_1_h_effect_AR} and below as $I_bI_s$ and $I_bC_s$ and represented by blue solid and dashed lines, respectively),  and perfectly conducting bottom wall with either perfectly conducting or insulting side walls (denoted as $C_bC_s$ and $C_bI_s$, red solid and dashed lines, respectively). The results presented are for low and high Hartmann numbers ($\Ha = 5.184$ and $103.625$, shown in the left and right-hand frames, respectively), with the top frames representing relatively low air-to-mercury flow-rate ratios, $Q_{21}=0.1$, and the bottom frames are for higher $Q_{21}$ values.  It can be observed that once the aspect ratio, $AR$, becomes less than $10$, the conductive-liquid holdup starts deviating significantly from the two-plate limit studied in \cite{Barmak25} for insulating or ideally conductive bottom walls.
The trends of the holdup variation with the $AR$ in fully conducting ($C_bC_s$, solid red line) and fully insulating ($I_b I_s$, solid blue line) ducts are similar. In both cases modest effect of the aspect ratio is observed for $AR\gtrsim3$, where the values converge and gradually approach the asymptotic limit predicted by the two-plate (TP) model values (indicated by red and blue diamonds). In general, for low-$\Ha$ flows, the holdup decreases with increasing aspect ratio across most of the practical range of $Q_{21}$ ($\leq100$). In contrast, at high $\Ha$ and $Q_{21}$ the opposite trend emerges, where the holdup increases with aspect ratio at high  values (Fig.\ \ref{Fig: 4_1_h_effect_AR}d).

The results indicate that when the bottom wall is insulating, the side-wall conductivity has a negligible influence on the holdup (Fig.\ \ref{Fig: 4_1_h_effect_AR}, solid and dashed blue lines). A small difference is found only in Fig.\ \ref{Fig: 4_1_h_effect_AR}d for high $\Ha=103.625$ and $Q_{21}=100$, where the holdup becomes larger for the ducts with the conducting side walls for $AR$ lower than 1. On the other hand, with conductive bottom wall, replacing the conducting side walls by insulating walls strongly influences the holdup (compare solid and red dashed lines). In the latter, the variation of the holdup with $AR$ is nonmonotonic with a minimum value achieved in range of $AR$ between 0.1 and 2. The exception is the case of high $\Ha$ and $Q_{21}$ (bottom right frame), where the holdup grows with aspect ratio, but is significantly lower than for fully conducting duct.

\begin{figure}[h!]
	\centering
	\subfloat[]{\includegraphics[width=0.33\textwidth,clip]{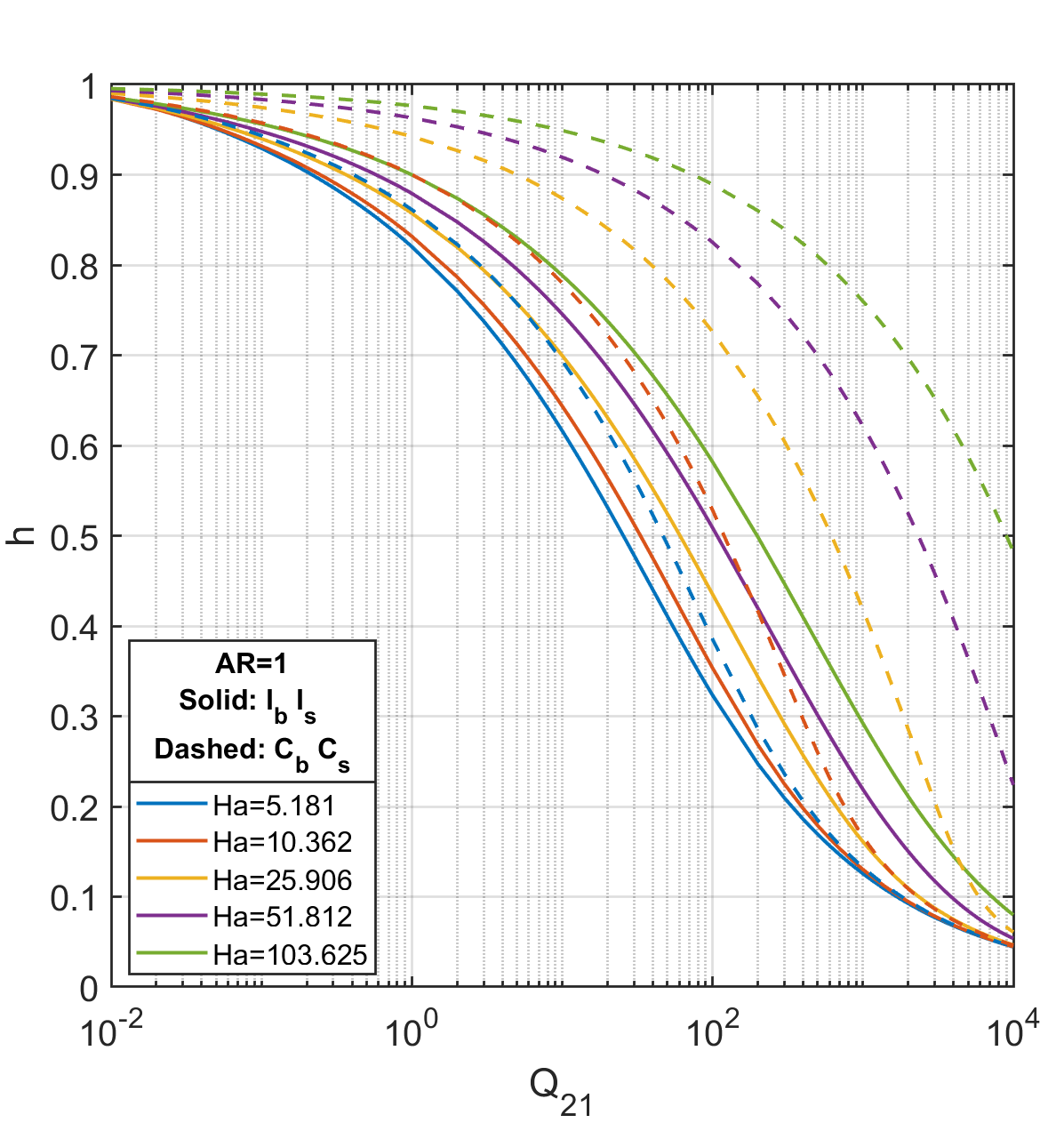}}
	\subfloat[]{\includegraphics[width=0.33\textwidth,clip]{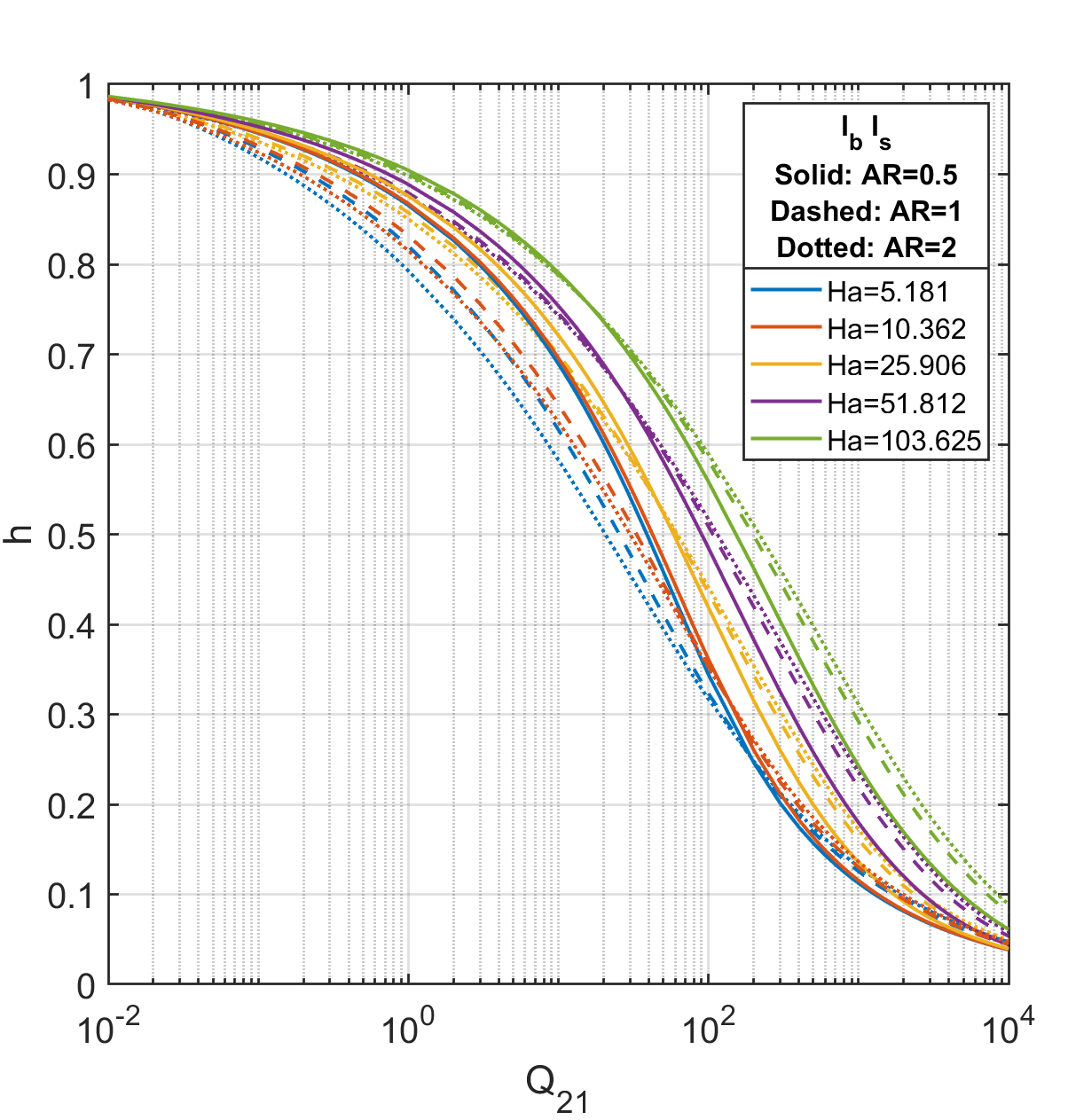}}
	\subfloat[]{\includegraphics[width=0.33\textwidth,clip]{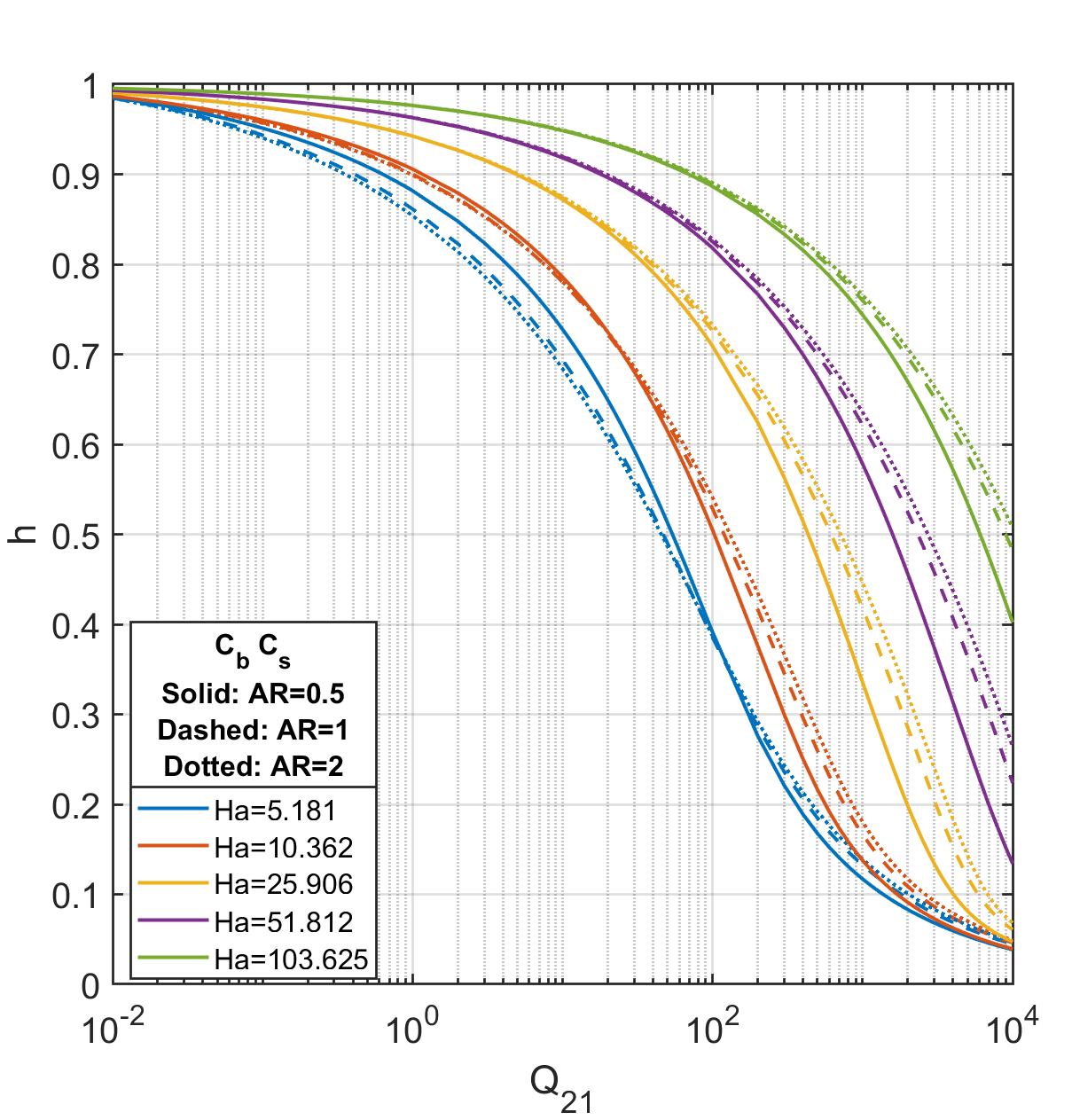}}
	\caption{\label{Fig: 4_3_h_effect_Q21}Effect of $\Ha$ and $Q_{21}$ on the mercury holdup (a) Fully insulating ($I_bI_s$, solid line) and perfectly conducting ($C_bC_s$, dashed line) square ducts. (b,c) Aspect ratio effect $AR=0.5,1,2$: (b) Fully insulating ducts  (c) Perfectly conducting ducts.}
\end{figure}

\begin{figure}[h!]
	\centering
	\subfloat[]{\includegraphics[width=0.49\textwidth,clip]{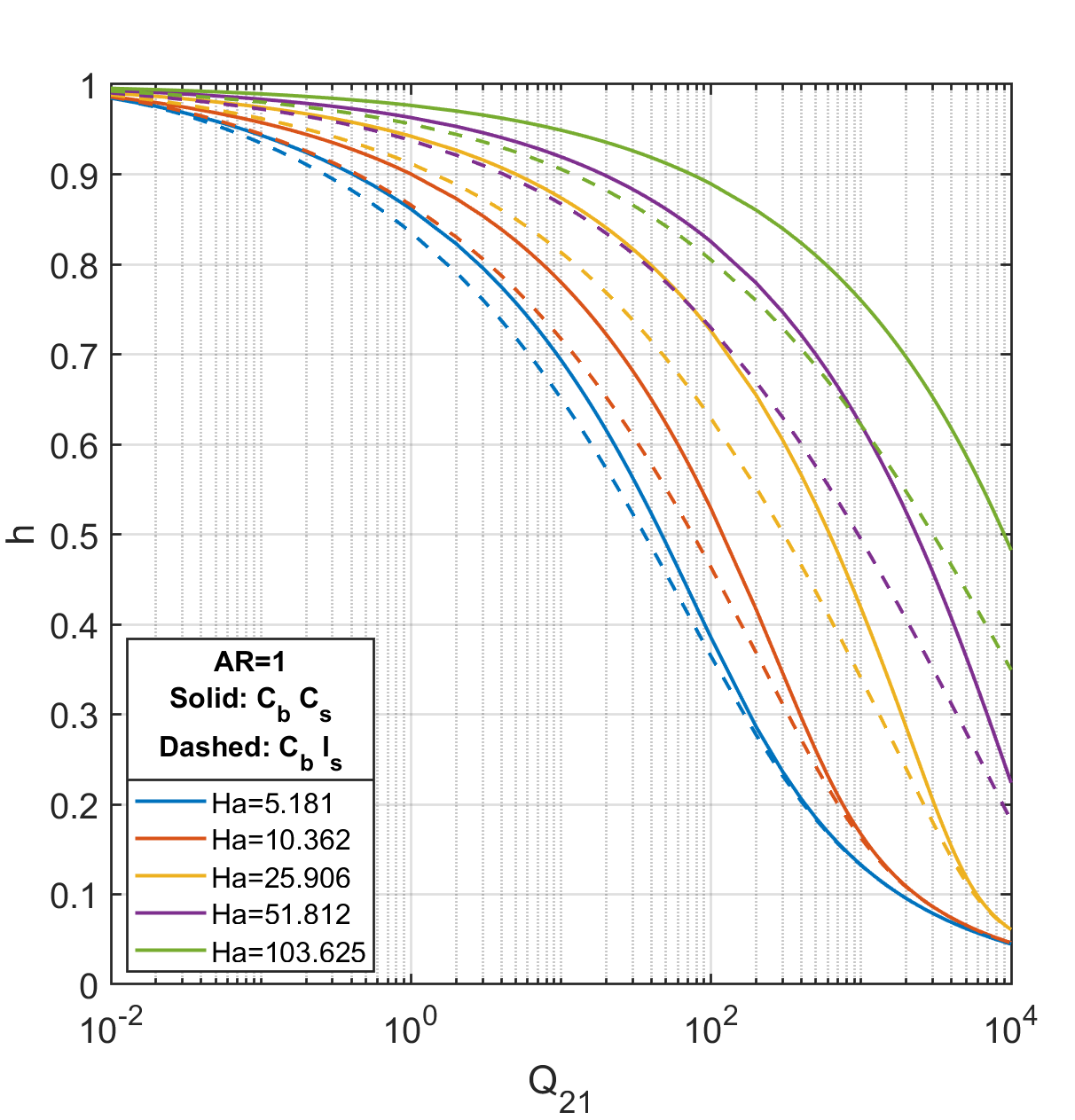}}
	\subfloat[]{\includegraphics[width=0.49\textwidth,clip]{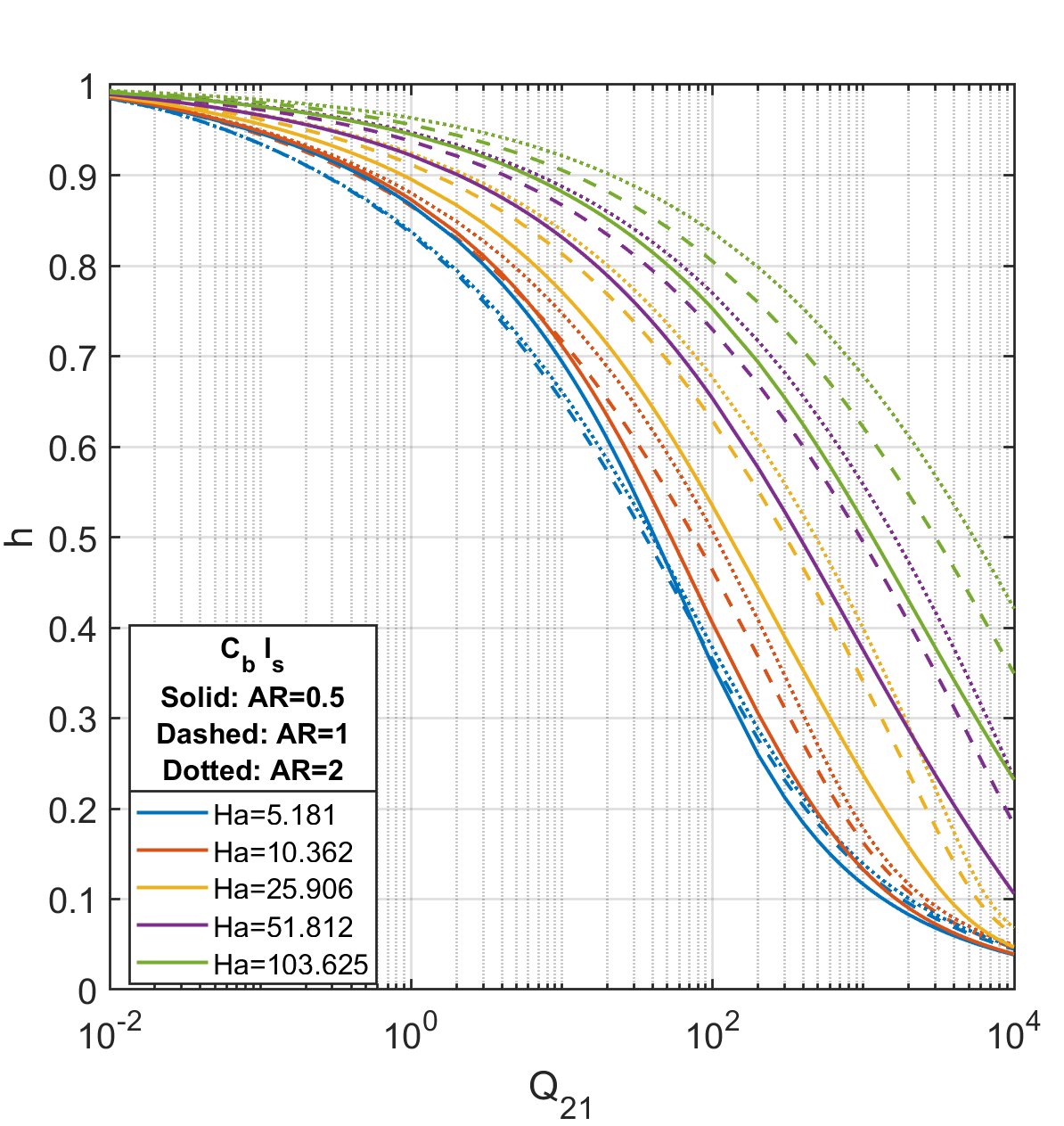}}
	\caption{\label{Fig: 4_4_h_effect_Q21_Cb}Effect of $\Ha$ and $Q_{21}$ on the mercury holdup in ducts with perfectly conducting bottom wall and insulating side walls, ($C_bI_s$) (a) Comparison with perfectly conducting ($C_bC_s$, solid line, $C_bI_s$, dashed line) square ducts. (b) Aspect ratio effect, $AR=0.5,1,2$ in $C_bI_s$ ducts.}
\end{figure}

The dependence of holdup on the flow-rate ratio and Hartman number is examined in Fig. \ref{Fig: 4_3_h_effect_Q21}. Figure \ref{Fig: 4_3_h_effect_Q21}a compares the holdup in $I_bI_s$ and $C_bC_s$ square  ducts ($AR=1$). Given the walls' conductivity and Hartmann number, the holdup is uniquely determined by the flow-rate ratio and decreases with increasing $Q_{21}$. Since the Lorentz force slows down the conductive liquid, its holdup increases with increasing $\Ha$, regardless of walls conductivity. However, for the same $\Ha$ and $Q_{21}$, the retarding Lorentz force in an insulated duct is smaller (see discussion with reference to Figs.\ \ref{Fig: 4_6_b_contours} and \ref{Fig: 4_7_bmax_effect_Q21} below), and therefore a smaller flow area (i.e., smaller holdup) is required to sustain the same flow rate of the conductive liquid compared with that in a conducting duct. The effect of wall conductivity on holdup becomes more pronounced at higher $\Ha$. In fact, in the conducting-duct case, mercury occupies nearly the entire flow cross-section even when $Q_{21}>1$. Comparing the results obtained in a square duct with two other aspect ratios corresponding to narrow ($AR=0.5$) and wide ($AR=2$) ducts (Fig. \ref{Fig: 4_3_h_effect_Q21}b and \ref{Fig: 4_3_h_effect_Q21}c for insulating and perfectly conducting walls, respectively), reveals that there is a range of flow rate ratios for every $\Ha$, for which  the holdup is almost the same for all aspect ratios considered (i.e., intersection of curves of the same color). This region shifts towards a lower value of $Q_{21}$ with an increase of Ha. Below this $Q_{21}$ value, the holdup decreases with $AR$, while above it, the trend is reversed, and the holdup is larger in wider ducts. 

The effect of flow-rate ratio on the holdup is further elaborated in Fig.\ \ref{Fig: 4_4_h_effect_Q21_Cb} for the case of conducting bottom wall, $C_b C_s$ and $C_bI_s$, for which the side-wall conductivity is found to significantly affect the two-phase flow characteristics. Figure\ \ref{Fig: 4_4_h_effect_Q21_Cb}a shows the case of a square duct, where the holdup is shown to be lower with insulated side walls, particularly at high Ha, where the influence of side-wall conductivity extends over a broader range of $Q_{21}$. The possible nonmonotonic dependence of holdup on the aspect ratio in $C_bI_s$ ducts around $AR\approx1$ for relatively low $\Ha$ (demonstrated in Fig. \ref{Fig: 4_3_h_effect_Q21}a,c above) is observed also in Fig.\ \ref{Fig: 4_4_h_effect_Q21_Cb}b. 

\begin{figure}[h!]
	\centering
	\subfloat[$h=0.82, U_{2|\text{max}}=9.251$]{\includegraphics[width=0.49\textwidth,clip]{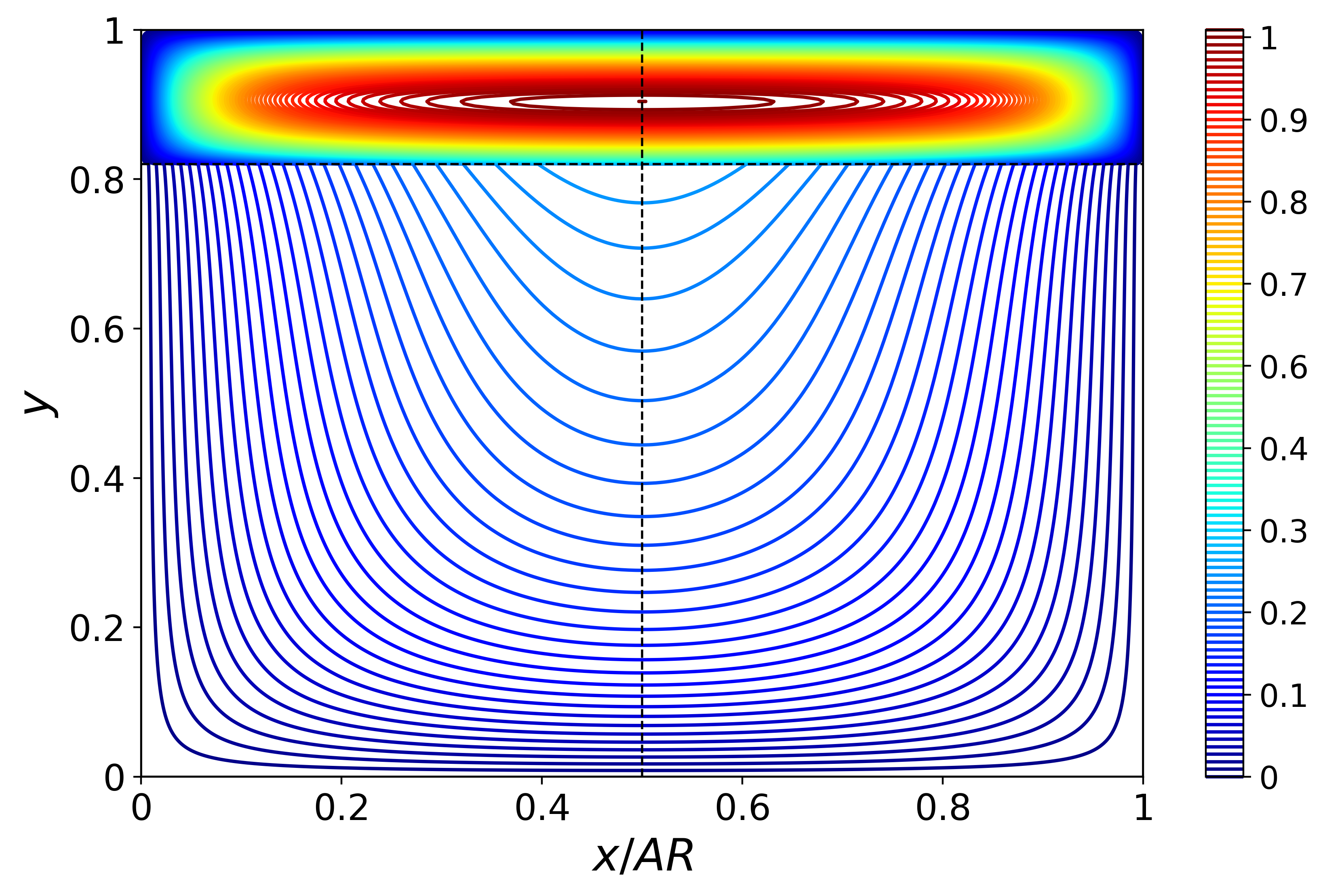}}
	\subfloat[$h=0.90,U_{1|\text{max}}=1.408  (U_{2|\text{max}}=15.747)$]{\includegraphics[width=0.49\textwidth,clip]{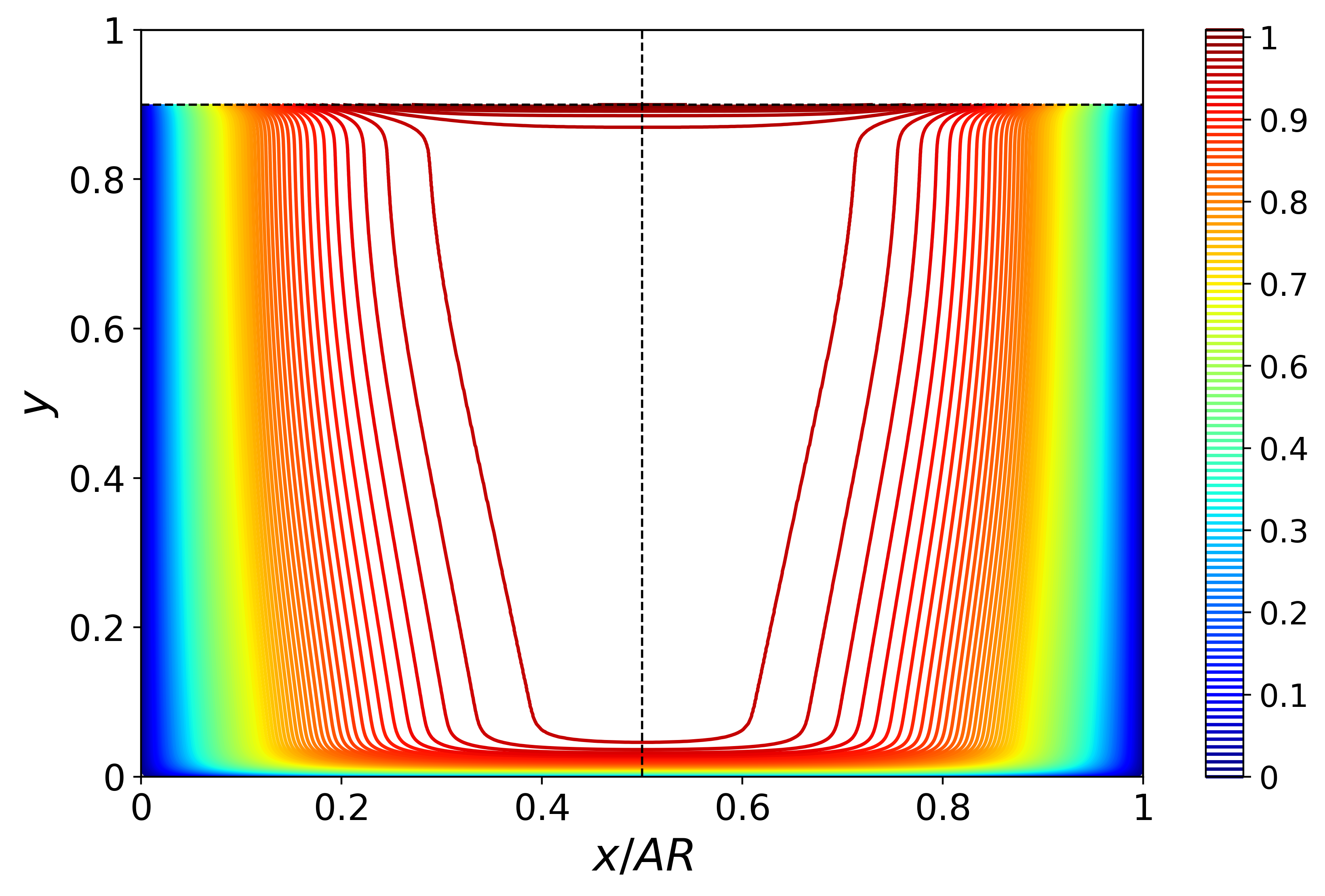}}
	\\
	\subfloat[$h=0.86, U_{2|\text{max}}=11.713$]{\includegraphics[width=0.49\textwidth,clip]{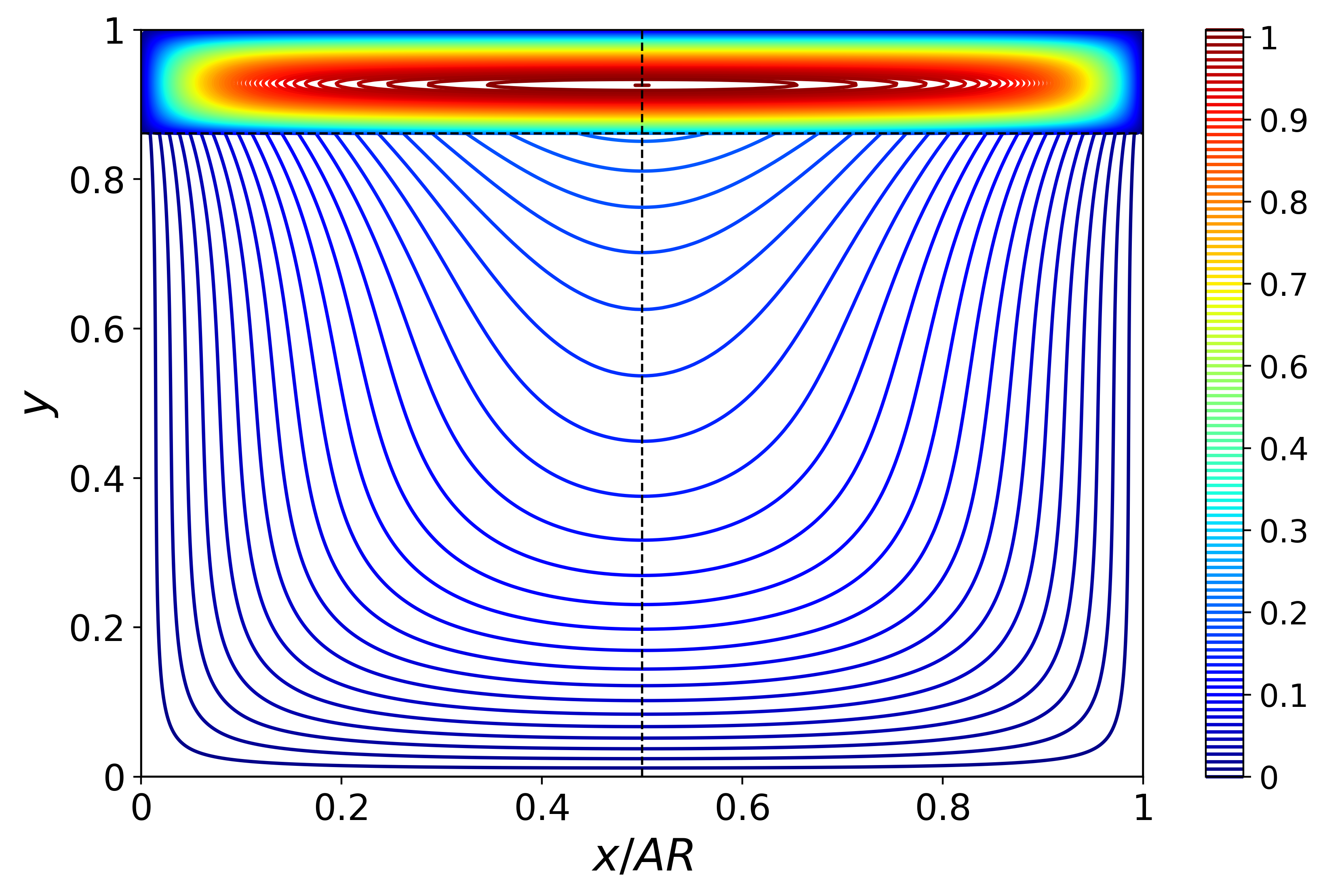}}
	\subfloat[$h=0.98,U_{1|\text{max}}=2.621  (U_{2|\text{max}}=64.237)$]{\includegraphics[width=0.49\textwidth,clip]{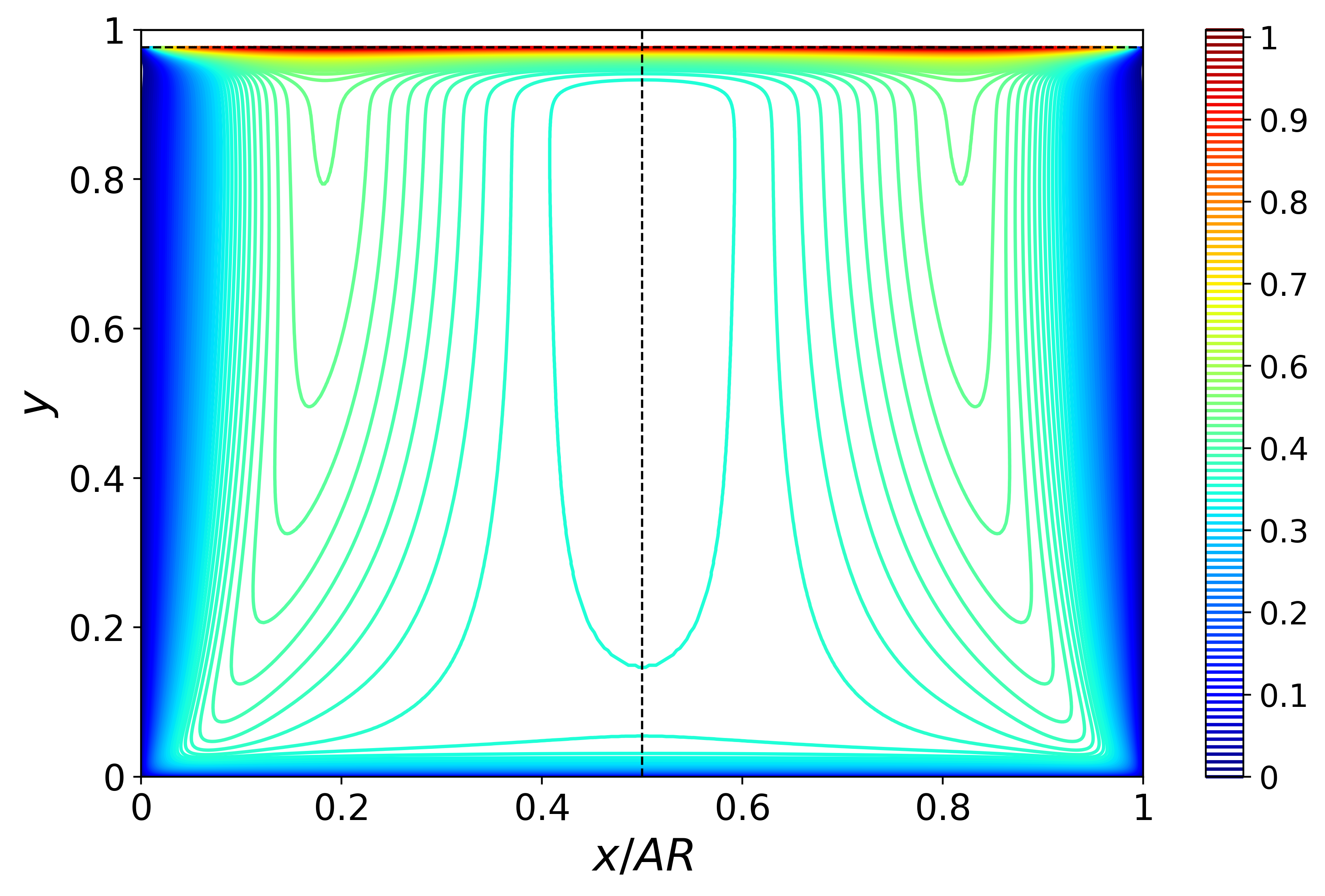}}
	\\
	\subfloat[Mercury flow, $\Ha=103.625$]{\includegraphics[width=0.49\textwidth,clip]{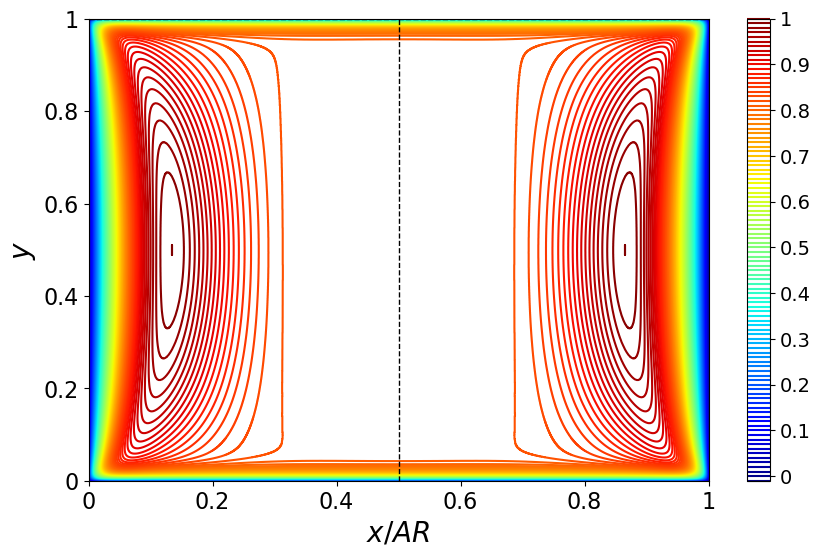}}
	\qquad
	\subfloat[Single-phase vs. Interfacial velocity]{\includegraphics[width=0.4\textwidth,height=0.38\textwidth,clip]{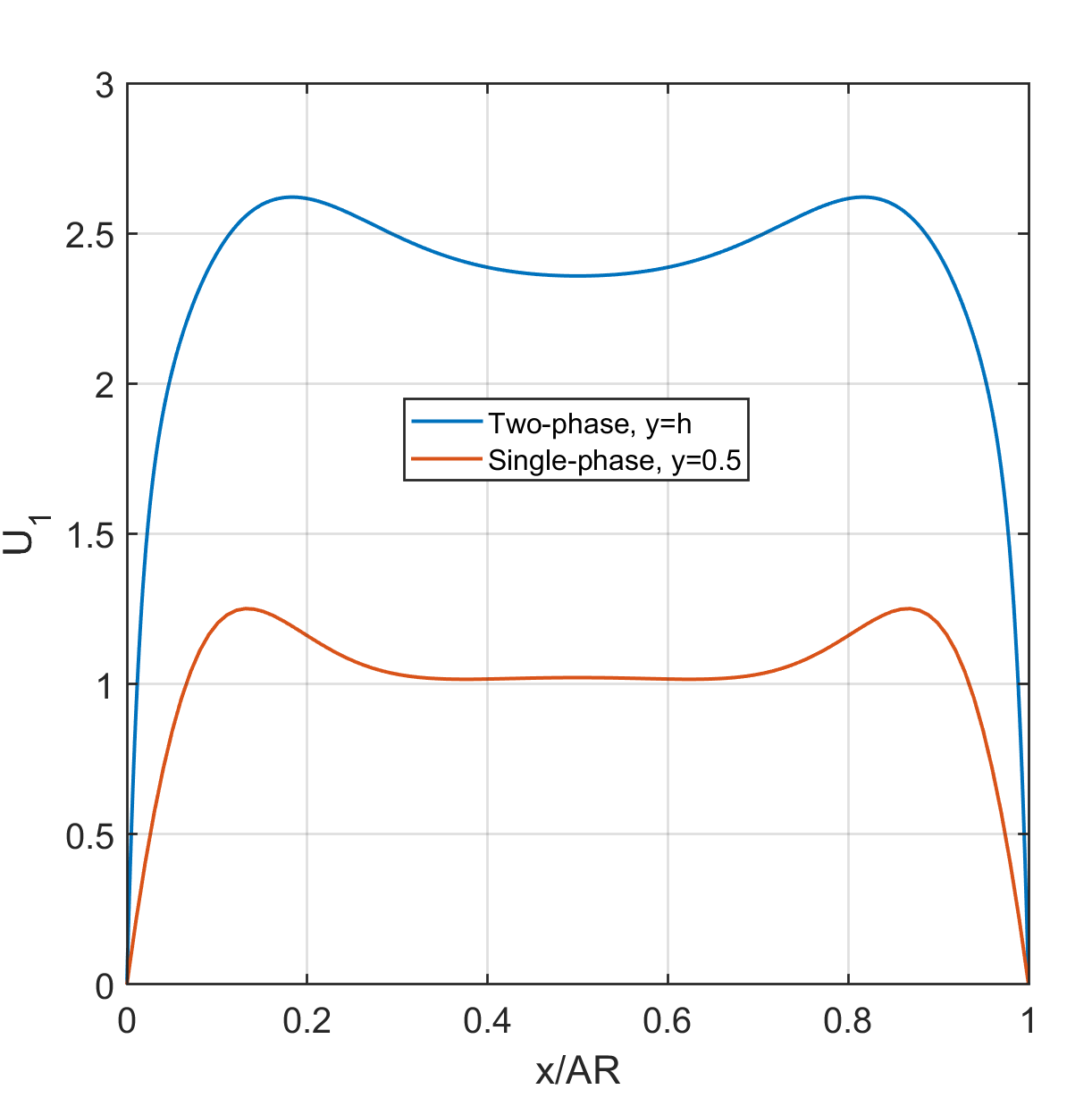}}
	\caption{\label{Fig: 4_5_U_contours}Velocity contours in a square duct for $Q_{21} = 1$: (a,b) all walls insulating, $I_bI_s$. (c,d) all wall conducting, $C_bC_s$. (a,c) $\Ha=5.181$, the velocities are scaled by the maximal velocity, which is located in the air layer. (b,d) $\Ha=103.625$, only the velocity contours in the mercury are shown (scaled by its maximal velocity). The corresponding $U_\text{max}$ values used for scaling are dimensionless (normalized by $U_{1S}$ ). (e,f) Single phase mercury flow in a conducting square duct, $\Ha=103.625$. (e) Velocity contours and (f) comparison of the single-phase velocity profile flow along the horizontal centerline ($y=0.5$) with the interfacial velocity profile ($y=h$) in mercury-air flow.}
\end{figure}

\begin{figure}[h!]
	\centering
	\subfloat[$\Ha=5.181, h=0.82, b_\text{max}/\Ha=0.135$]{\includegraphics[width=0.49\textwidth,clip]{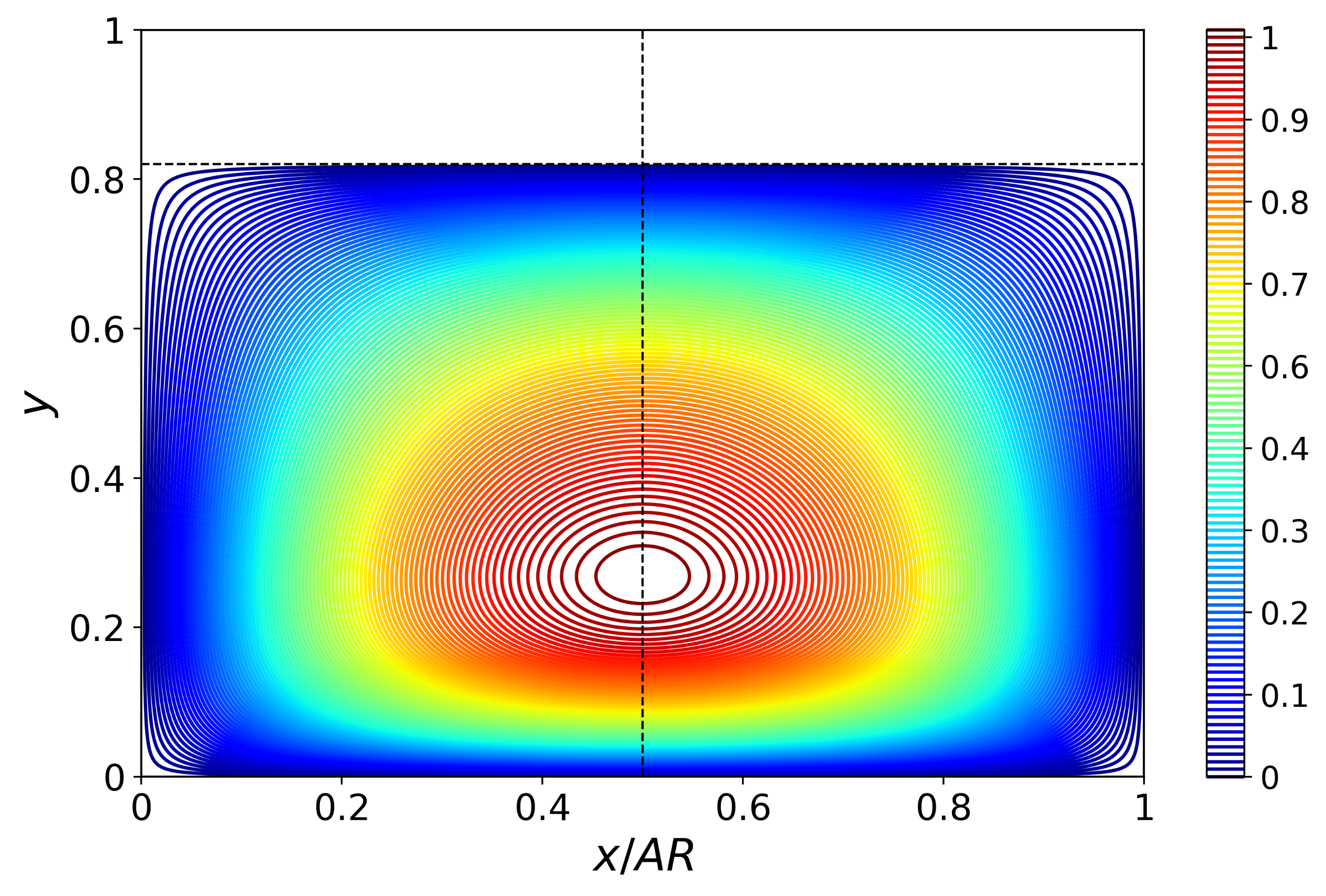}}
	\subfloat[$\Ha=103.625, h=0.90, b_\text{max}/\Ha=0.012$]{\includegraphics[width=0.49\textwidth,clip]{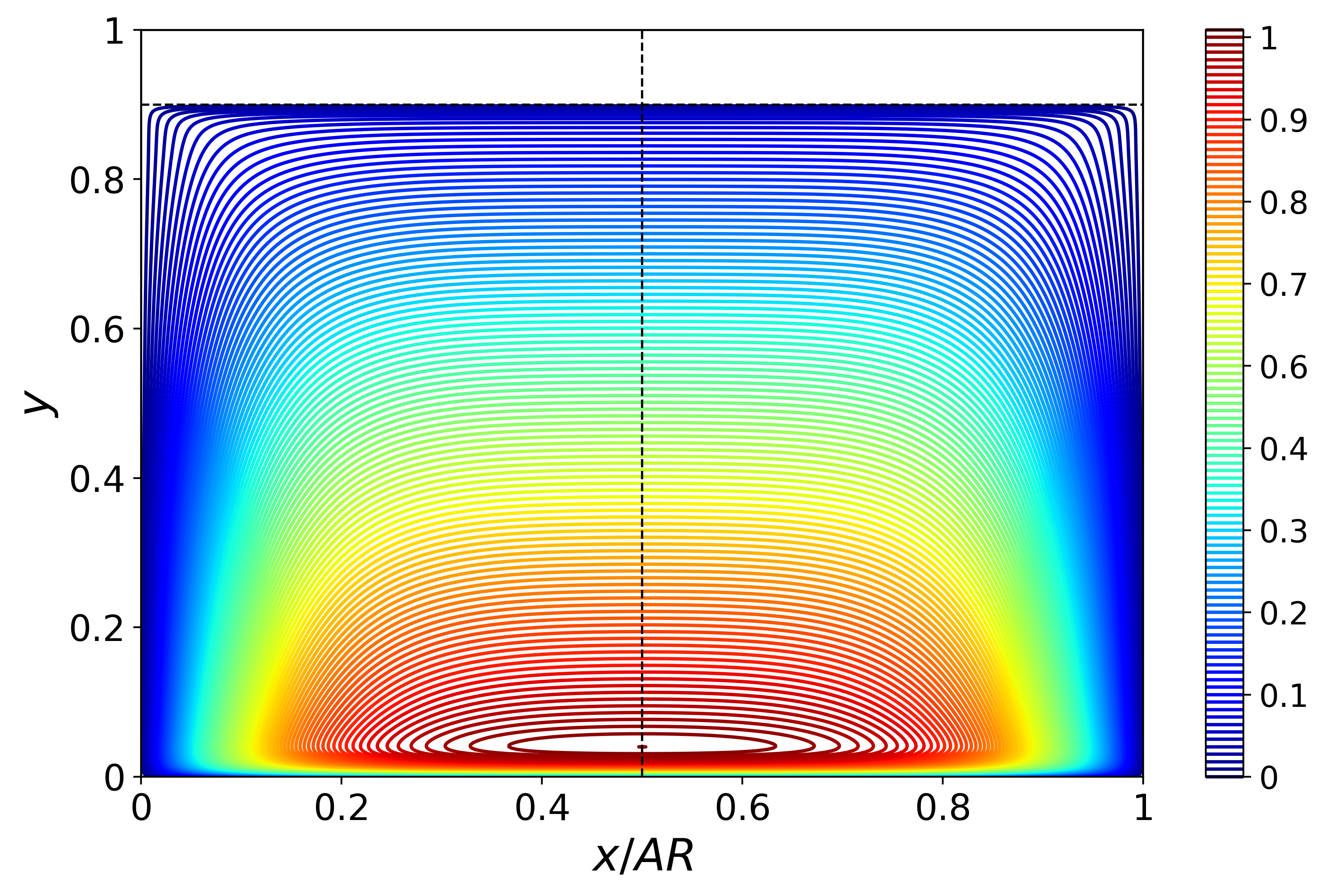}}
	\\
	\subfloat[$\Ha=5.181, h=0.86, b_\text{max}=1.047$]{\includegraphics[width=0.49\textwidth,clip]{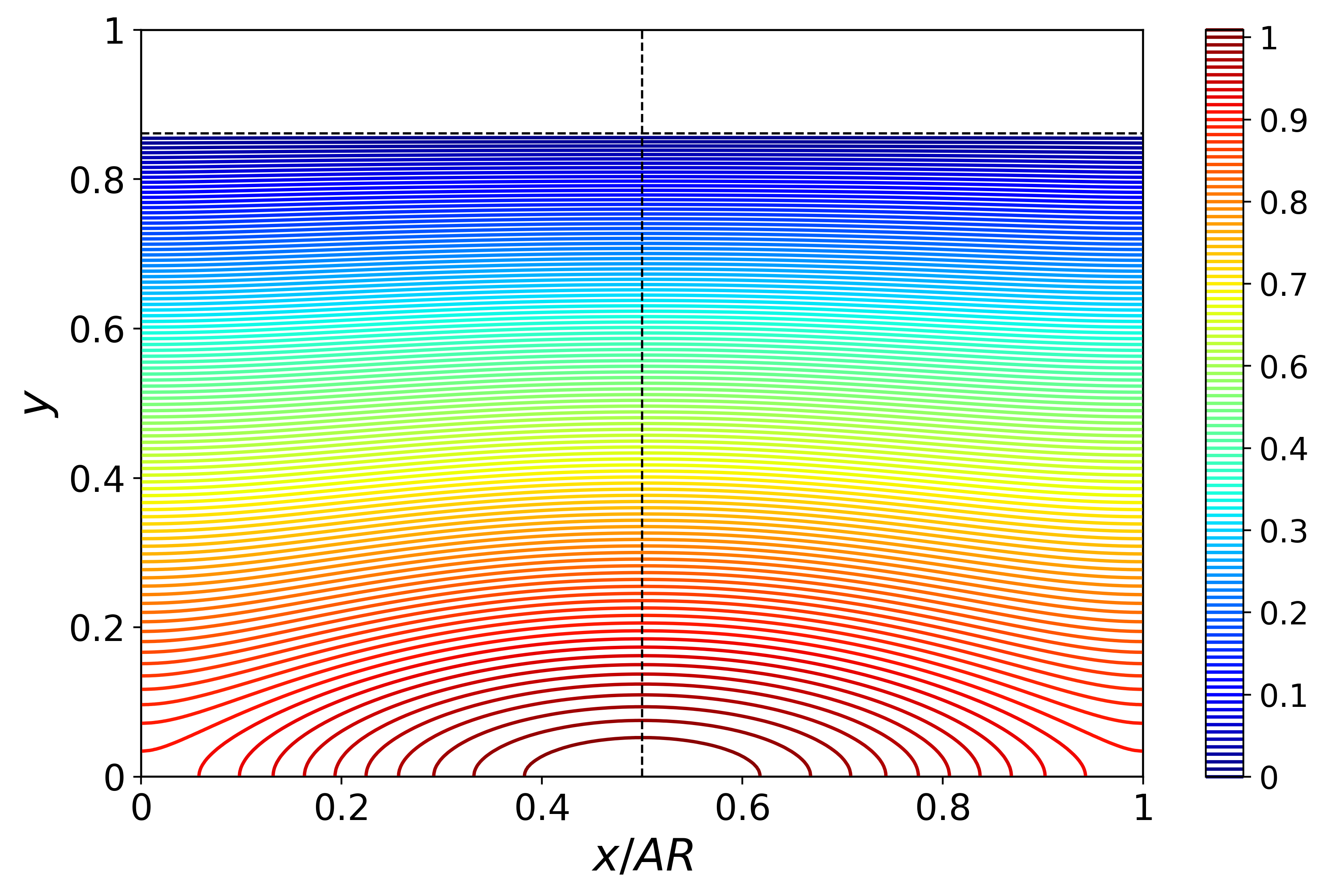}}
	\subfloat[$\Ha=103.625, h=0.86, b_\text{max}/\Ha=1.002$]{\includegraphics[width=0.49\textwidth,clip]{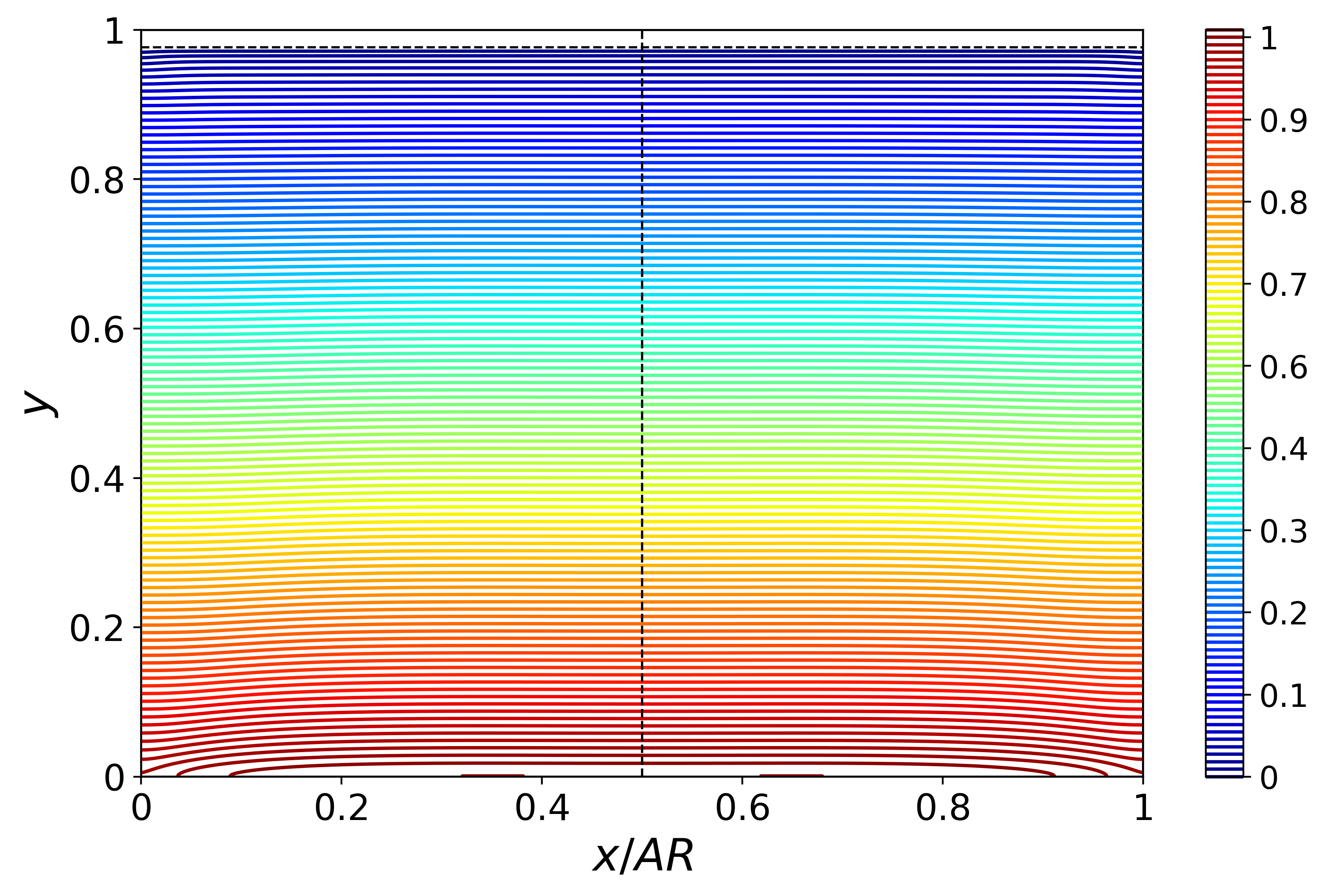}}
	\caption{\label{Fig: 4_6_b_contours}Contours of the induced magnetic field, $b/\Ha$ (scaled by its maximal value), corresponding to Fig.\ \ref{Fig: 4_5_U_contours}, in a square duct for $Q_{21} = 1$: (a,b) all walls insulating, $I_bI_s$. (c,d) all wall conducting, $C_bC_s$. }
\end{figure}
Velocity contours in square ducts cross section are shown in Fig.\ \ref{Fig: 4_5_U_contours} for a particular flow rate ratio, $Q_{21}=1$. Two cases of wall conductivity, i.e., all walls insulating (Fig.\ \ref{Fig: 4_5_U_contours}a,b) and all walls perfectly conducting (Fig.\ \ref{Fig: 4_5_U_contours}c,d), are compared for low and high Hartmann numbers (left- and right-hand frames, respectively). In the left-hand frames, velocity contours are normalized by the maximum velocity, which is reached in the air layer. Since the air is less viscous and nonconductive, it flows much faster than the mercury, yielding contour patterns similar to those in single-phase Poiseuille flow in a high-aspect-ratio channel, $\displaystyle AR=1/(1-h)$. In $C_bC_s$ ducts, the mercury holdup is larger than in $I_bI_s$ ducts at all $\Ha$. Therefore, the maximal air velocity (for the same flow rate ratio) is larger in this case, and so are the velocities at the interface, where the mercury is dragged by the air. This effect is even more pronounced for higher Hartmann numbers (Fig.\ \ref{Fig: 4_5_U_contours}b,d; only the mercury velocity scaled by its maximal value is shown). For $\Ha=103.625$, the mercury velocity contours appear significantly different for insulating and conducting ducts. A noticeable difference is the regions of higher velocity observed near the side walls ("jet"-like contour pattern), where the local velocity exceeds the velocity in the core. However, because the maximum mercury velocity in the conductive duct is nearly twice that in insulating one, the core velocities are, in fact, comparable in Fig.\ \ref{Fig: 4_5_U_contours}b and Fig.\ \ref{Fig: 4_5_U_contours}c. The Hartmann boundary layer at the bottom wall (i.e., perpendicular to the external magnetic field) and the thicker Shercliff boundary layers at the side walls, which are clearly visible in those figures, are also similar.  It is worth noting, however, that, while in $I_bI_s$ ducts the maximum mercury velocity occurs at the center of the interface, in $C_bC_s$ ducts, it is shifted closer to the side wall. As a result, the core velocity is less uniform in $C_bC_s$ ducts and the Shercliff boundary layer appears thinner near the channel bottom, whereas in $I_bI_s$ ducts their thickness remains nearly uniform.

Two off-center maxima in the velocity profile are also observed in single-phase flow of a conductive liquid in a perfectly conducting duct (e.g., Fig.\ \ref{Fig: 4_5_U_contours}e). In this case, the maxima are located symmetrically about the duct centerline in the $x$-direction, whereas in gas–liquid flow they are always situated at the interface. For the same $\Ha$ and $AR$, the maximum velocities in the two-phase flow are consistently higher than those obtained in the corresponding single-phase flow, but its overshoot compared to velocity at $x/AR=0.5$ is similar (see Fig.\ \ref{Fig: 4_5_U_contours}f). In both single-phase and two-phase flows, increasing $\Ha$ strengthens the influence of the conducting side walls, widening the range of $AR$ for which two maxima occur, shifting them toward the side walls, and increasing their magnitude. For the same $\Ha$ value, however, the transition from a single centerline maximum to two off-center maxima occurs at higher $AR$ in the two-phase flow.  For example, at $\Ha=10.362$ and $\Ha=103.625$, this transition occurs in single-phase mercury flow at $AR\approx1.2$ and $AR\approx0.43$, respectively, compared to $AR\approx1.7$ and $AR\approx0.53$ in the mercury--air flow.

Contours of the induced magnetic field, $b/Ha$ (i.e., dimensional induced magnetic field scaled by $B_{0|y}$ and $\Rey_m$), which represent the streamlines of the induced electrical current, are depicted in Fig.\ \ref{Fig: 4_6_b_contours}. They are associated with the velocity fields shown in Fig.\ \ref{Fig: 4_5_U_contours}. The values shown are normalized by $b_\text{max}/Ha$.  In case of the insulating bottom wall (Fig.\ \ref{Fig: 4_6_b_contours}a,b), the induced magnetic field contours resemble those obtained in single phase flow of a conductive fluid in a duct of somewhat larger aspect ratio $AR=1/h$ (see, e.g., \cite{Muller01}). However, the location of the maximal value is not in the center of the conductive layer but shifted towards the bottom wall. The direction of the Lorentz force is defined by a sign of the vertical gradient of the induced magnetic field, $\partial b/\partial y$, and varies across the conductive phase. It is positive in a lower part of the layer, i.e., $\partial b/\partial y>0$, where $b$ grows from 0 at the insulating wall to its maximum, so that the (local) Lorentz force and the pressure gradient act in the same direction to balance the viscous shear stresses. The maximal value of $\partial b/\partial y$ is reached at the bottom wall, while its minimal value ($<0$) is observed at the interface. At high $\Ha$, the maximum value of $b$ is close to the bottom wall, hence in a larger part of the conductive layer  $\partial b/\partial y<0$ and the Lorentz force acts opposite to the flow direction and is balanced by the pressure gradient.  Near the side walls, on the other hand, $\partial b/\partial y$ is close to zero across the whole conducting layer. Hence, the Lorentz force, which accelerates the flow over most of the bottom wall, becomes ineffective near the side walls, and the viscous shear in the Shercliff boundary layers are balanced by the pressure gradient.

\begin{figure}[h!]
	\centering
	\subfloat[]{\includegraphics[width=0.33\textwidth,clip]{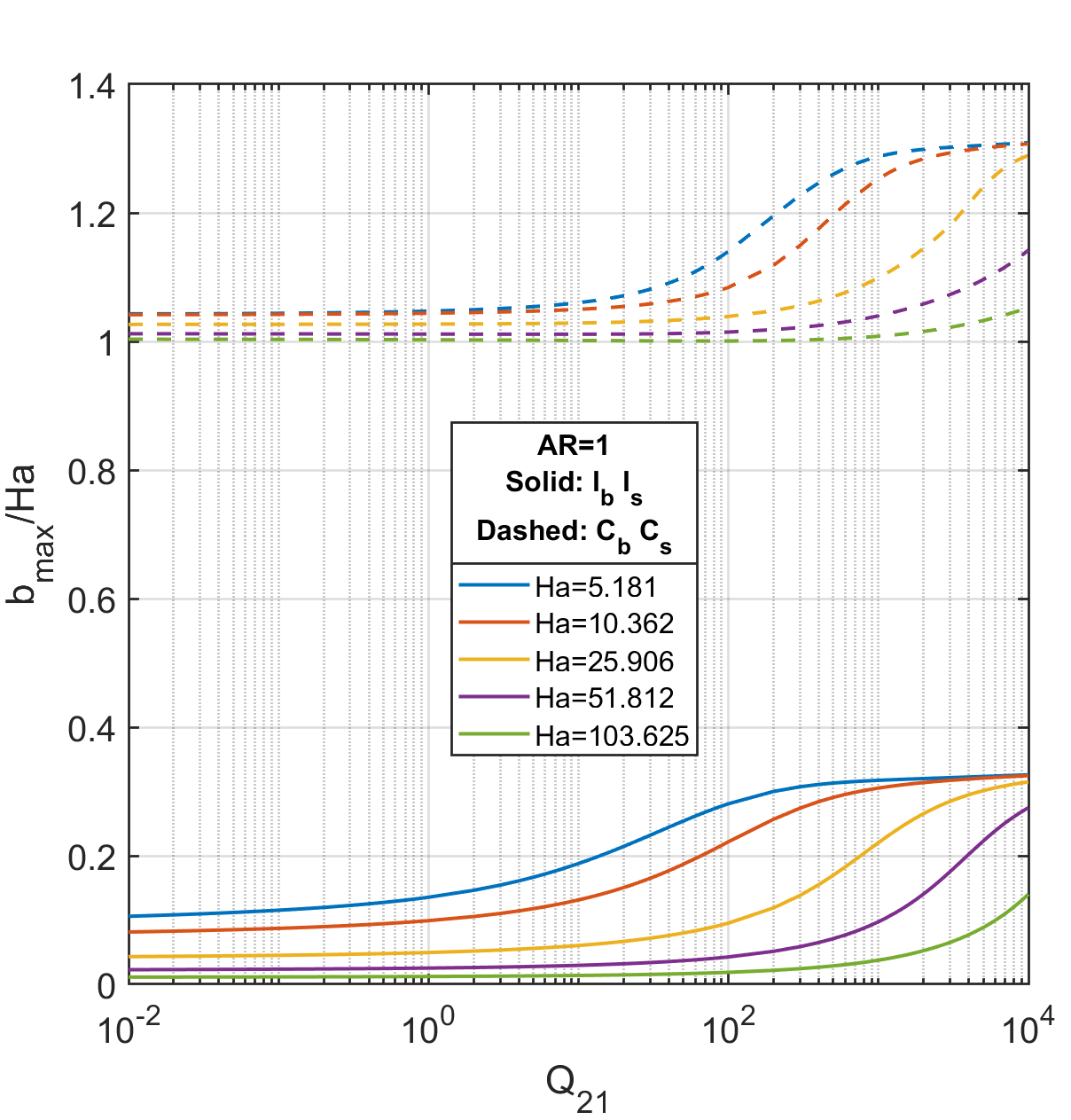}}
	\subfloat[]{\includegraphics[width=0.33\textwidth,clip]{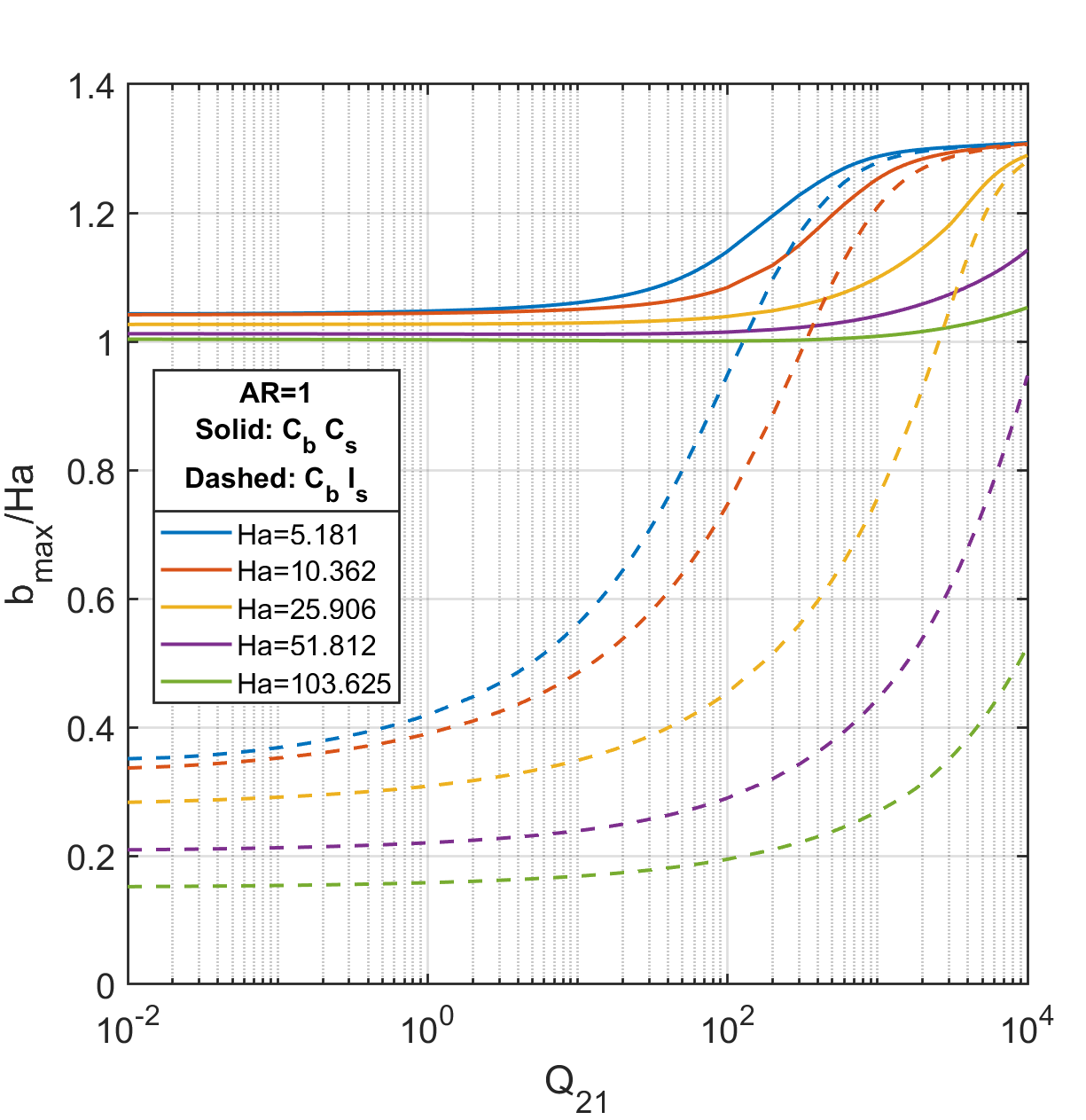}}
	\subfloat[]{\includegraphics[width=0.33\textwidth,clip]{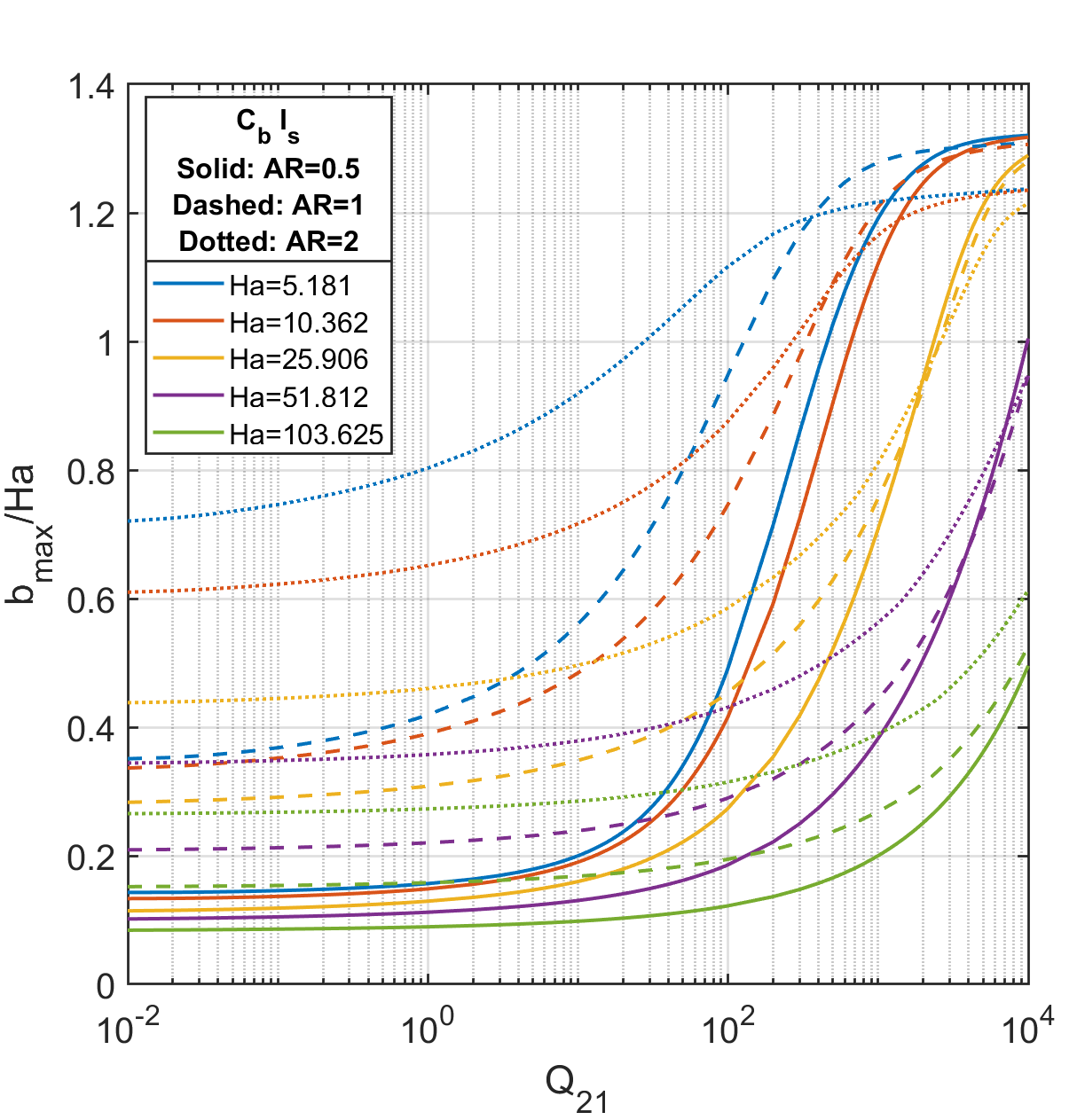}}
	\caption{\label{Fig: 4_7_bmax_effect_Q21}Effect of $\Ha$ and $Q_{21}$ on the maximal value of the induced magnetic filed, $b_\text{max}/Ha$. (a) Comparison of $I_bI_s$ and $C_bC_s$ square ducts. (b) Side wall effect in square ducts with a conductive bottom wall. (c) $AR$ effect in $C_bI_s$ ducts: $AR=0.5,1,2$.}
\end{figure}

Contours of the induced magnetic field for perfectly conducting walls are shown in Fig.\ \ref{Fig: 4_6_b_contours}c,d. Since the interface is the only non-conducting boundary of the conductive layer, the electric currents cannot cross it and they are closed through the zero-resistance side and bottom walls, producing field lines perpendicular to the boundaries. The pattern is analogous to single-phase flow \cite{Muller01} in the lower half of a duct with conducting walls, where $b=0$ along the horizontal centerline due to the problem symmetry. For large $\Ha$, the isolines are nearly parallel to the bottom wall, with corner distortions introduced by the side walls. The induced magnetic field intensity, $b/\Ha$, attains its maximum ($\approx1$) at the bottom wall and decreases almost linearly ($\partial b/\partial y \approx const <0$, except at the bottom surface where $\partial b/\partial y=0$), and vanishes at the insulating interface. Consequently, the Lorentz force acts opposite to the flow throughout the mercury layer. The slightly higher velocity compared to the core velocity (in the jet-like region near the side walls, Fig.\ \ref{Fig: 4_5_U_contours}c) arises from the weaker negative $\partial b/\partial y$ in these regions. The locally reduced Lorentz force results in the higher local velocity under the same pressure gradient set in the flow.
\begin{figure}[h!]
	\centering
	\subfloat[$h=0.84, U_{2|\text{max}}=9.997$]{\includegraphics[width=0.33\textwidth,clip]{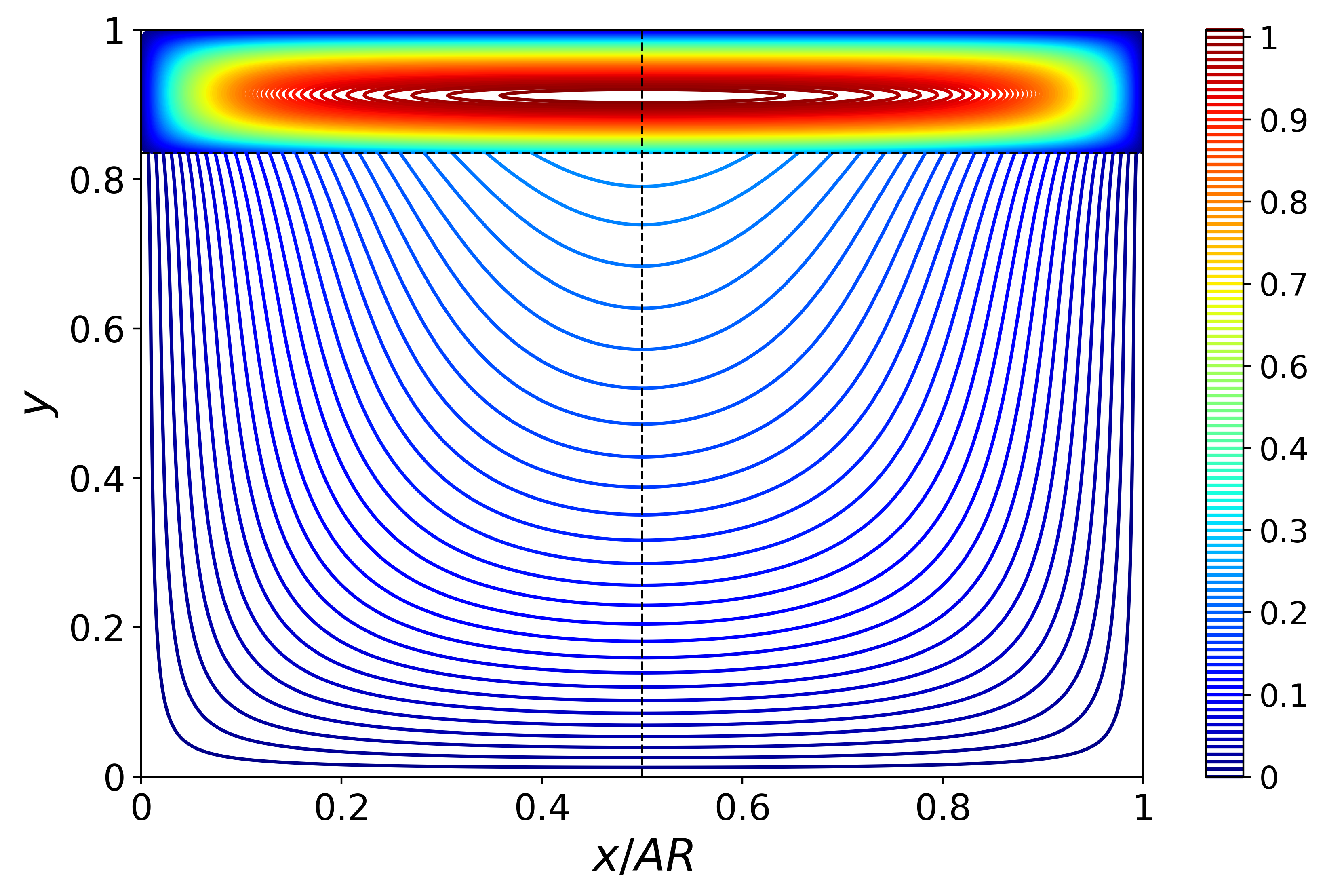}}
	\subfloat[$h=0.96$, $U_{1|\text{min}} = -0.087$, \\ $U_{2|\text{max}} = 35.654$]{\includegraphics[width=0.33\textwidth,clip]{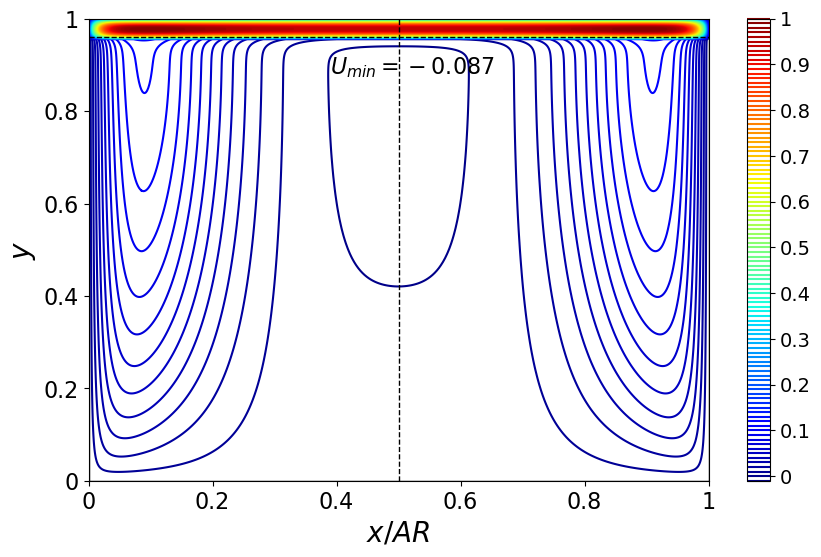}}
	\subfloat[$h=0.96, -0.087\le U_1 \le 4.367$]{\includegraphics[width=0.33\textwidth,clip]{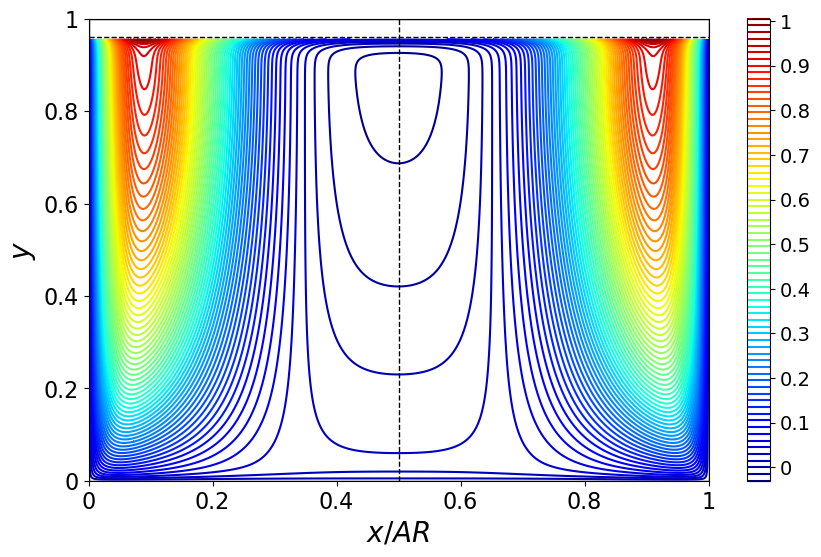}}
	\caption{\label{Fig: 4_8_U_contours_CbIs}Velocity contours in a square duct with a conducting bottom wall and insulating side walls, $C_bI_s$, $Q_{21}=1$. (a) $\Ha= 5.181$; (b,c) $\Ha=103.625$. (c) Only the mercury velocity contours are shown. In each subfigure, the velocity is scaled by its corresponding maximum value.}
\end{figure}
\begin{figure}[h!]
	\centering
	\subfloat[$h=0.84, b_\text{max}/\Ha = 0.419$]{\includegraphics[width=0.49\textwidth,clip]{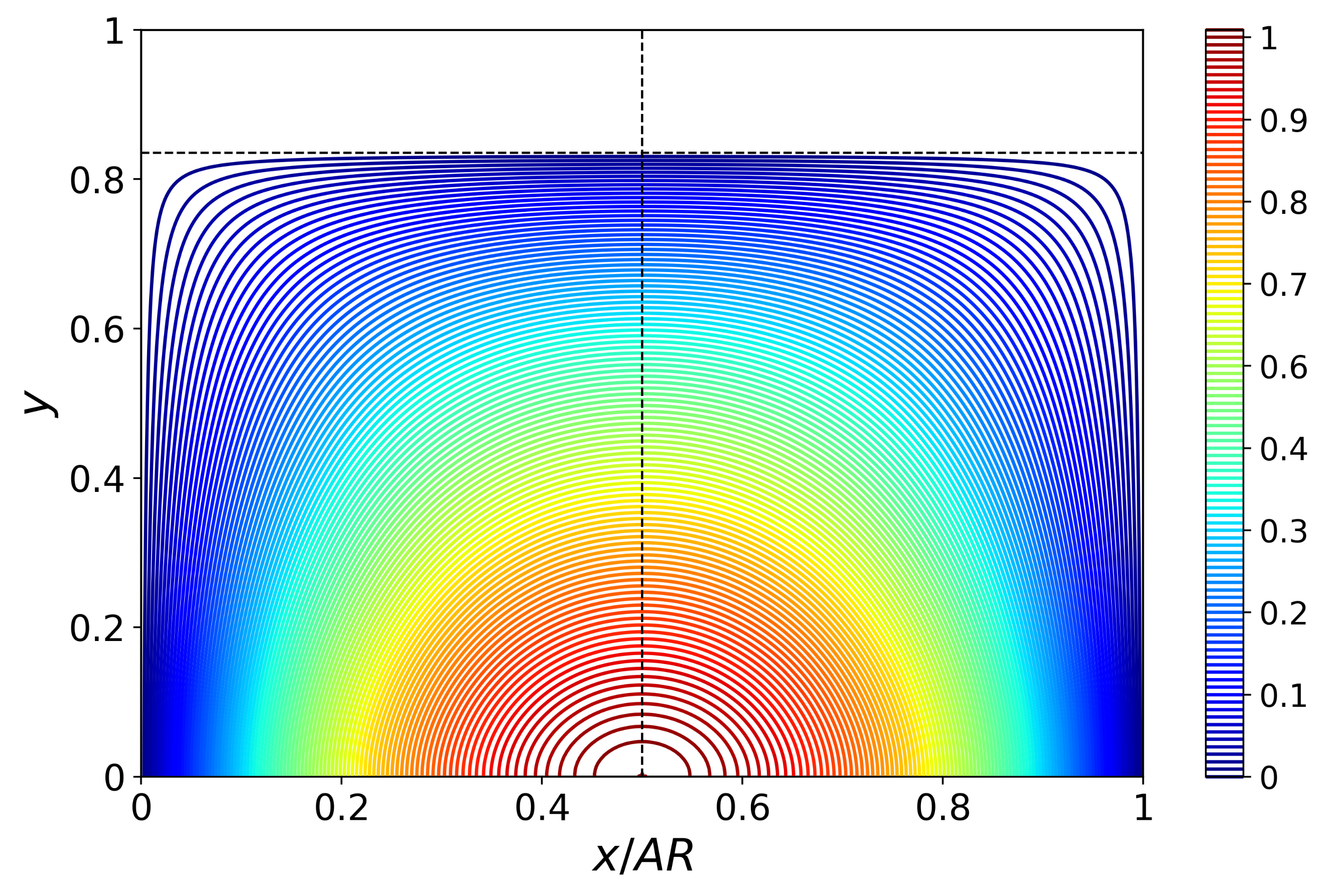}}
	\subfloat[$h=0.96, b_\text{max}/\Ha = 0.158$]{\includegraphics[width=0.49\textwidth,clip]{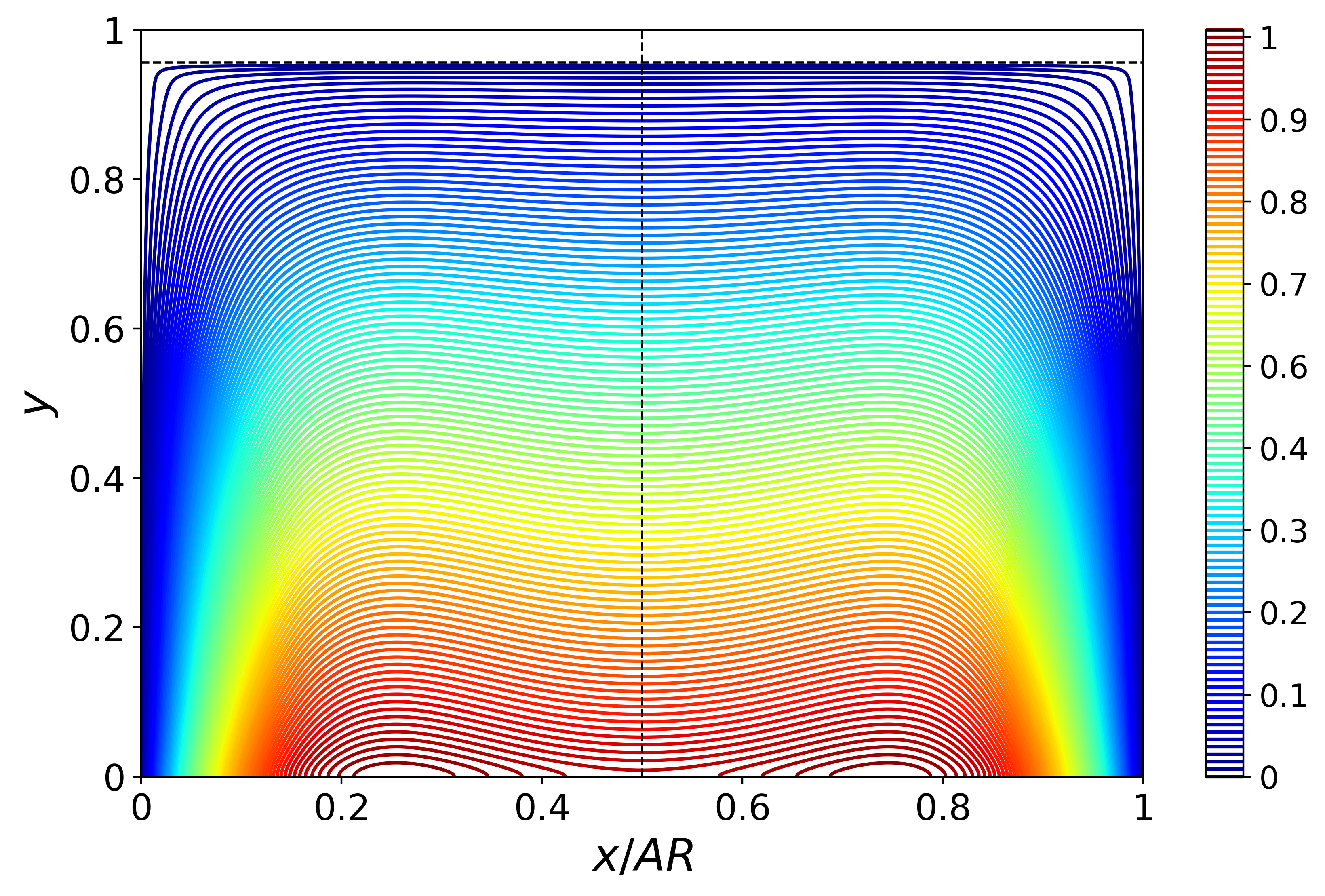}}
	\caption{\label{Fig: 4_9_b_contours_CbIs}Contours of the induced magnetic field, $b/\Ha$, in a $C_bI_s$ square duct, $Q_{21} = 1$: (a) $\Ha=5.181$; (b) $\Ha=103.625$.}
\end{figure}

The variation of the normalized maximal value of the induced magnetic field, $b_\text{max}/Ha$ with flow rate ratio, $Q_{21}$, in a square duct is shown in Fig.\ \ref{Fig: 4_7_bmax_effect_Q21}a for different $\Ha$. In a  $C_bC_s$ duct, $b_\text{max}/Ha$ is located at the center of the bottom wall and remains of order $1$ over a wide range of $Q_{21}$ (up to approximately $10$) and $\Ha$, but start growing for $Q_{21}>10$. For the whole range of $Q_{21}$, higher Hartman numbers result in higher $b_\text{max}/Ha$. With insulating bottom wall ($I_bI_s$ or $I_bC_s$, the latter not shown), the variation $b_\text{max}/Ha$ with $\Ha$ and $Q_{21}$ is similar, though the values are significantly lower. As was illustrated in Fig.\ \ref{Fig: 4_6_b_contours}a,b, with the increase of $\Ha$, the maximum of the induced magnetic field shifts off the center of the conductive layer toward the bottom wall, and, for high $\Ha$, it reaches the edge of the Hartmann boundary layer. The results indicate both in insulating and perfectly conducting ducts, the reduction of $b_\text{max}/Ha$ is almost proportional to $\Ha$ for high $\Ha$ values, and it becomes very small in case of the insulating duct, in particular for  $Q_{21}<10$. However, even for small $\Ha$, analysis of the results (not shown) indicates that  dimensional induced magnetic field scaled by $B_{0|y}$ is an order of magnitude smaller than $\Rey_m$). As a result of the very low magnetic Prandtl number of mercury ($\Pran_m=1.383\times 10^{-7}$),  $\Rey_m\ll1$ even for laminar flows of relatively high Reynolds numbers, and the  dimensional induced magnetic intensity, is, in fact,  negligible compared to $B_{0|y}$.  
\begin{figure}[h!]
	\centering
	\subfloat[]{\includegraphics[width=0.33\textwidth,clip]{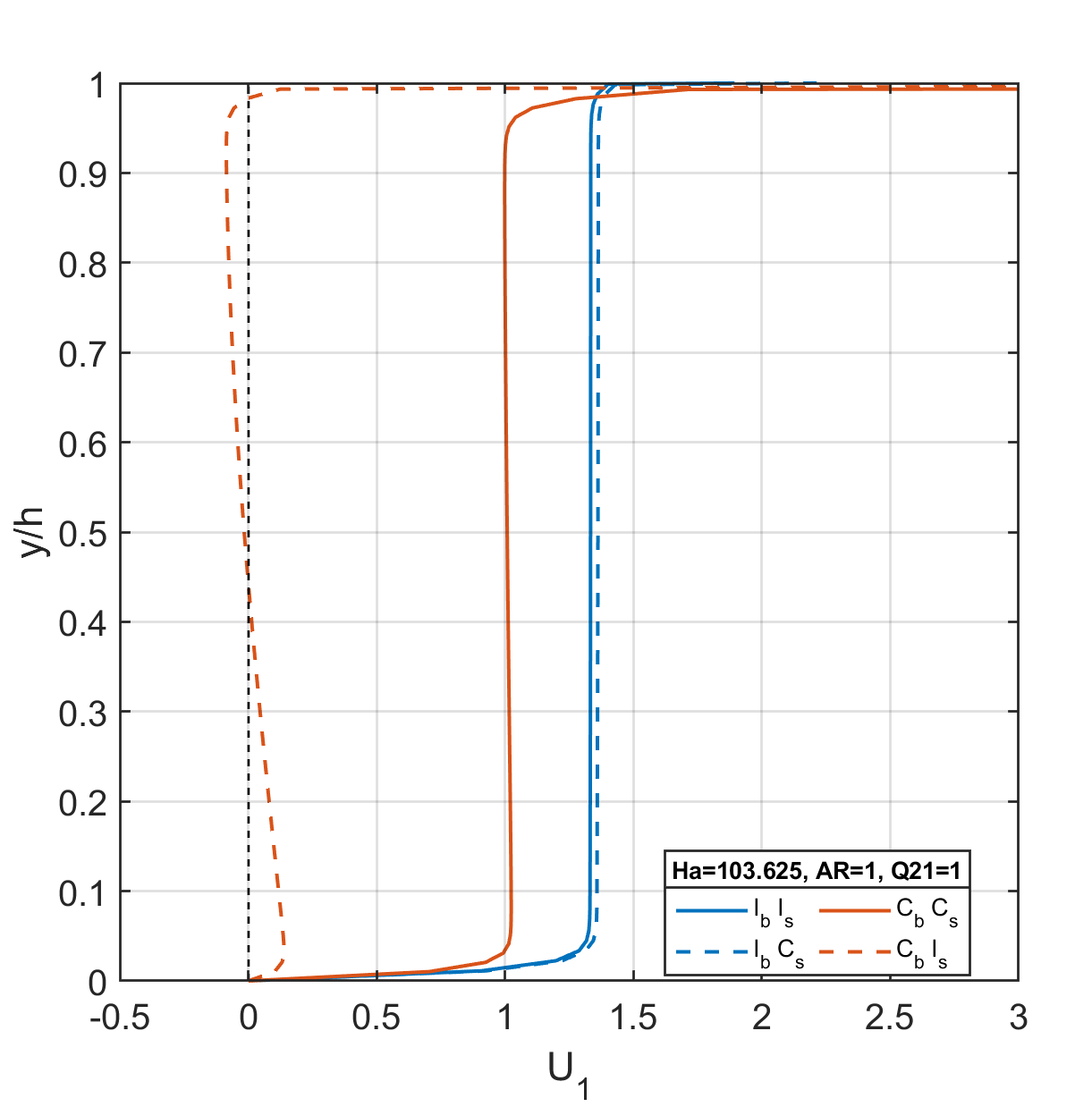}}
	\subfloat[]{\includegraphics[width=0.33\textwidth,clip]{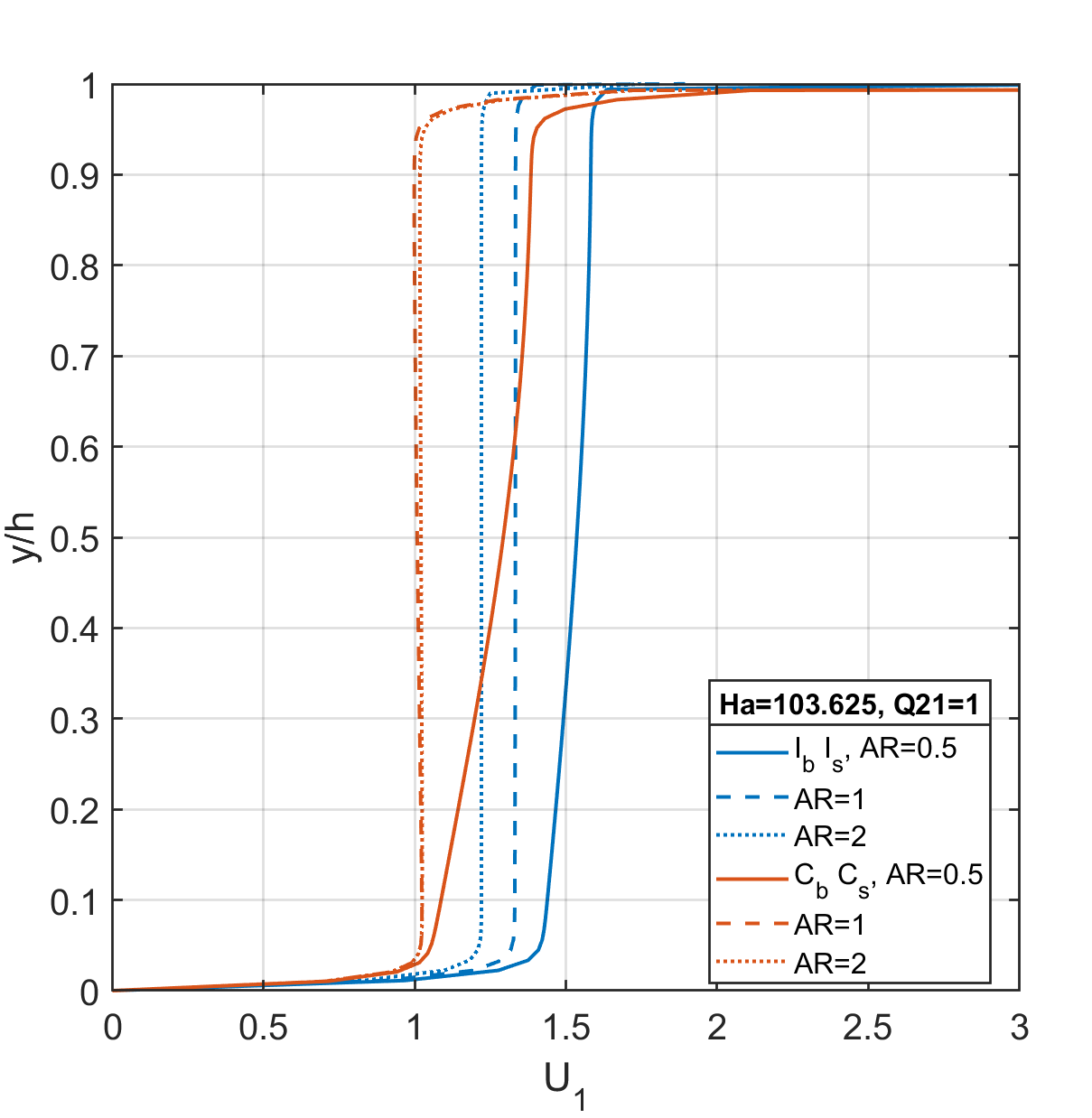}}
	\subfloat[]{\includegraphics[width=0.33\textwidth,clip]{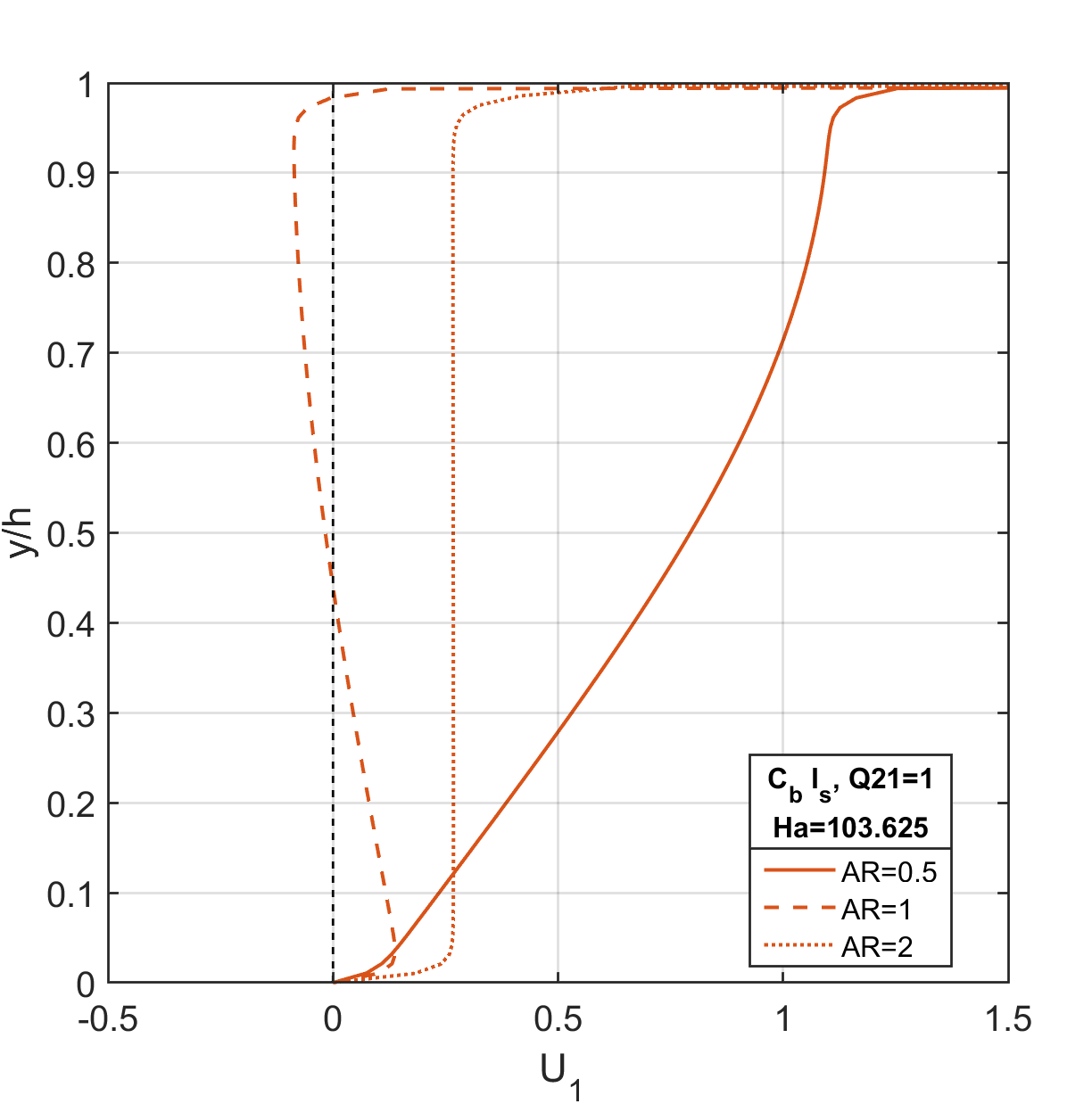}}
	\caption{\label{Fig: 4_10_U_centerline}Centerline mercury velocity profiles: effect of wall conductivities and aspect ratio, $Q_{21}=1$, $\Ha=103.625$: (a) $AR=1$; (b) $AR=0.5,1,2$ for fully insulating, $I_bI_s$, and perfectly conducting, $C_bC_s$, ducts. (c) $C_bI_s$ ducts with $AR=0.5,1,2$.}
\end{figure}
It was found that the effect of aspect ratio on maximal intensity of the induced magnetic field is rather weak, resulting in only slightly lower values for wider ducts (not shown), when the bottom wall is insulating (regardless of the side-wall conductivity) or when all walls are conducting. In contrast, when the bottom wall is conducting, replacing the conducting side walls with the insulating ones (Fig.\ \ref{Fig: 4_7_bmax_effect_Q21}b), leads to significant reduction in $b_\text{max}/Ha$ and a stronger effect of the duct aspect ratio (Fig.\ \ref{Fig: 4_7_bmax_effect_Q21}c). 

In fact, when the bottom wall is conductive, the side-wall conductivity has a strong effect on the velocity distribution (Fig.\ \ref{Fig: 4_8_U_contours_CbIs}a-c). For low $\Ha$ (Fig.\ \ref{Fig: 4_8_U_contours_CbIs}a), the velocity field still resembles that obtained with conducting side walls (Fig.\ \ref{Fig: 4_5_U_contours}a), but for high Ha the behavior changes dramatically. As shown in Fig.\ \ref{Fig: 4_8_U_contours_CbIs}c, two high-velocity jets develop near the side walls. The maximum velocity in each of the jets is reached at the interface. Accordingly, the overall maximum is found above the jet maximum, in the gas layer, and not on the vertical centerline (Fig.\ \ref{Fig: 4_8_U_contours_CbIs}b). Almost the entire mass of the mercury flows in these two jets, while the central core of the conductive layer is almost stagnant while weak backflow is obtained in its upper part. The jets form in the regions of low $b$ and $\partial b/\partial y$ caused by the insulating side walls and the interface (see Fig.\ \ref{Fig: 4_9_b_contours_CbIs}). In these regions, the Lorentz force vanishes, and the flow is driven by the (constant) pressure gradient, which is set in the flow to balance the wall and interfacial shear stresses and the Lorentz force at the conductive bottom wall. It is worth noting that backflow in the duct center and jets near the insulating side walls is obtained also in single- phase flow of a conductive fluid for this configuration of wall conductivities  (see, e.g., Fig. 3.4 in \cite{Muller01}).

With an insulating bottom wall and conducting side walls ($I_b C_s$), the flow field closely resembles the fully insulating case ($I_b I_s$). This is illustrated in Fig.\ \ref{Fig: 4_10_U_centerline}a, by presenting the velocity profiles along the duct centerline for high $\Ha$ ($=103.625$) and $Q_{21}=1$. Compared to a conducting duct, the holdup is in both cases is lower (0.9 instead of 0.98) and the core velocity is higher, essentially unaffected by the side-wall conductivity. The mercury velocity peaks at the interface and exhibits a profile resembling MHD Couette--Poiseuille flow, except for spanwise variations at the mercury--air interface. In conducting ducts of $AR\ge 1$ (Fig.\ \ref{Fig: 4_10_U_centerline}b) the core velocity for high $\Ha$ nearly equals the mercury superficial velocity (i.e. $U_1\approx1$). In $I_b I_s$ and $I_b C_s$ ducts, however, the core velocity in narrower ducts (e.g., $AR = 0.5$) is larger and increases almost linearly across the mercury core.
\begin{figure}[h!]
	\centering
	\subfloat[$h=0.95, U_{1|\text{max}}=2.349$]{\includegraphics[width=0.49\textwidth,clip]{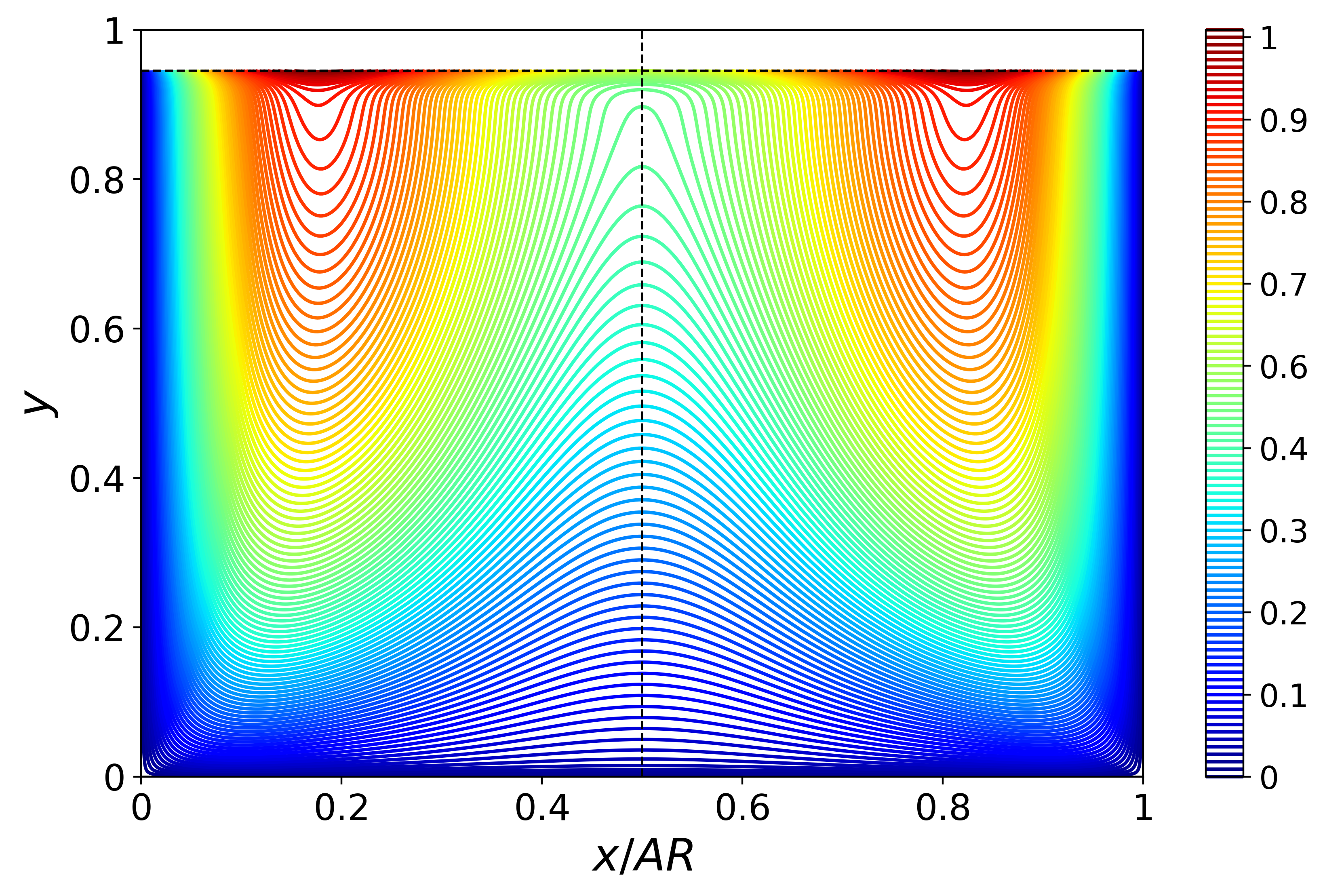}}
	\subfloat[$h=0.96, -0.015\le U_1 \le 7.375$]{\includegraphics[width=0.49\textwidth,clip]{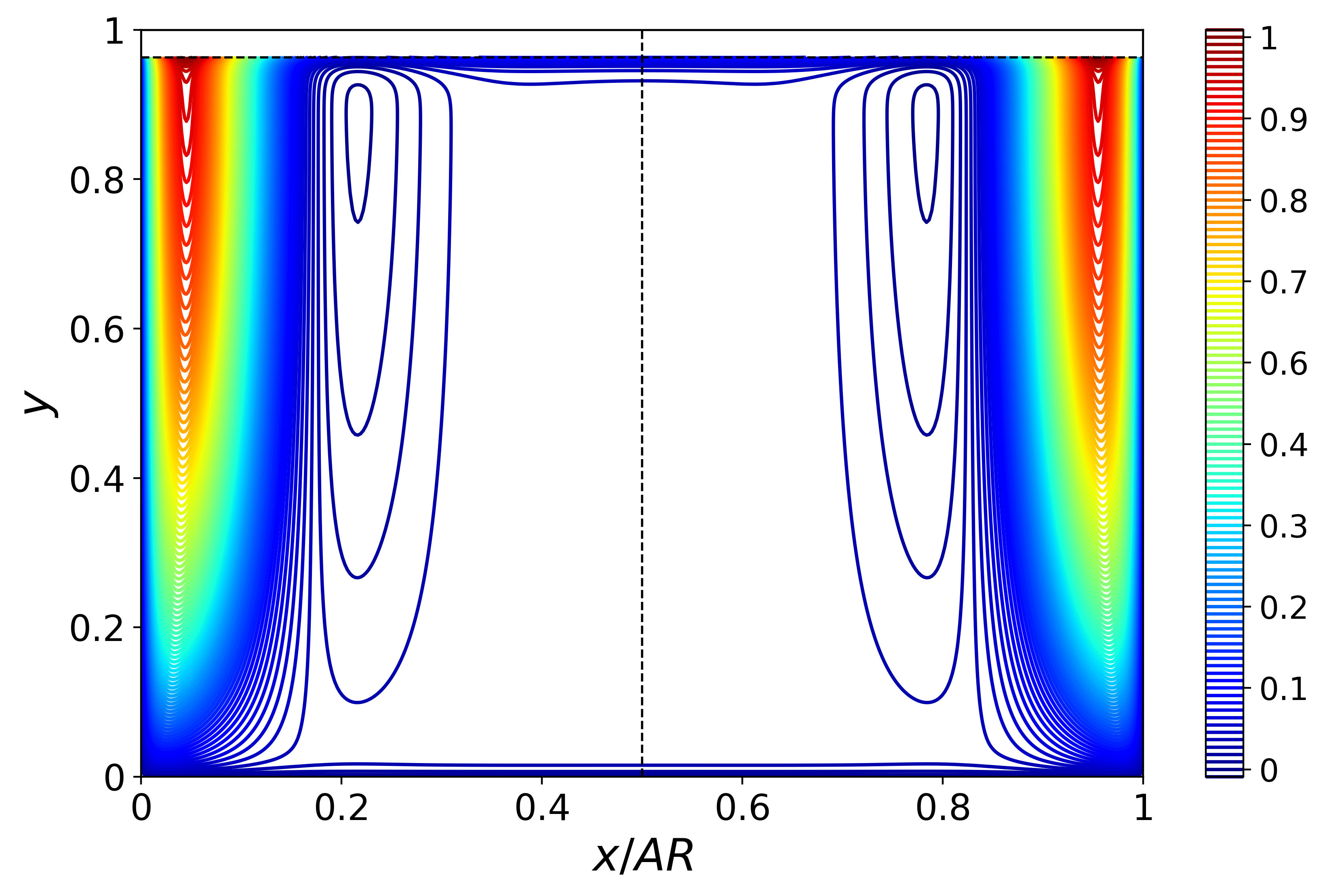}}
	\caption{\label{Fig: 4_11_U_contours_CbIs_nonsquare}Velocity contours of mercury (scaled by its maximum value) in $C_bI_s$ rectangular ducts, $Q_{21}=1$, $\Ha=103.625$: (a) $AR=0.5$, (b) $AR=2$.}
\end{figure}

\begin{figure}[h!]
	\centering
	\subfloat[]{\includegraphics[width=0.33\textwidth,clip]{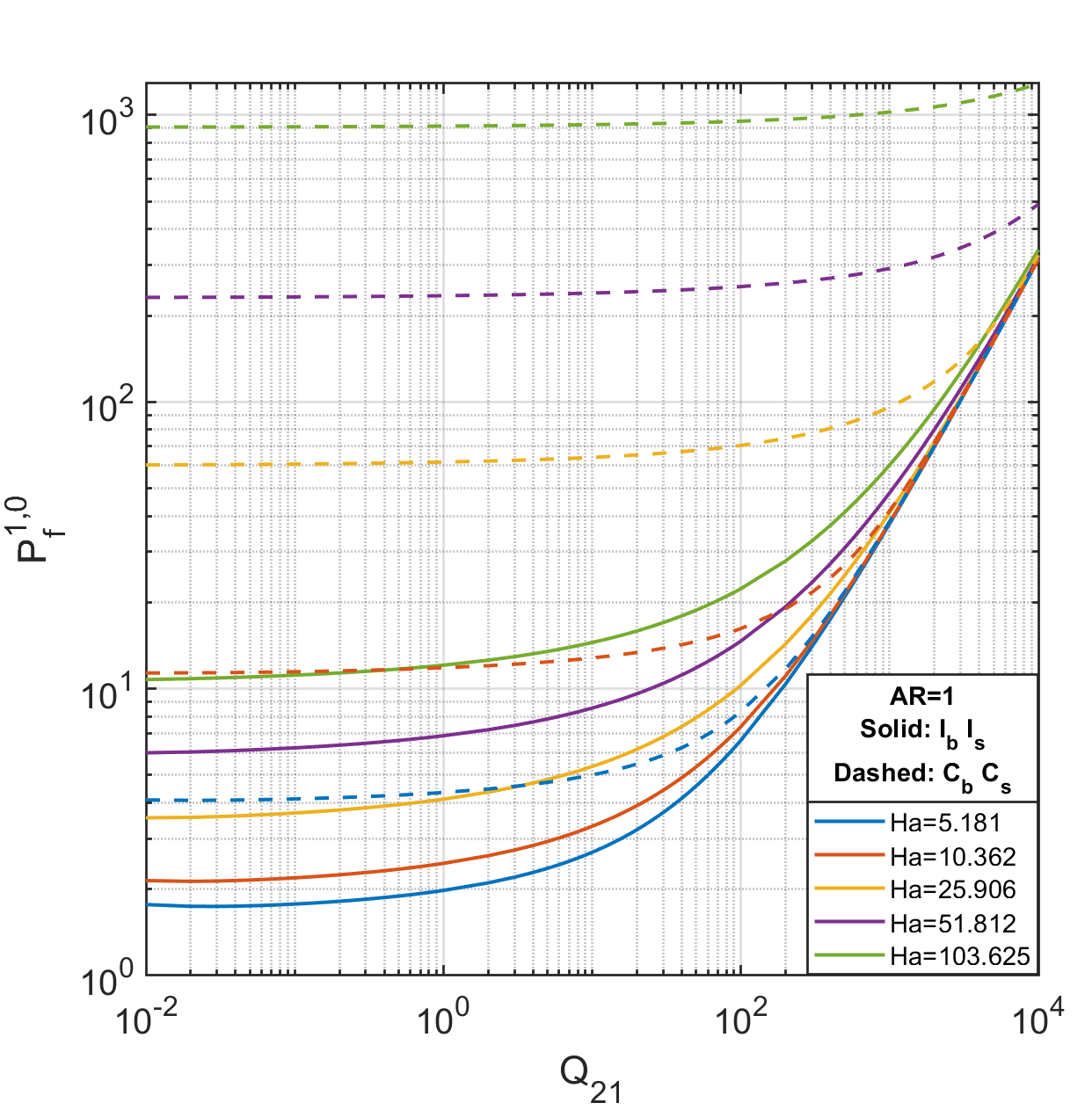}}
	\subfloat[]{\includegraphics[width=0.33\textwidth,clip]{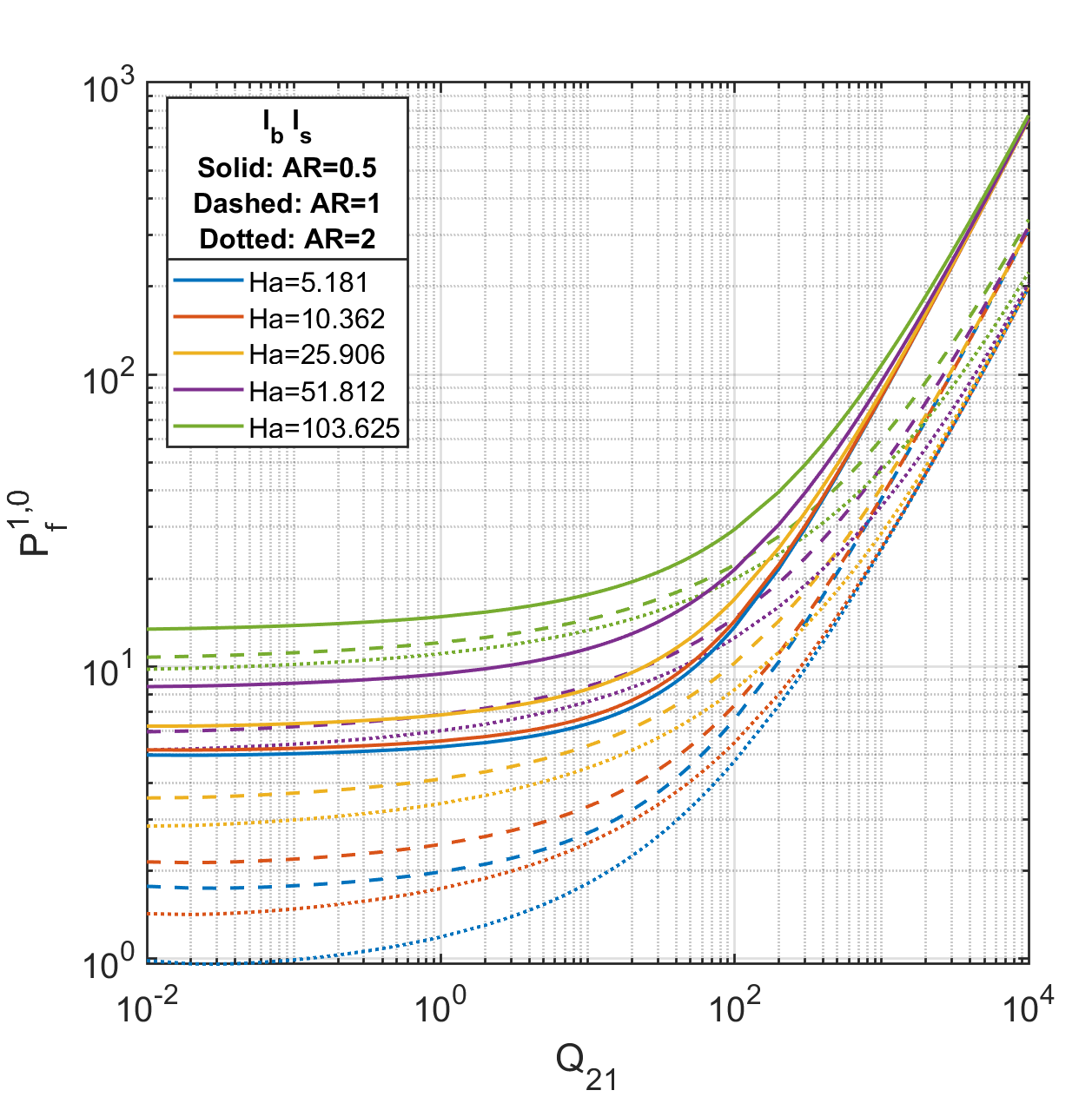}}
	\subfloat[]{\includegraphics[width=0.33\textwidth,clip]{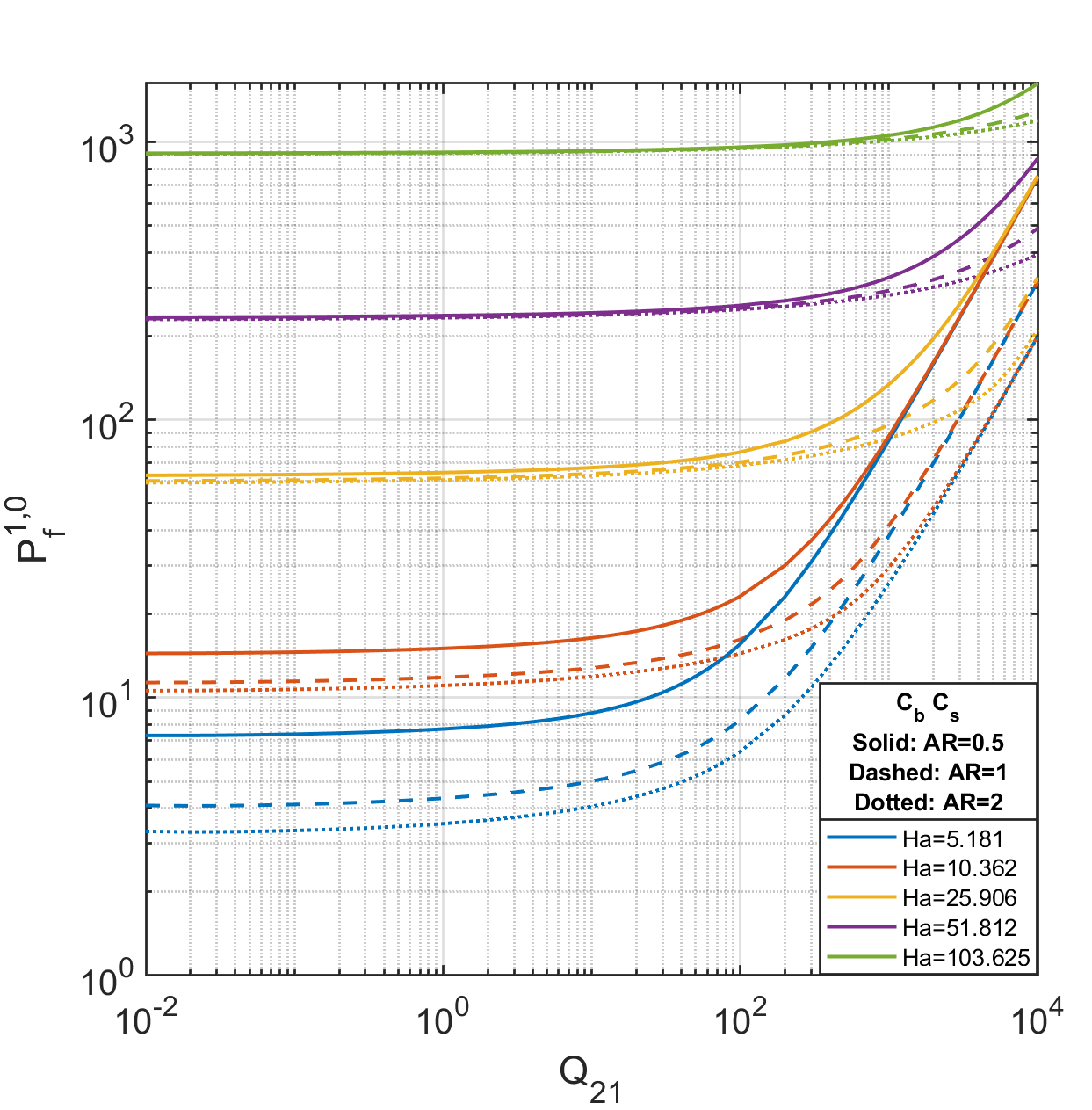}}
	\caption{\label{Fig: 4_12_P_effect_Q21}Effect of $\Ha$ and $Q_{21}$ on the pressure gradient scaled by that of the single-phase flow of mercury between two plates (with $\Ha=0$), $P_f^{1,0}$ (a) Comparison between $I_bI_s$ (solid line) and $C_bC_s$ (dashed line) square ducts. (b,c) Effect of aspect ratio, $AR=0.5,1,2$: (b) Insulating ducts (c) Conducting ducts.}
\end{figure}
Figure\ \ref{Fig: 4_10_U_centerline}a demonstrates the pronounced effect of replacing conducting with insulating side walls on the centerline velocity profile when the bottom wall is conductive. In a $C_b I_s$ square duct, the mercury centerline velocity is nearly zero in the core of the conductive layer, with backflow in its the upper part and positive velocity near the bottom, indicating axial circulation. This backflow in the duct centerline is found to vanish both in wider ($AR = 2$) and in narrower ($AR = 0.5$) ducts (Fig.\ \ref{Fig: 4_11_U_contours_CbIs_nonsquare}). Velocity contours (Fig.\ \ref{Fig: 4_11_U_contours_CbIs_nonsquare}b) reveal that in rectangular ducts of $AR=2$ the central backflow splits into two weaker off-centre regions, in proximity to the near-wall jets, whose intensity increases with AR ($max(U_1)=7.38$ vs. $max(U_1)=4.38$ in the square duct). In fact, two off-centre backflow regions are obtained at any $AR>1$. In narrow ducts of $AR\lesssim0.7$ (e.g., $AR=0.5$ in Fig.\ \ref{Fig: 4_10_U_centerline}c and Fig.\ \ref{Fig: 4_11_U_contours_CbIs_nonsquare}a) backflow is not found and the jets are weaker (e.g., $max(U_1)=2.35$). Hence, for stabilizing the stratified flow pattern, narrow ducts may be preferable, since axial backflow may trigger entrance disturbances and destabilize the stratified flow pattern.
\begin{figure}[h!]
	\centering
	\subfloat[]{\includegraphics[width=0.49\textwidth,clip]{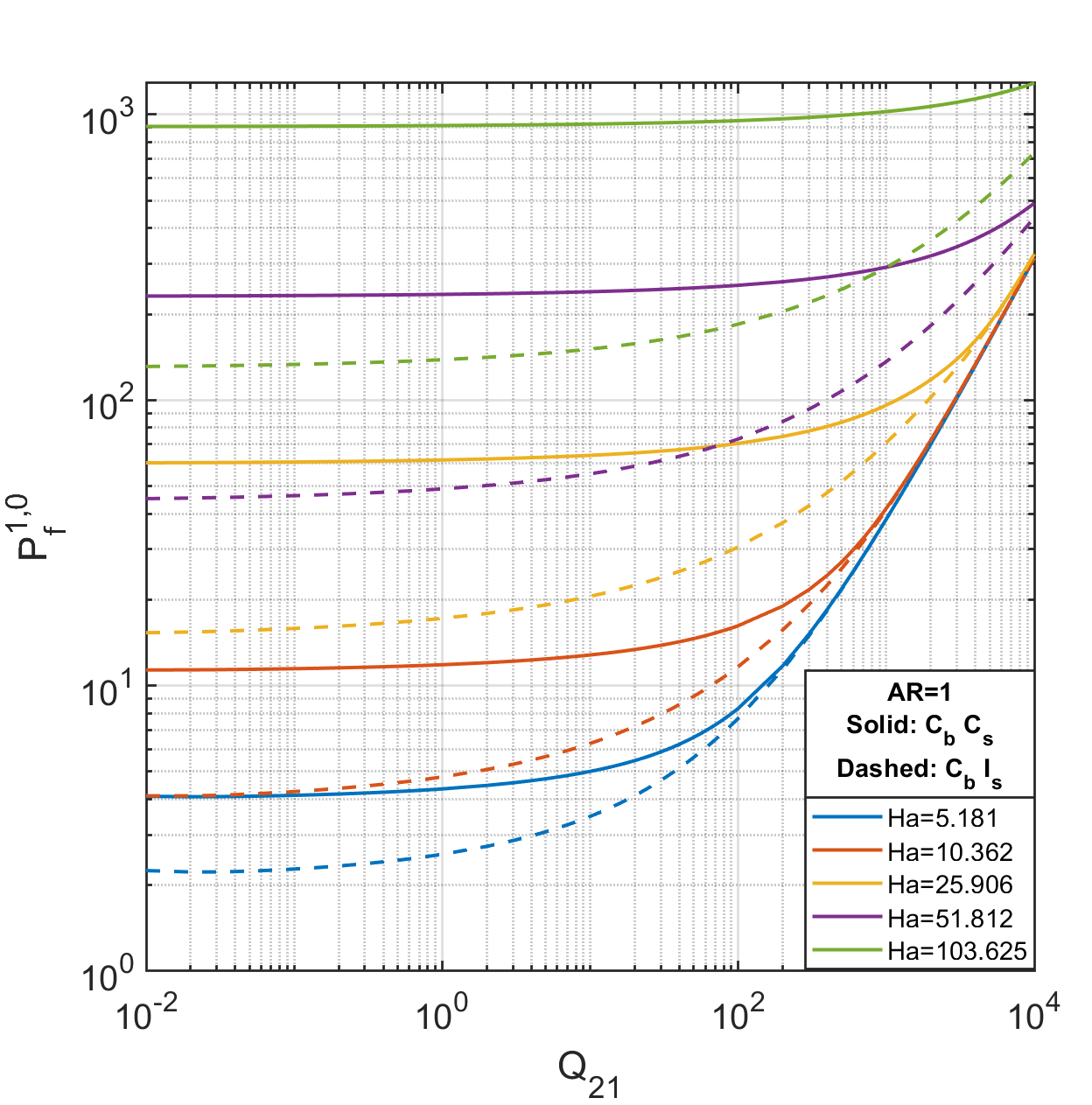}}
	\subfloat[]{\includegraphics[width=0.49\textwidth,clip]{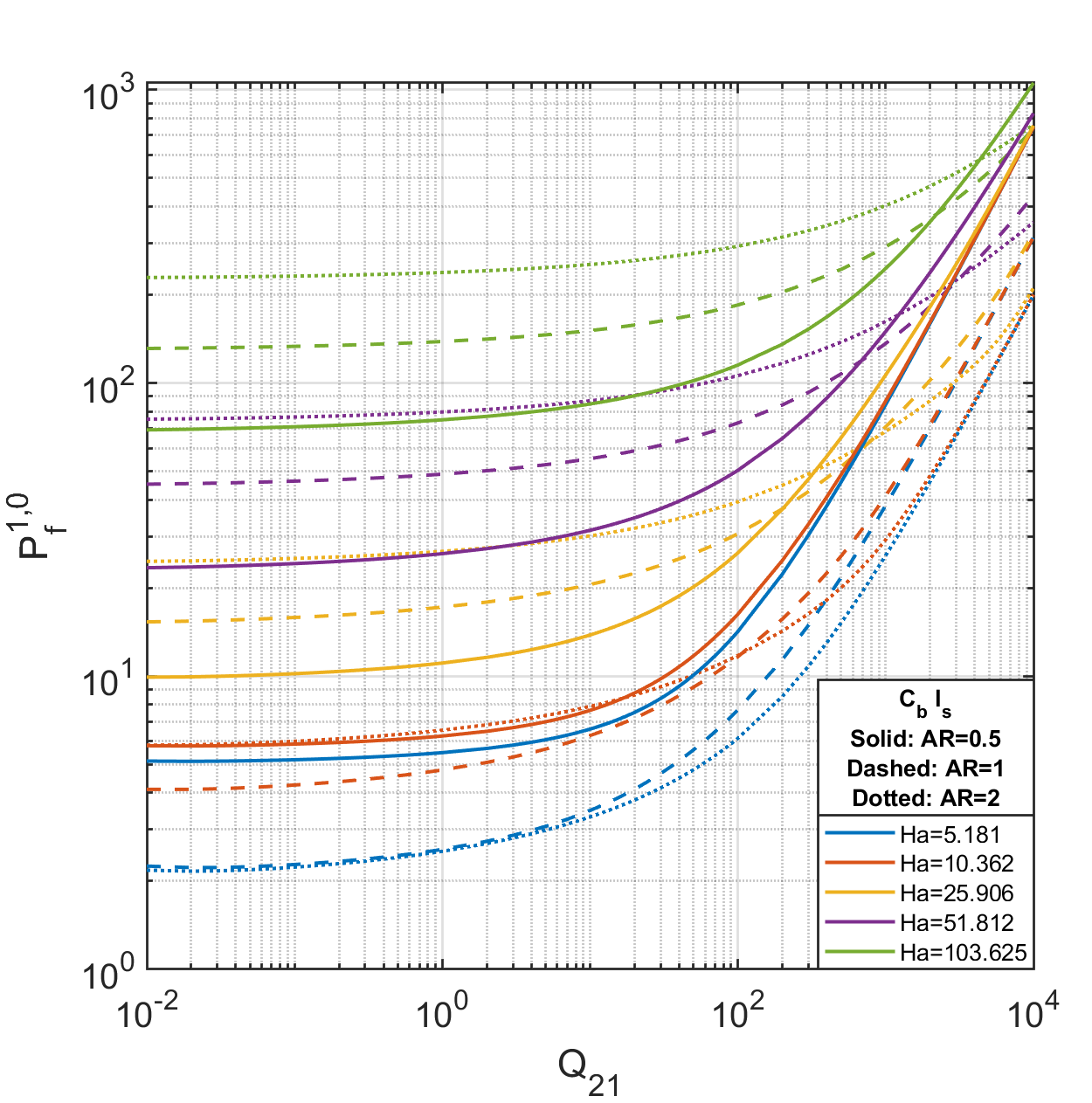}}
	\caption{\label{Fig: 4_13_P_effect_Q21_Cb}Effect of the side wall conductivity in ducts with a conducting bottom wall on $P_f^{1,0}$. (a) Effect of $\Ha$ and $Q_{21}$ in a square duct. (b) Effect of aspect ratio, $AR=0.5,1,2$, in $C_bI_s$.}
\end{figure}
The combined effects of wall conductivities on conductive phase holdup, induced magnetic field, and velocity profiles also alter the pressure gradients required to sustain specified fluid flow rates. The dimensionless pressure drop, $P_f^{1,0}=-G/12$ , is shown in Figs.\ \ref{Fig: 4_12_P_effect_Q21} and \ref{Fig: 4_13_P_effect_Q21_Cb} as a function of flow rate ratio, $Q_{21}$. It represents the pressure gradient normalized with respect to that for single-phase flow of the conductive phase between two infinite plates without an external magnetic field ($\Ha=0$). As illustrated in Fig.\ \ref{Fig: 4_12_P_effect_Q21}a, in both perfectly conducting and fully insulated square ducts,$P_f^{1,0}>1$ and increases with $\Ha$ and with the gas flow rate (i.e., with $Q_{21}$), particularly for $Q_{21}>10$. However, for the same $\Ha$, the pressure gradient is consistently lower in insulated ducts across the entire range of $Q_{21}$. This reduction is mainly due to the much weaker Lorentz force when the bottom wall is insulating, which can even aid the pressure gradient in balancing viscous shear stresses near the bottom wall (where $\partial b/\partial y >0$, see Fig.\ \ref{Fig: 4_6_b_contours}a,b). Figures\ \ref{Fig: 4_12_P_effect_Q21}b and \ref{Fig: 4_12_P_effect_Q21}c show that the pressure gradient drops with increasing the aspect ratio for both insulated and conducting ducts, respectively, with more pronounced reduction in the former case. In that case, replacing the insulating side walls with conductive ones has no effect on the pressure gradient. By contrast, when the side walls of a conducting duct are replaced with insulating walls, the pressure gradient decreases significantly for all $\Ha$, with nearly an order of magnitude reduction at high $\Ha$ and low $Q_{21}$ , although the reduction persists across the entire considered range of  $Q_{21}$ (see Fig.\ \ref{Fig: 4_13_P_effect_Q21_Cb}a). Furthermore, in $C_b I_s$ ducts and high $\Ha$, the pressure gradient increases with $AR$ (see Fig.\ \ref{Fig: 4_13_P_effect_Q21_Cb}b).
\begin{figure}[h!]
	\centering
	\subfloat[]{\includegraphics[width=0.4\textwidth,clip]{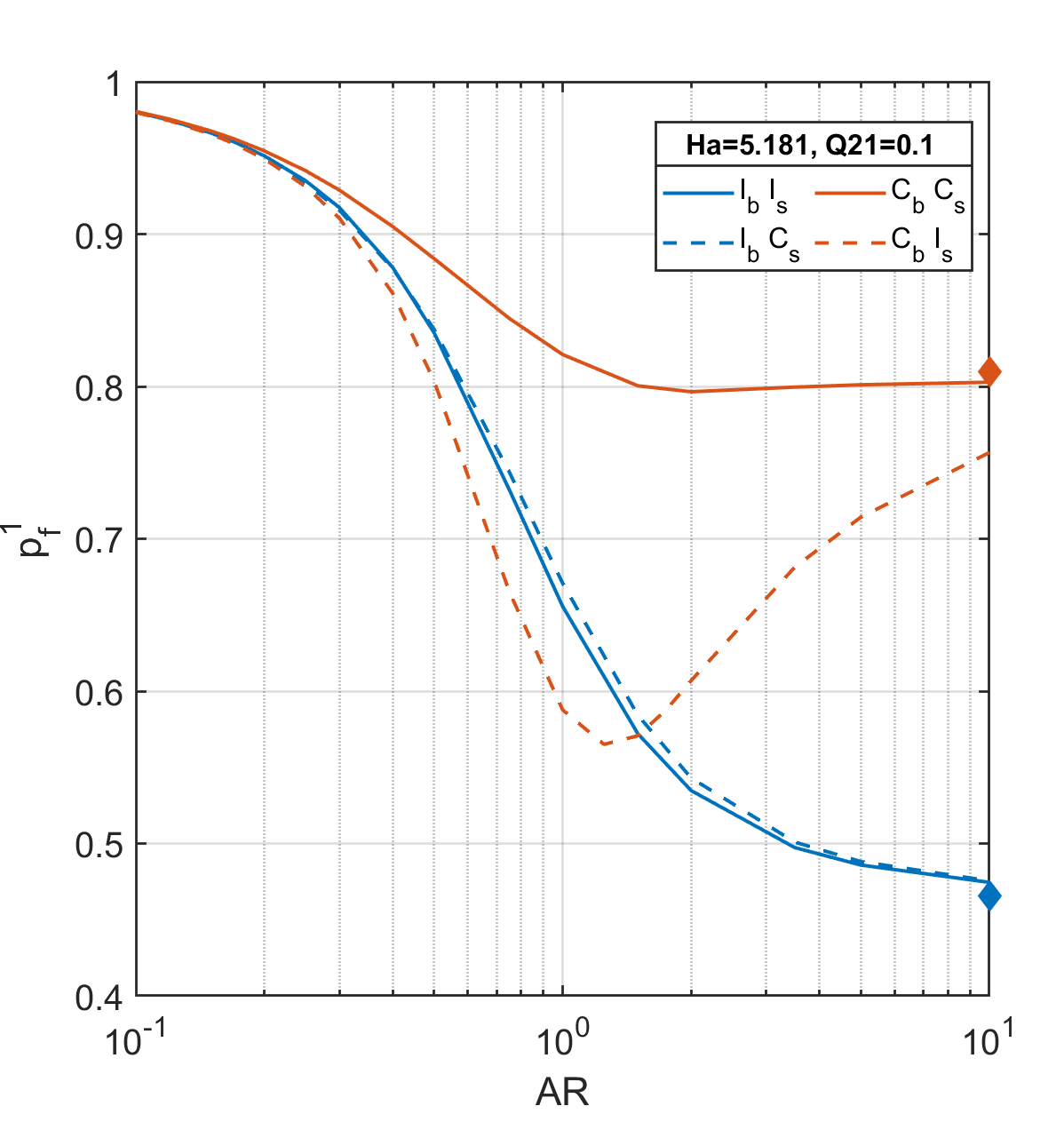}}
	\qquad
	\subfloat[]{\includegraphics[width=0.4\textwidth,clip]{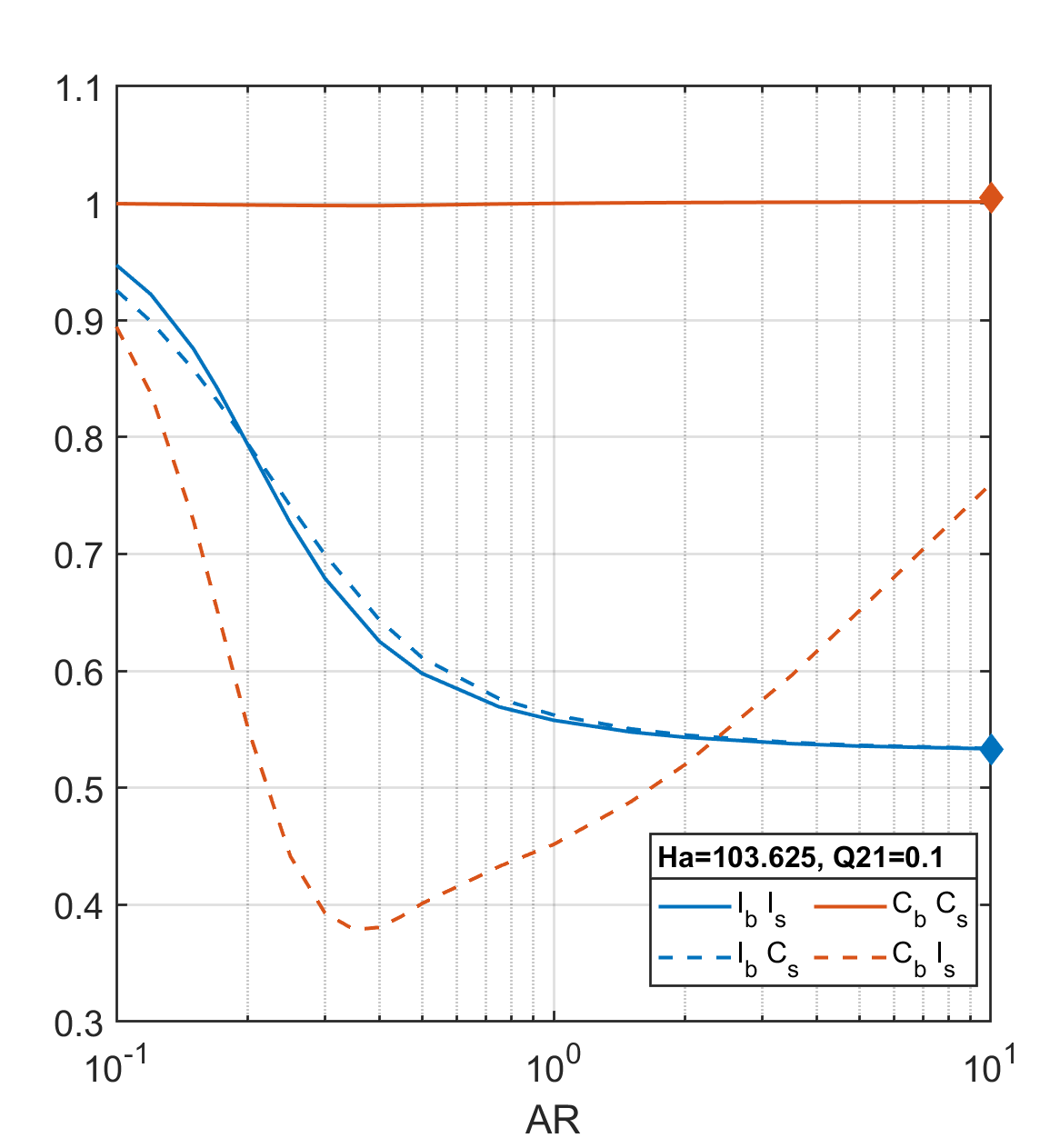}}
	\caption{\label{Fig: 4_14_P_effect_AR}Effect of the duct aspect ratio and wall conductivities on the lubrication factor, $P_1^f$: (a) $\Ha=5.181$; (b) $\Ha=103.625$. The red (blue) diamond represents holdup value for two-plate geometry with a conducting (insulating) bottom wall, respectively.}
\end{figure}
The effect of air flow on the pressure gradient can be quantified by the pressure gradient factor $P_f^1$, defined as the ratio between the axial pressure gradient in two-phase flow and that in single-phase flow of the conductive fluid in the same duct for the same $\Ha$ and same wall conductivities. The corresponding values of $G$ obtained in single-phase flow are provided in Appendix\ \ref{Sec: appendix_single_phase}. Values of $P_f^1<1$ indicate a lubrication effect of the air on the mercury flow. The corresponding pumping power savings are represented by the power factor, $\Po=P_f^1(1+Q_{21})$.   

The influence of duct wall conductivity on the variation of $P_f^1$ with $AR$ is demonstrated in Fig.\ \ref{Fig: 4_14_P_effect_AR} for low and high Ha numbers ($Q_{21}=0.1$). The air lubrication effect in insulated ducts strengthens with increasing aspect ratio, particularly for $AR\lesssim2$, where substantial reductions in $P_f^1$ are observed at both low and high $\Ha$ in wider rectangular insulated ducts (Fig.\ \ref{Fig: 4_14_P_effect_AR}). In contrast, in conducting ducts at high $\Ha=103.625$, $P_f^1\approx1$ and almost independent of AR. However, replacing the conductive side walls with insulating ones leads to a drastic reduction in. In such ducts, the influence of $AR$ on $P_f^1$ is non-monotonic: increasing AR enhances the lubrication effect in narrow ducts ($AR\ll1$), but reduces it in wide ducts ($AR>1$), due to the increased intensity of the side-wall jet flow in wider ducts. At high $\Ha$($=103.625$ Fig.\ \ref{Fig: 4_14_P_effect_AR}b), the minimum occurs in narrow ducts of $AR\approx0.35$, reaching values of $P_f^1$ even lower than those attainable in fully insulated ducts of any $AR$. 
\begin{figure}[h!]
	\centering
	\subfloat[]{\includegraphics[width=0.49\textwidth,clip]{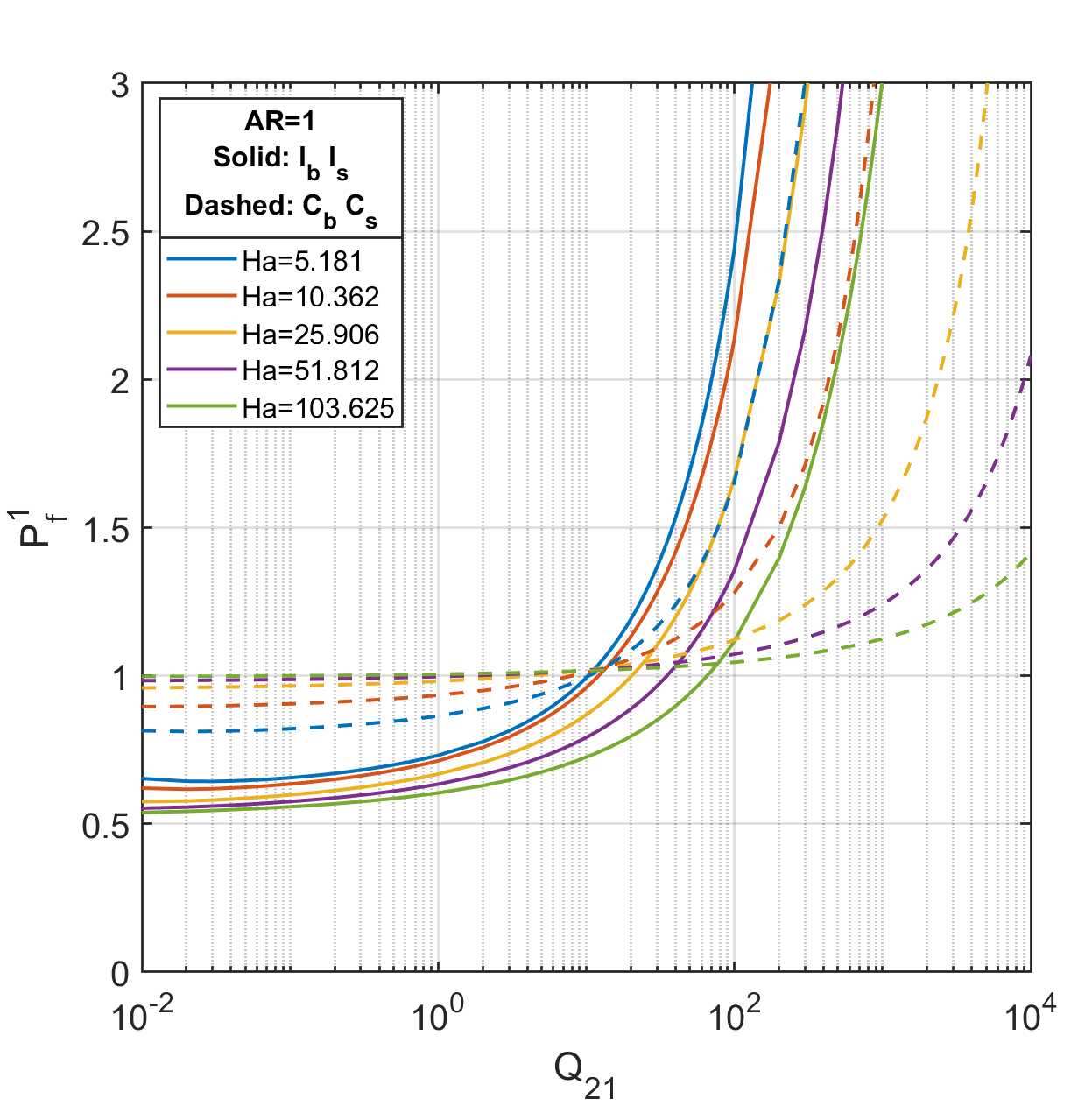}}
	\subfloat[]{\includegraphics[width=0.49\textwidth,clip]{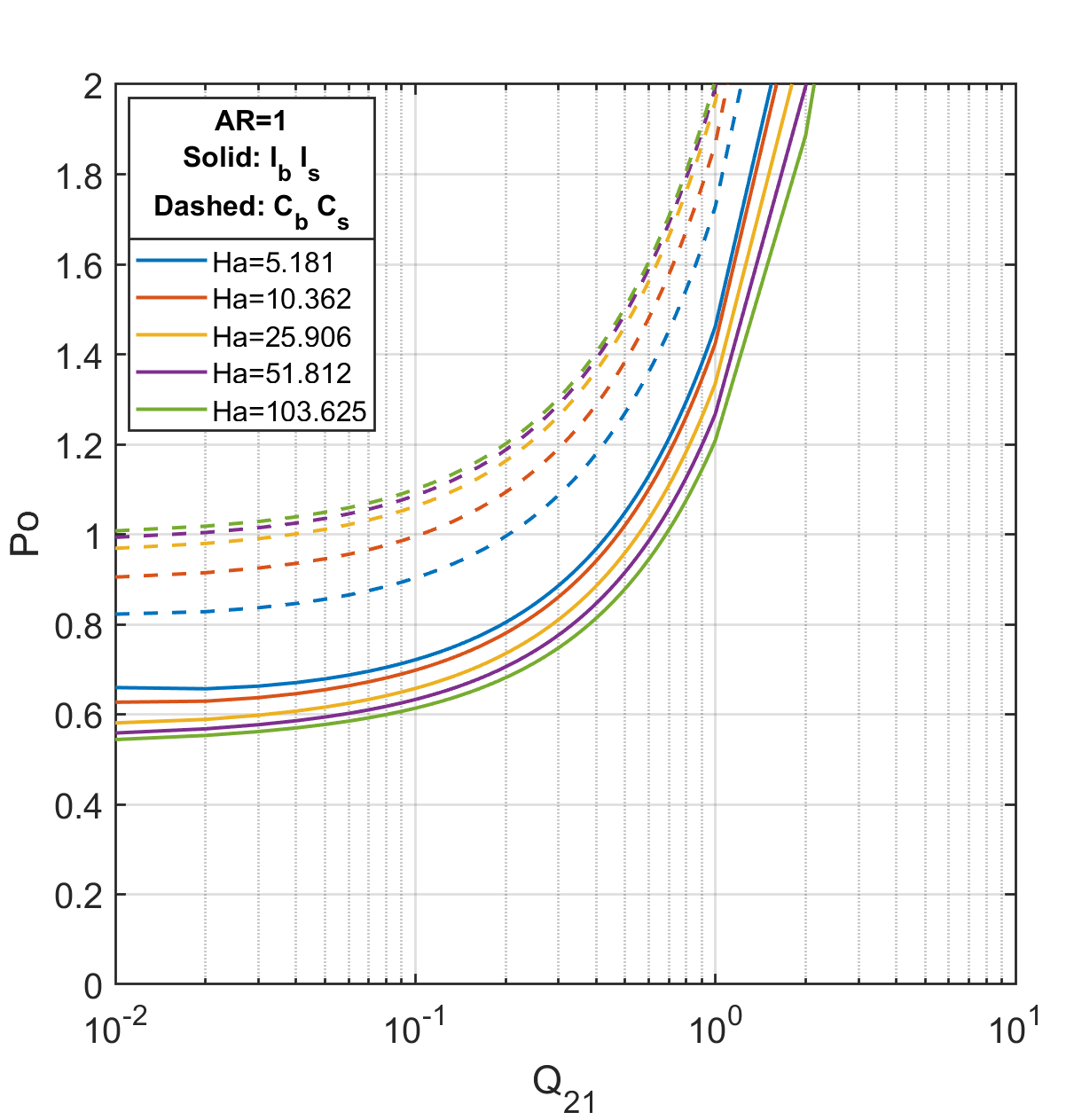}}
	\caption{\label{Fig: 4_15_P1f_effect_Q21}Effect of $\Ha$ and $Q_{21}$ on (a) the lubrication factor, $P_1^f$, and (b) the power factor, $\Po$. Comparison of fully insulating, $I_bI_s$ (solid line) and perfectly conducting, $C_bC_s$ (dashed line) square ducts.}
\end{figure}
The variation of $P_f^1$ and the power factor, $\Po$, with $Q_{21}$ for and different Ha in square ducts with different wall conductivity configurations is shown in Figs.\ \ref{Fig: 4_15_P1f_effect_Q21} and \ref{Fig: 4_16_P1f_effect_Q21_Cb}.  Note that, consistent with their definition, all curves converge to $P_f^1=1$, $\Po=1$ at $Q_{21}=0$ (outside the range displayed in Fig.\ \ref{Fig: 4_15_P1f_effect_Q21}).  In insulated square ducts (Fig.\ \ref{Fig: 4_15_P1f_effect_Q21}a, solid lines), air lubrication is evident for all $\Ha$ and $Q_{21}$ up to $\approx10$, extending to higher $Q_{21}$ for higher $\Ha$. The lubrication effect strengthens with increasing $\Ha$, reaching up to $\approx50\%$ reduction in the pressure gradient when a small air flow rate is introduced ($Q_{21}\approx 0.01$). However, beyond the value of $Q_{21}$ corresponding to $P_f^1=1$, further increase in $Q_{21}$ (e.g., increase of gas flow rate) leading to a steep rise in the pressure gradient is observed due to the increased wall shear stresses in the gas layer. In a perfectly conductive square duct (Fig.\ \ref{Fig: 4_15_P1f_effect_Q21}a, dashed lines), the lubrication effect is weaker and limited to low $\Ha$. Unlike the trend in insulated ducts, here the air lubrication decreases as $\Ha$ increases. The corresponding values of the pumping power factor, $\Po$), are presented in Fig.\ \ref{Fig: 4_15_P1f_effect_Q21}b. In insulated ducts, $\Po$ is largely insensitive to $\Ha$, and pumping power savings can be achieved by introducing air flow up to  $Q_{21}\approx 1$. In conducting ducts, however, the potential for pumping power savings is much lower, restricted to low $\Ha$ and a narrower range of $Q_{21}$ (up to $\approx 0.1$).
\begin{figure}[h!]
	\centering
	\subfloat[]{\includegraphics[width=0.49\textwidth,clip]{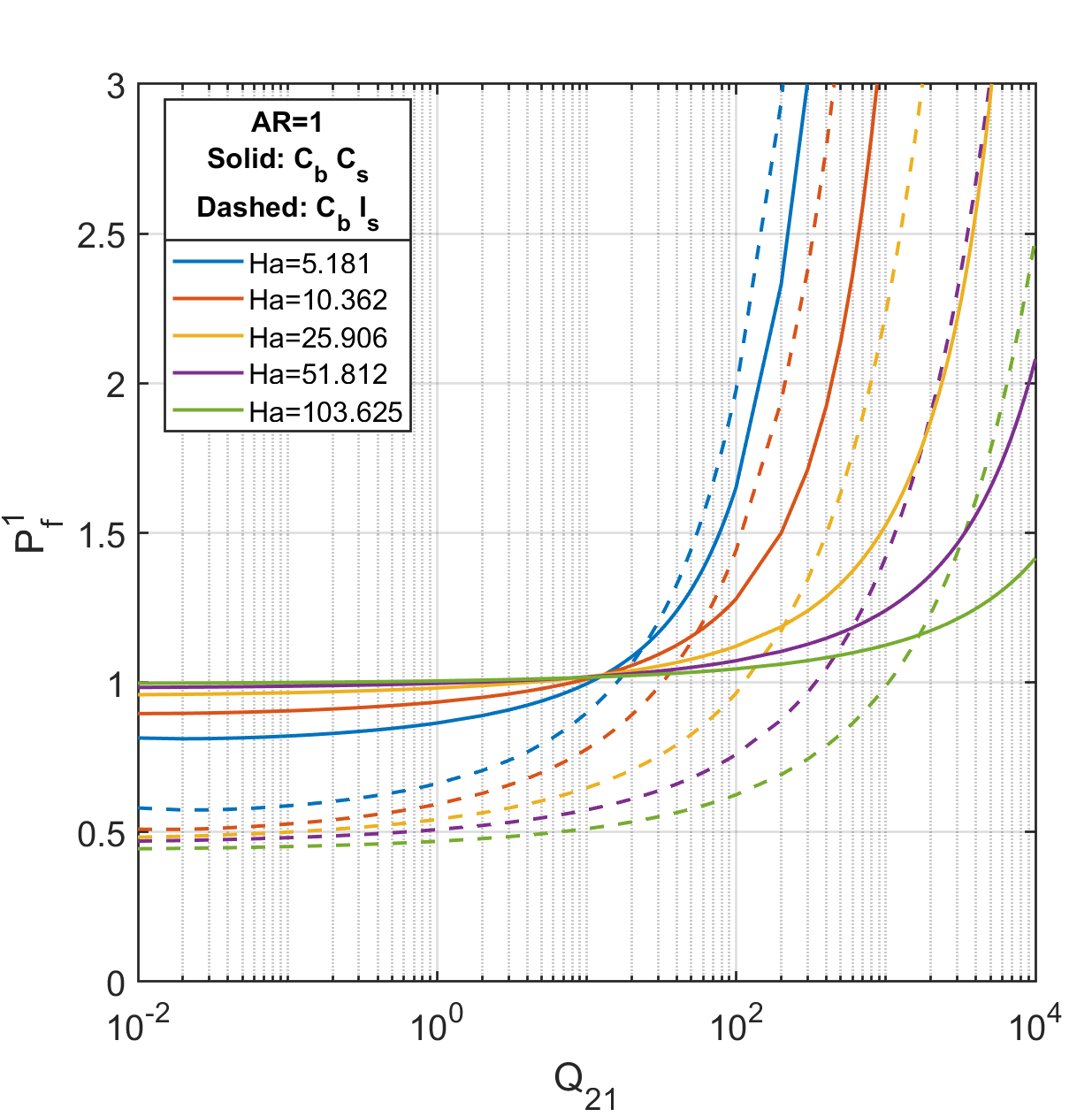}}
	\subfloat[]{\includegraphics[width=0.49\textwidth,clip]{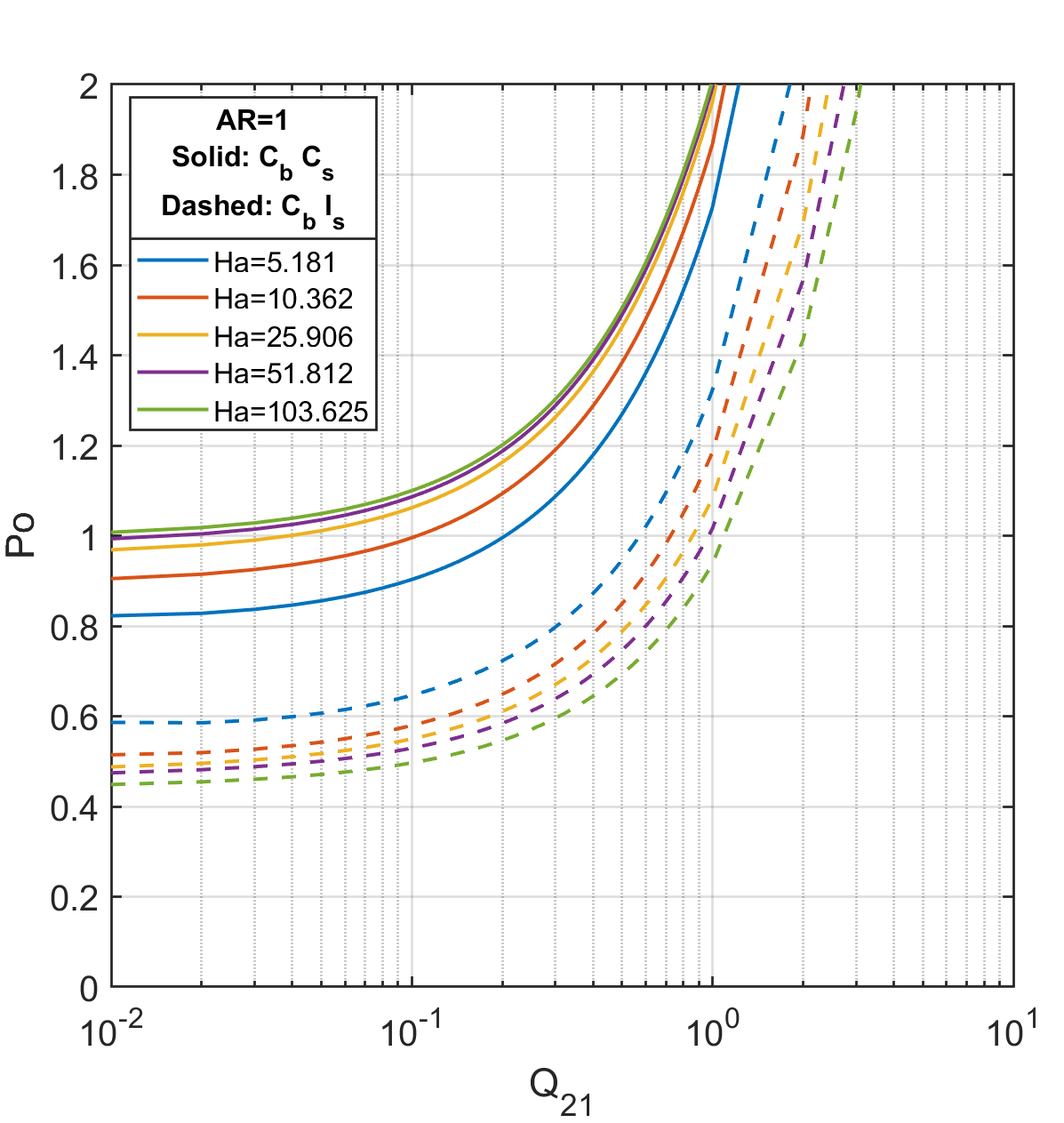}}
	\caption{\label{Fig: 4_16_P1f_effect_Q21_Cb}Effect of side walls conductivity on (a) $P_f^{1}$ and (b) $\Po$ for various $\Ha$ and $Q_{21}$. Comparison of $I_bI_s$ (solid line) and $C_bI_s$ (dashed line) square ducts.}
\end{figure}
The effect of replacing the conductive side walls of a square duct with insulating walls on the $P_f^1$ values is shown in Fig.\ \ref{Fig: 4_16_P1f_effect_Q21_Cb}a. A substantially stronger lubrication effect is observed in this configuration, with $P_f^1$ values comparable to those of a fully insulated duct at low $Q_{21}$ (cf. Fig.\ \ref{Fig: 4_15_P1f_effect_Q21}a). Moreover, at high $\Ha$, the lubrication range extends to even larger $Q_{21}$. As illustrated in Fig.\ \ref{Fig: 4_16_P1f_effect_Q21_Cb}b, this wall replacement also enhances the potential for pumping power savings, with $\Po$ values similar those obtained in a fully insulating square ducts (cf. Fig.\ \ref{Fig: 4_15_P1f_effect_Q21}b).

\subsection{Horizontal external magnetic field} \label{Sec: x-problem} 

\begin{figure}[h!]
	\centering
	\subfloat[]{\includegraphics[width=0.49\textwidth,clip]{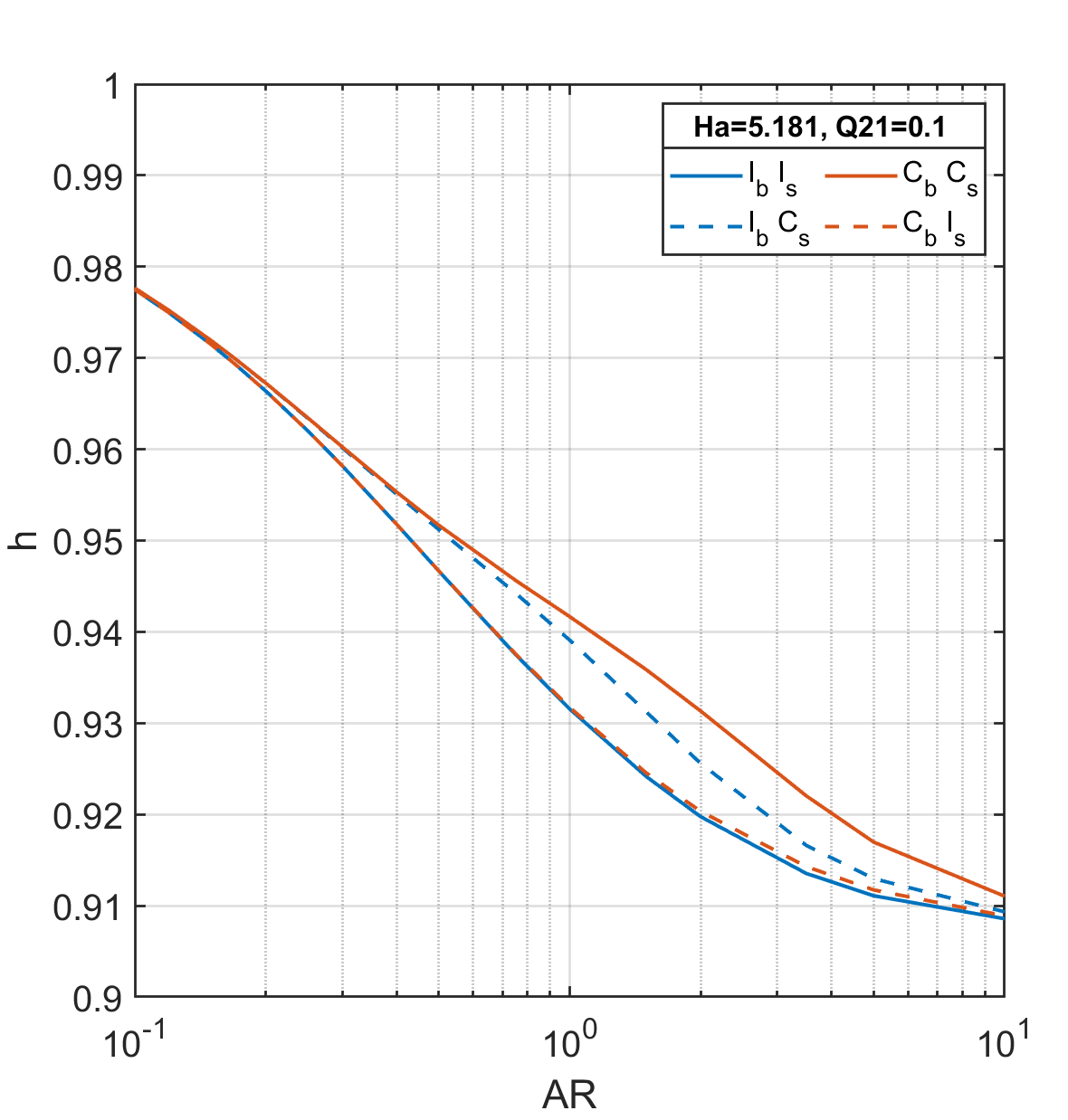}}
	\subfloat[]{\includegraphics[width=0.49\textwidth,clip]{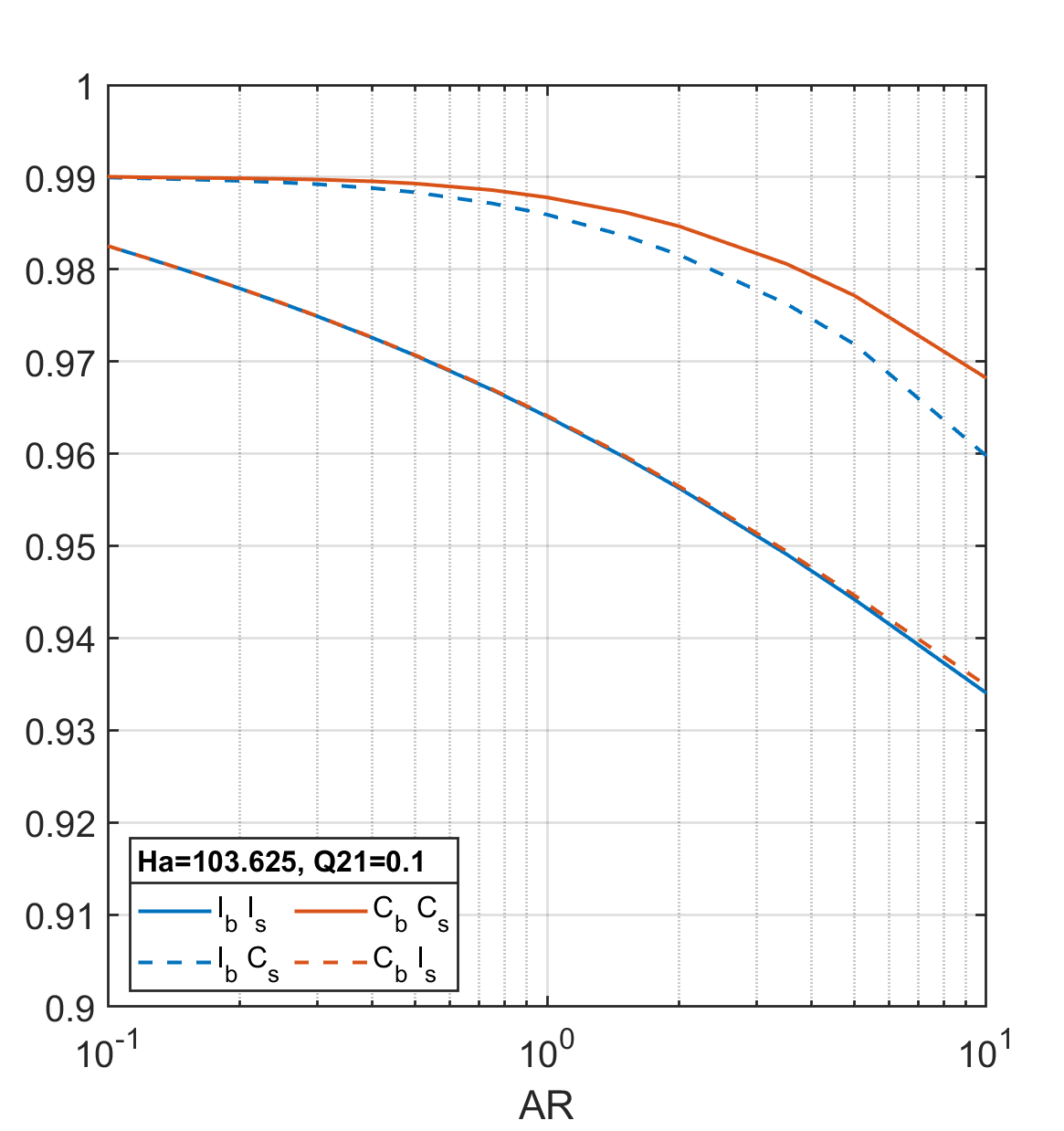}}
	\caption{\label{Fig: 4_17_h_effect_AR_hor}Effect of $AR$ on holdup for different wall conductivities for $Q_{21}=0.1$, (a) $\Ha=5.181$, (b) $\Ha=103.625$.}
\end{figure}

When the external magnetic field is applied in the horizontal direction ($B_{0|x}\ne0$,$B_{0|y}=0$, denoted below as $B_{0|x}$) the side walls of the duct (perpendicular to $B_0$) act as the Hartmann walls, while the bottom one corresponds to the Shercliff wall. Figure\ \ref{Fig: 4_17_h_effect_AR_hor} illustrates the variation of holdup with the aspect ratio ($AR$) for different combinations of bottom- and side-wall conductivities, at low and high Hartmann numbers ($\Ha=5.181$ and $103.625$, respectively) for $Q_{21}=0.1$. 
Comparison of Fig.\ \ref{Fig: 4_17_h_effect_AR_hor}a ($\Ha=5.181$) and Fig.\ \ref{Fig: 4_17_h_effect_AR_hor}b ($\Ha=103.625$) shows the expected increase of holdup with Ha , caused by the stronger retarding Lorentz force. Similar to the case of vertical magnetic field, also when a horizontal field is applied, the holdup in a perfectly conducting duct ($C_b C_s$) is higher than that obtained in a fully insulating duct ($I_b I_s$). However, in the case of insulating side walls, the conductivity of bottom wall has almost no influence on the conductive liquid holdup (see solid blue and dashed red curves in Fig.\ \ref{Fig: 4_17_h_effect_AR_hor}), and thus on the pressure gradient. On the other hand, conducting side walls combined with insulating bottom wall reduce the holdup compared to a fully conducting duct. Another interesting observation is reduction of the holdup with increase in the aspect ratio for all wall-conductivity combinations, which is in contrast to the trends observed under vertical magnetic field (see Fig.\ \ref{Fig: 4_1_h_effect_AR}). 
\begin{figure}[h!]
	\centering
	\subfloat[]{\includegraphics[width=0.49\textwidth,clip]{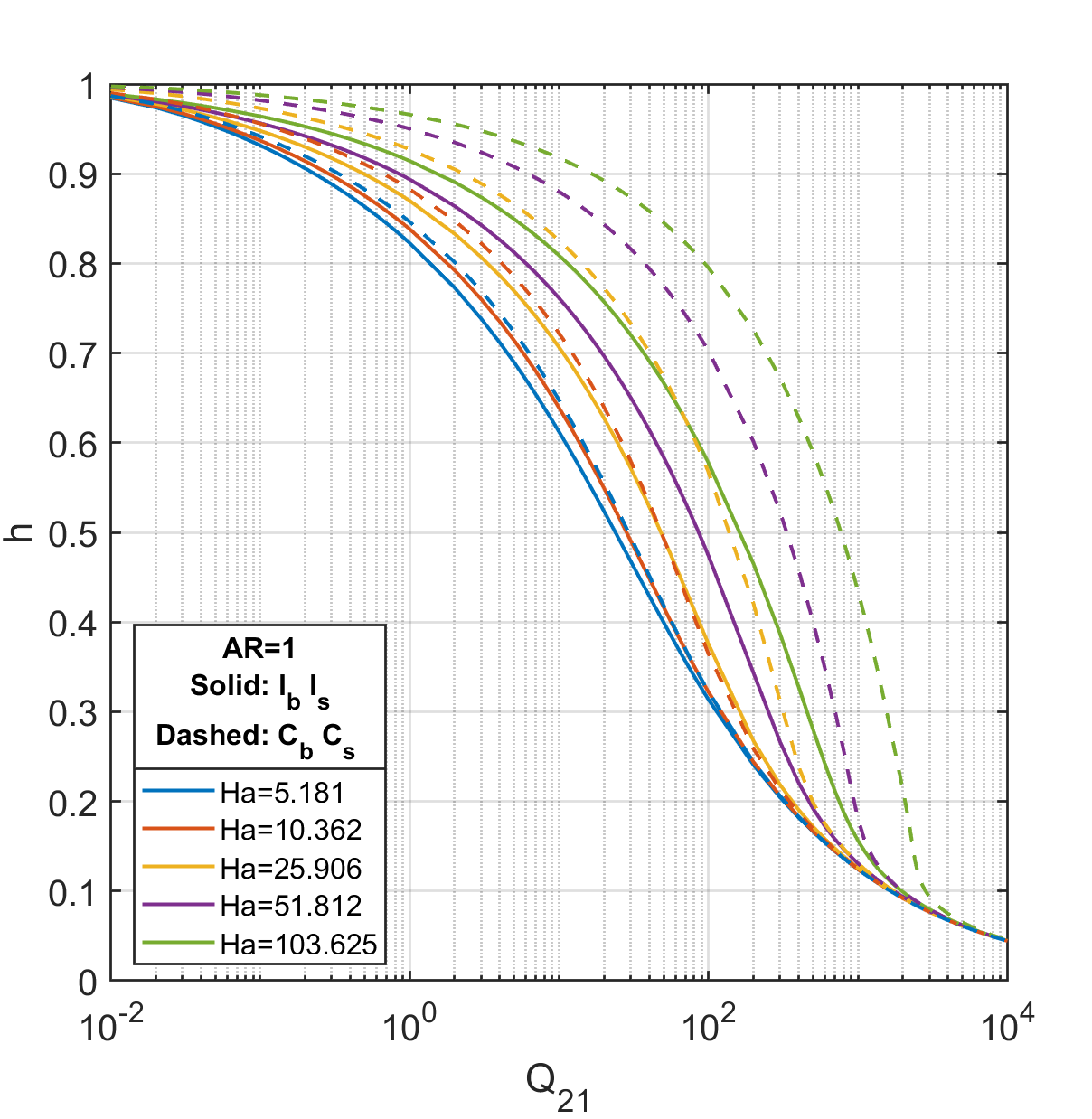}}
	\subfloat[]{\includegraphics[width=0.49\textwidth,clip]{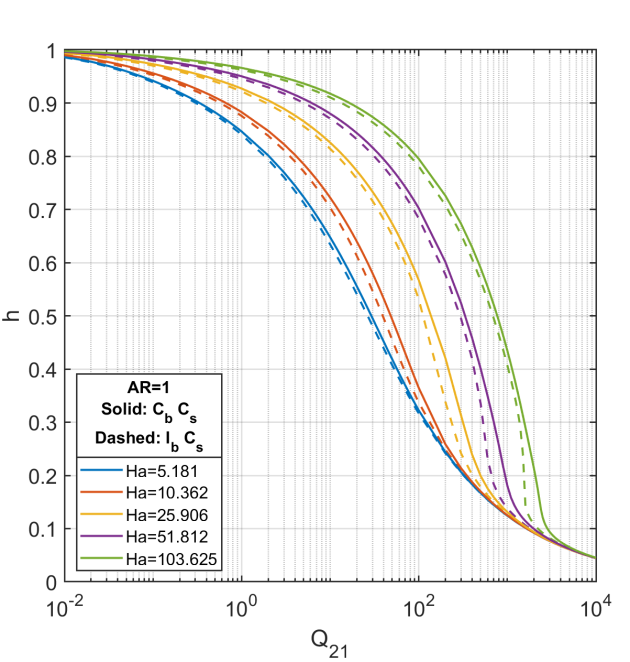}}
	\caption{\label{Fig: 4_18_h_effect_Q21_hor}Variation of the holdup vs. $Q_{21}$ for various $\Ha$. Comparison between (a) fully insulating, $I_bI_s$, and perfectly conducting, $C_bC_s$, ducts, and between (b) perfectly conducting,  $C_bC_s$, and insulating bottom and conducting side walls $I_bC_s$ ducts (solid and dashed lines, respectively).}
\end{figure}

The influence of the wall conductivities on the holdup is further analyzed in Fig.\ \ref{Fig: 4_18_h_effect_Q21_hor} for a square duct ($AR=1$) for different Ha over a wide range $Q_{21}$. Figure\ \ref{Fig: 4_18_h_effect_Q21_hor}a compares the holdup in a fully insulating (solid lines) and perfectly conducting (dashed lines) ducts, while Figure\ \ref{Fig: 4_18_h_effect_Q21_hor}b presents the difference between the holdup in conducting ducts (solid lines) and that obtained with an insulating bottom wall (dashed lines). Comparison of Fig.\ \ref{Fig: 4_18_h_effect_Q21_hor}a with Fig. \ref{Fig: 4_3_h_effect_Q21}a shows that in the case of square insulating ducts the holdup is almost unaffected by the direction of the magnetic field. However, Figure\ \ref{Fig: 4_18_h_effect_Q21_hor}a shows that although the holdup in a $C_b C_s$ duct is consistently higher than that obtained in a $I_b I_s$ duct, the impact of the wall conductivity is lower than that in the case of a vertical field, $B_{0|y}$ (cf. Fig. \ref{Fig: 4_3_h_effect_Q21}a). As shown in Fig.\ \ref{Fig: 4_18_h_effect_Q21_hor}a, the holdup converges to the same low values at high $Q_{21}$ at all $\Ha$, whereas for $B_{0|y}$ and high $\Ha$, the conductive fluid occupies a substantial part of the duct even at high $Q_{21}$. Figure\ \ref{Fig: 4_18_h_effect_Q21_hor}b further shows that ducts with an insulating bottom and conducting side walls ($I_b C_s$) exhibit slightly lower holdup than fully conducting ducts for all $\Ha$ and over almost the whole range of $Q_{21}$ considered. The reduced sensitivity of the holdup to the conductivity of the walls under $B_{0|x}$ is due to independence of the effective length scale of the conductive liquid holdup. Under $B_{0|y}$ and the same $\Ha$, the effective field strength increases as the holdup decreases, thereby amplifying sensitivity to wall conductivity.
\begin{figure}[h!]
	\centering
	\subfloat[$h=0.91, U_{1|\text{max}}=2.786, U_{2|\text{max}}=18.117$]{\includegraphics[width=0.33\textwidth,clip]{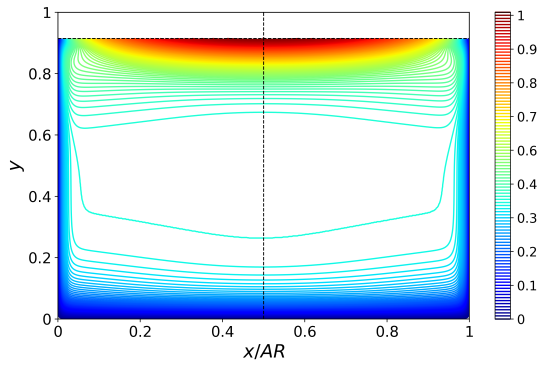}}
	\subfloat[$h=0.97, -0.361\le U_1\le 15.167, U_{2|\text{max}}=44.376$]{\includegraphics[width=0.33\textwidth,clip]{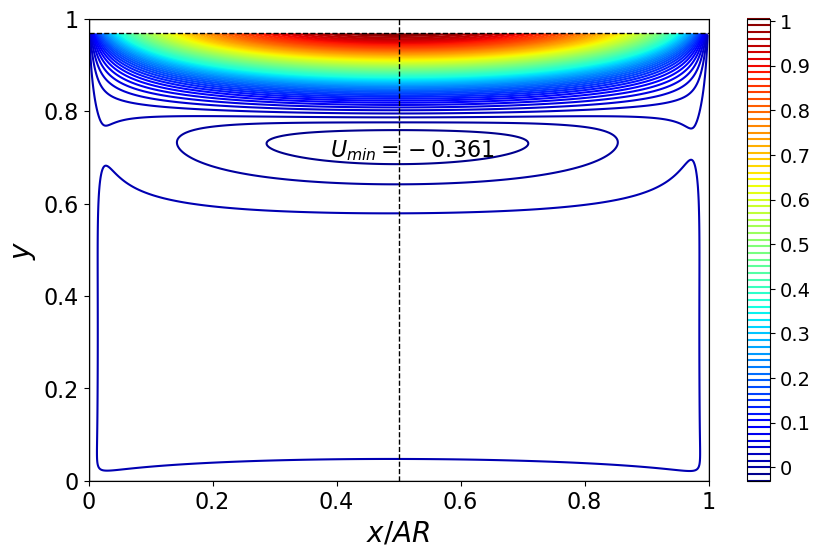}}
	\subfloat[$h=0.96, -0.296\le U_1\le 12.029, U_{2|\text{max}}=40.288$]{\includegraphics[width=0.33\textwidth,clip]{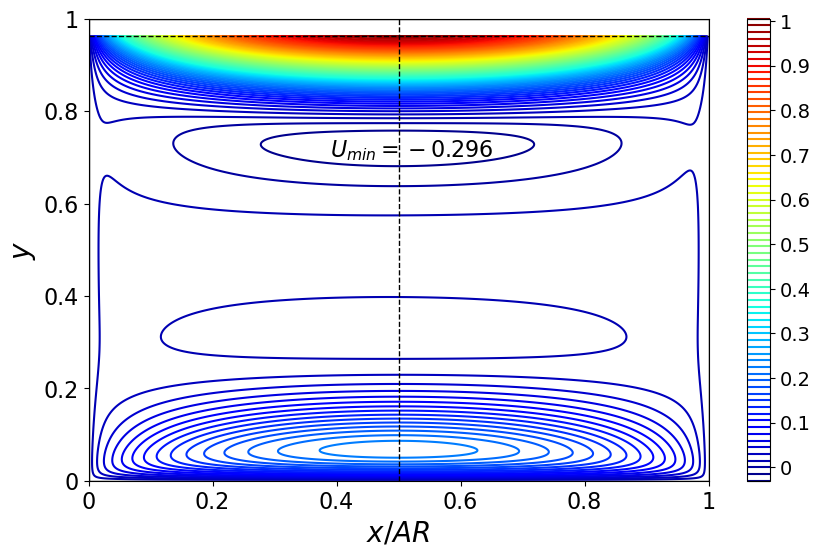}}
	\caption{\label{Fig: 4_19_U_contours_B_hor}Velocity contours of mercury (scaled by its maximal value) in a square duct for $Q_{21}=1,\Ha=103.625$: (a) fully insulating ducts $I_bI_s$, (b) perfectly conducting ducts $C_bC_s$, and (c) ducts with insulating bottom and conducting side walls $I_bC_s$.}
\end{figure}

\begin{figure}[h!]
	\centering
	\subfloat[]{\includegraphics[width=0.33\textwidth,clip]{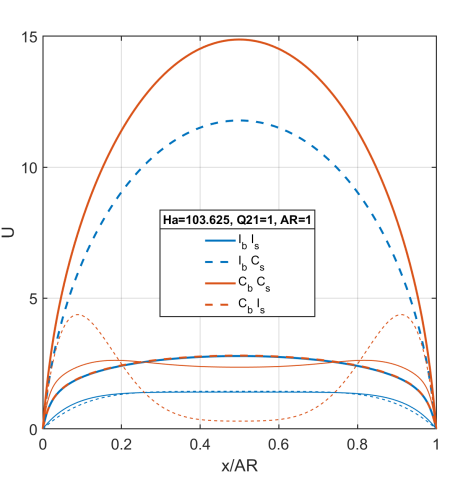}}
	\subfloat[]{\includegraphics[width=0.32\textwidth,clip]{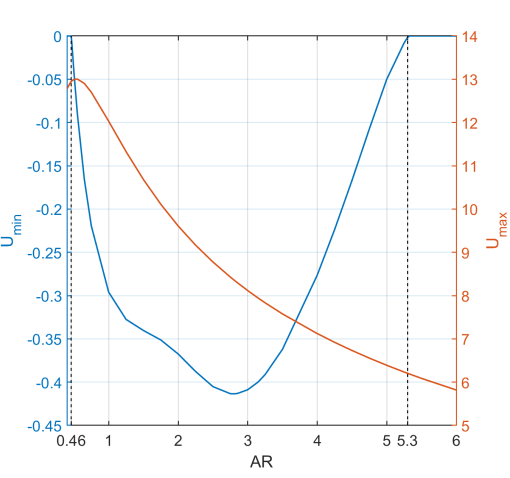}}
	\quad
	\subfloat[]{\includegraphics[width=0.32\textwidth,clip]{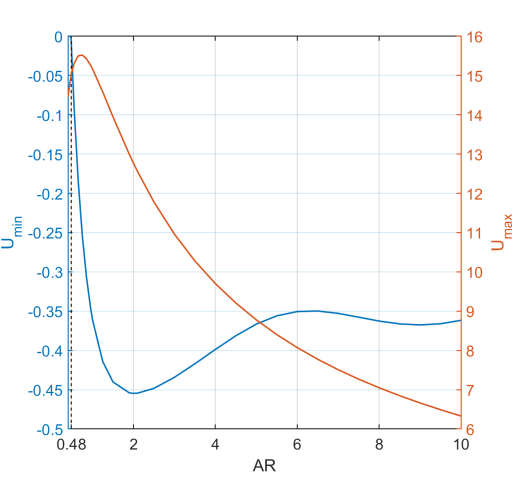}}
	\caption{\label{Fig: 4_20_Velocity_charact_B_hor}Characteristic velocities of the conductive fluid ($Q_{21}=1,\Ha=103.625$): (a) Interface velocity profile in square ducts. Comparison of the effects of walls conductivity under horizontal magnetic field (thick lines) and vertical magnetic field (thin lines). (b,c ) Effect of $AR$ on $U_\text{min}$ (in case of backflow) and $U_\text{max}$ in the mercury under a horizontal magnetic field for (b) insulating bottom and conducting side walls, $I_bC_s$, (c) all walls conducting. }
\end{figure}

The mercury velocity field in a square duct subjected to a horizontal magnetic field is shown in Fig.\ \ref{Fig: 4_19_U_contours_B_hor} for $\Ha=103.625$, $Q_{21}=1$. The same parameters were also used to study the velocity field under a vertical magnetic field (see Figs.\ \ref{Fig: 4_5_U_contours}c,d and \ref{Fig: 4_8_U_contours_CbIs}b,c). The velocity values are normalized by the maximum mercury velocity, which always occurs at the mercury--air interface, regardless of the wall conductivities or the magnetic field orientation. For the same wall conductivity configuration, however, the maximal velocity value is significantly higher for the case of $B_{0|x}$. The interfacial velocity for this case is shown in Fig.\ \ref{Fig: 4_20_Velocity_charact_B_hor}a. As can be seen in the figure, the maximum velocity is always located at the center of the interface. In contrast, under $B_{0|y}$ and a conducting bottom wall, the maximum interfacial velocity shifts off-centre toward the side walls. Due to the lower holdup obtained under $B_{0|x}$ ($C_b C_s$) ducts, the maximal velocity of the air is lower in this case, however, it is still significantly higher than that of the mercury. Accordingly, the air velocity profiles (not shown) resemble those obtained when a vertical magnetic field is applied. 

\begin{figure}[h!]
	\centering
	\subfloat[]{\includegraphics[width=0.33\textwidth,clip]{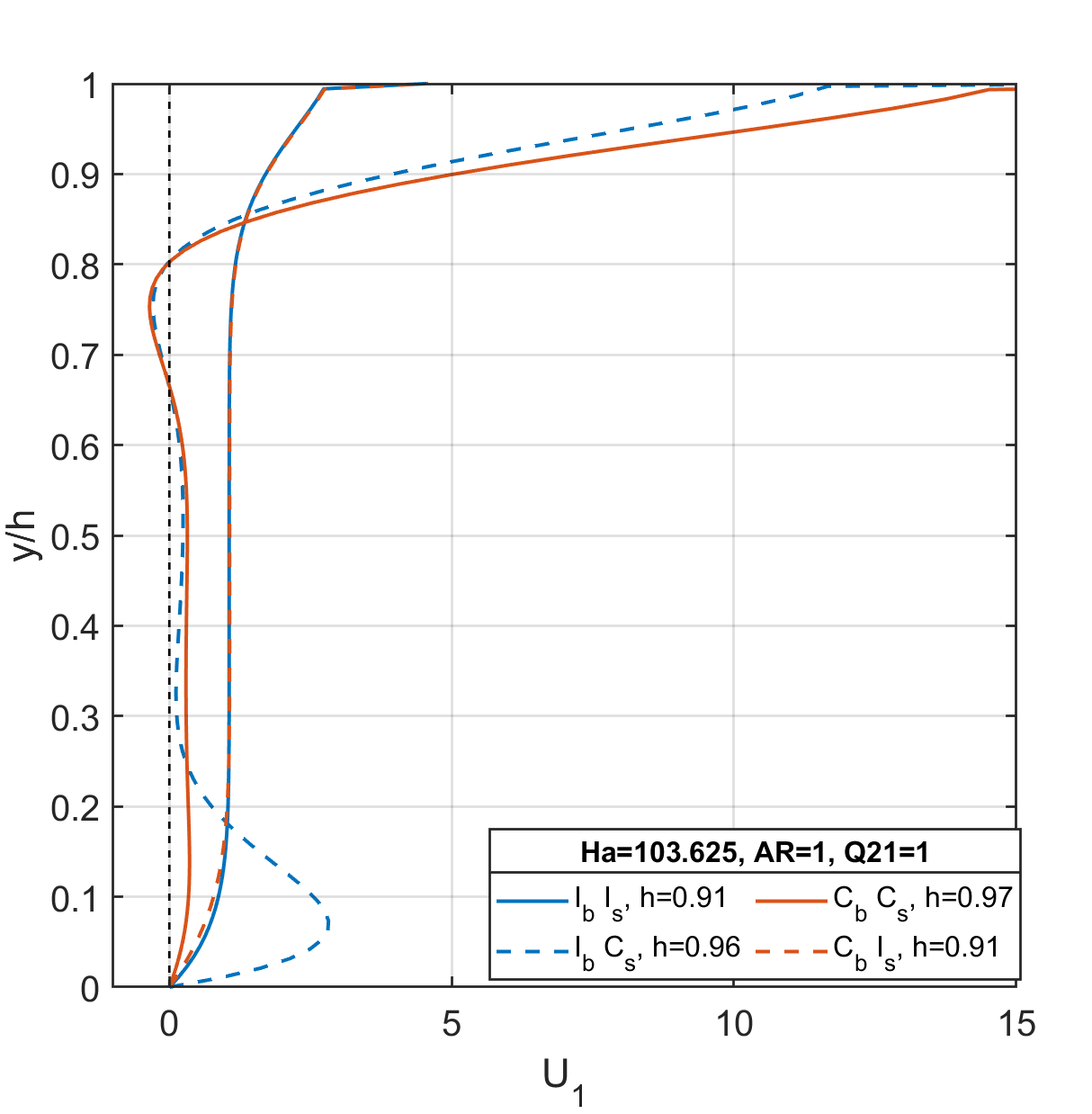}}
	\subfloat[]{\includegraphics[width=0.33\textwidth,clip]{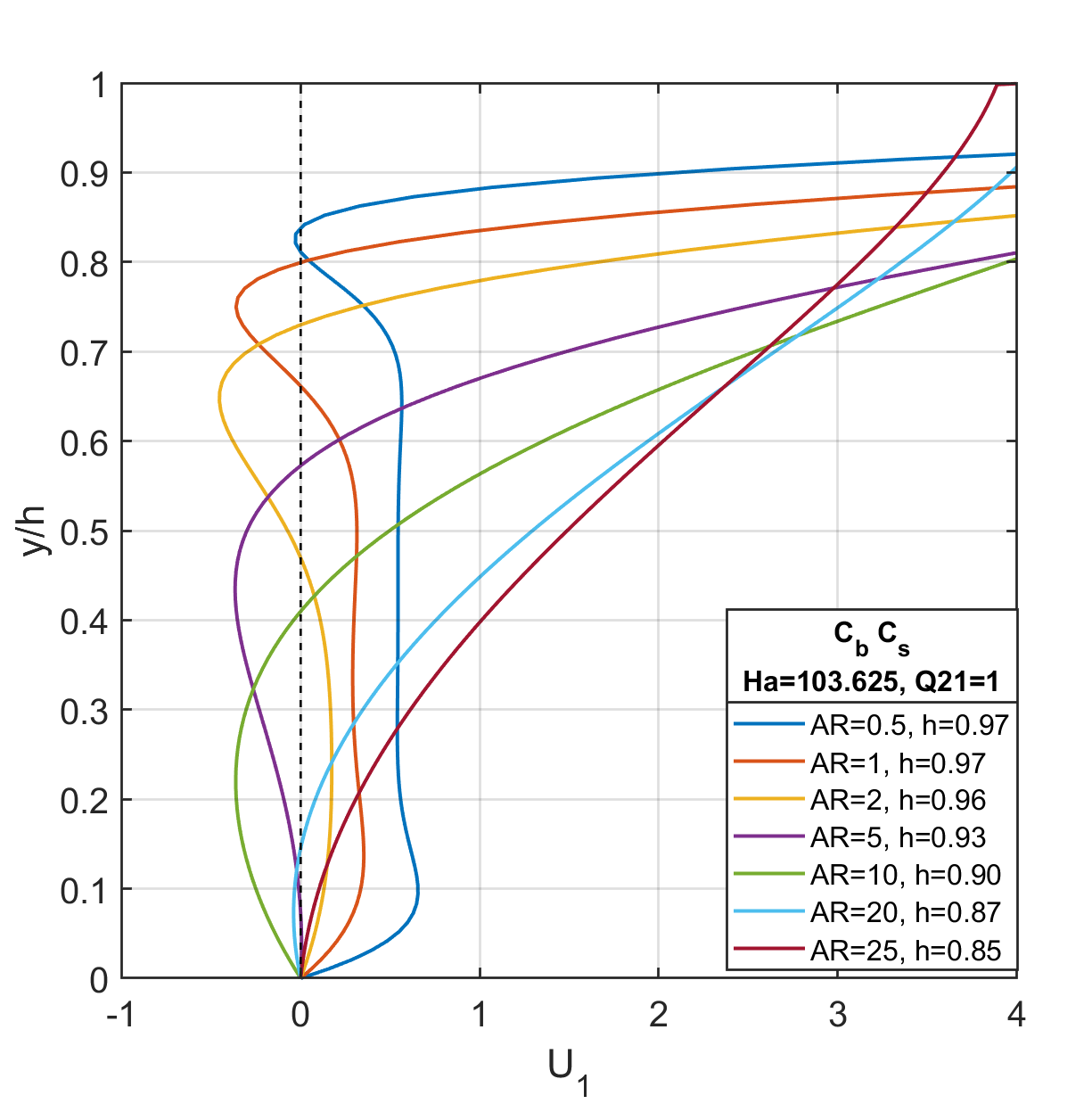}}
	\subfloat[]{\includegraphics[width=0.33\textwidth,clip]{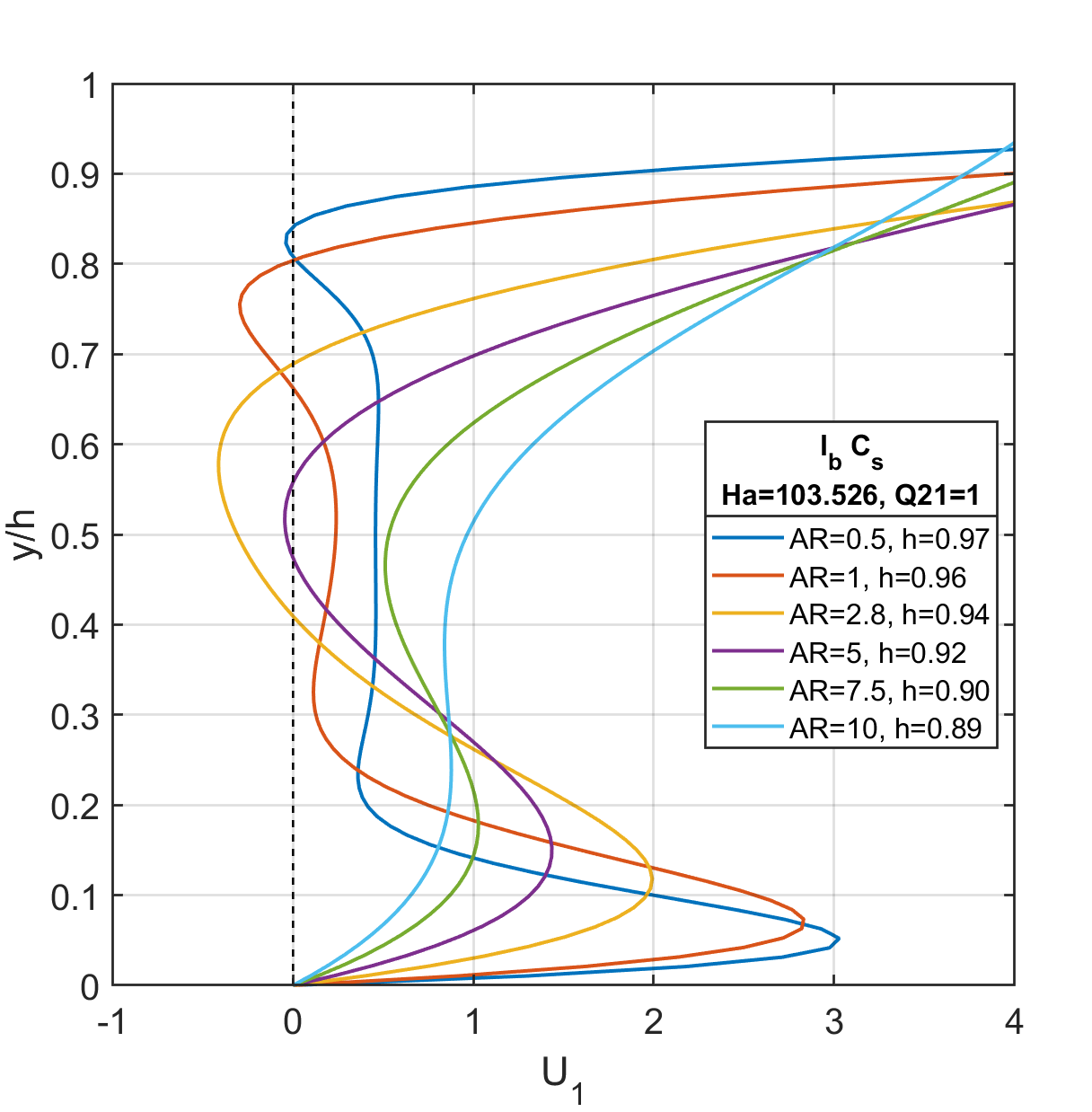}}
	\caption{\label{Fig: 4_21_Centerline_Velocity_B_hor}Centerline mercury velocity profiles for $Q_{21}=1$ and $\Ha=103.625$. (a) Effect of wall conductivities in a square duct,(b,c) effect of aspect ratio in (b) fully conducting ducts and (c) ducts with insulating bottom and conducting side walls. The maximal velocity in (b,c,) is at the interface (out of the range of those figures, see values in Figure\ \ref{Fig: 4_20_Velocity_charact_B_hor}c,b.}
\end{figure}

Comparing Fig.\ \ref{Fig: 4_19_U_contours_B_hor}a with Fig.\ \ref{Fig: 4_5_U_contours}b and considering the fact that the maximum velocity used for scaling the results for $B_{0|x}$ (i.e., $U_{1|\text{max}}=2.786$) is about twice that obtained for $B_{0|y}$ (i.e., $U_{1|\text{max}}=1.408$), indicates that in fully insulating ducts ($I_b I_s$), the mercury velocity field and the corresponding bulk velocity are similar in both cases. Replacing the bottom wall by a conducting one (i.e., $C_b I_s$), which was found to significantly affect the velocity field under $B_{0|y}$ (see Fig.\ \ref{Fig: 4_8_U_contours_CbIs}), results in a minor change of the velocity field under $B_{0|x}$. The velocity is practically the same as in Fig.\ \ref{Fig: 4_19_U_contours_B_hor}a and therefore not shown. This can however be observed in Fig.\ \ref{Fig: 4_21_Centerline_Velocity_B_hor}a where centerline velocity profiles for $I_b I_s$ and $C_b I_s$ practically overlap. 

The influence of magnetic field orientation is further highlighted by comparing the velocity fields in $C_b C_s$ square ducts (compare Fig.\ \ref{Fig: 4_19_U_contours_B_hor}b vs. Fig.\ \ref{Fig: 4_5_U_contours}c). Under $B_{0|x}$, the maximum (dimensionless) interfacial velocity reaches $14.875$, compared to only $2.62$ under $B_{0|y}$. This occurs despite the lower mercury holdup, and therefore lower air velocity associated with $B_{0|x}$. In this case, the high-velocity "jetting" region near the interface penetrates deeper into the mercury, with a compensating axial backflow region forming beneath it to satisfy the imposed flow rate. As a result, further from the interface, a bulk region of nearly uniform mercury velocity develops and occupies most of the layer, almost up to the bottom wall. This behavior is clearly illustrated in Fig.\ \ref{Fig: 4_21_Centerline_Velocity_B_hor}a, which shows the corresponding centerline velocity profile. 

When the side walls are conducting and the bottom wall is insulating ($I_b C_s$, Fig.\ \ref{Fig: 4_19_U_contours_B_hor}c), a jetting region appears also in the lower part of the duct. Consequently, the mercury bulk region and its velocity are further reduced, as shown by the corresponding centerline velocity profile ($I_b C_s$) in Fig.\ \ref{Fig: 4_21_Centerline_Velocity_B_hor}a. This velocity field is very different from that obtained with $B_{0|y}$, where the same wall configuration produced a profile similar to $I_b I_s$. A common feature, however, is that jetting occurs near insulating boundaries whenever conducting surface(s) are present. Under $B_{0|y}$  (Fig.\ \ref{Fig: 4_8_U_contours_CbIs}), two symmetric jets develop near the insulating side walls when the bottom wall is conducting. On the other hand, under $B_{0|x}$, jetting occurs near both the insulating bottom wall and the interface when the side walls are conducting. Here, the presence of the interface breaks the symmetry, and the bottom jet is much weaker than the dominant interfacial jet. 
\begin{figure}[h!]
	\centering
	\subfloat[$AR=0.5, -0.041<U_1<12.99, h=0.97$]{\includegraphics[width=0.49\textwidth,clip]{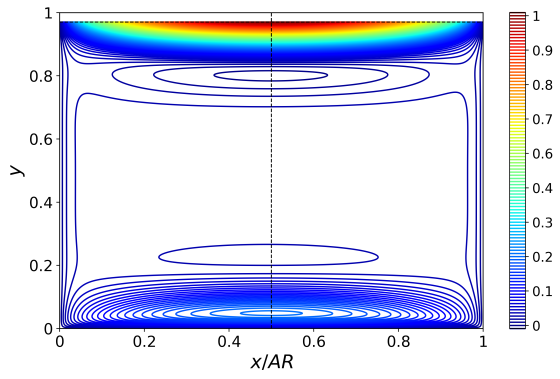}}
	\subfloat[$AR=2.8, -0.413<U_1<8.363, h=0.94$]{\includegraphics[width=0.49\textwidth,clip]{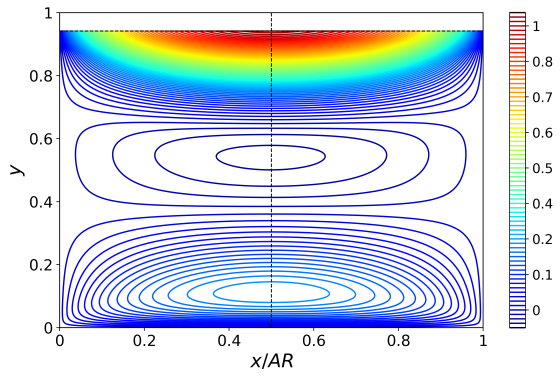}}
	\\
	\subfloat[$AR=5, -0.049<U_1<6.389, h=0.92$]{\includegraphics[width=0.49\textwidth,clip]{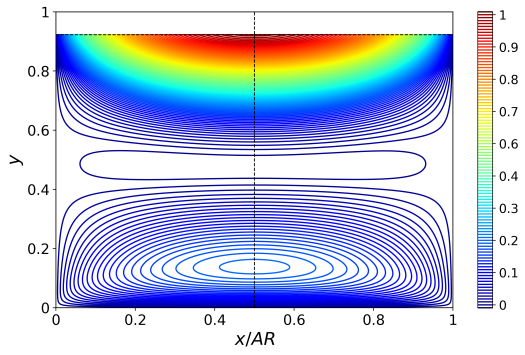}}
	\subfloat[$AR=5.25, -0.008<U_1<6.233, h=0.92$]{\includegraphics[width=0.49\textwidth,clip]{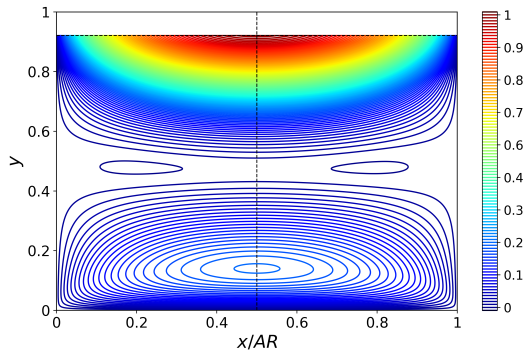}}
	\caption{\label{Fig: 4_22_Umercury_contours_hor}Effect of $AR$ on the mercury velocity field (scaled by its maximal value) in rectangular ducts with insulating bottom and conducting side walls, $I_bC_s$, for $Q_{21}=1$ and $\Ha=103.625$: (a) $AR=0.5$, (b) $AR=2.8$ (c) $AR=5$ (d) $AR=5.25$.}
\end{figure}
\begin{figure}[h!]
	\centering
	\subfloat[$h=0.82, \big|b/\Ha\big|\le 0.098$]{\includegraphics[width=0.4\textwidth,clip]{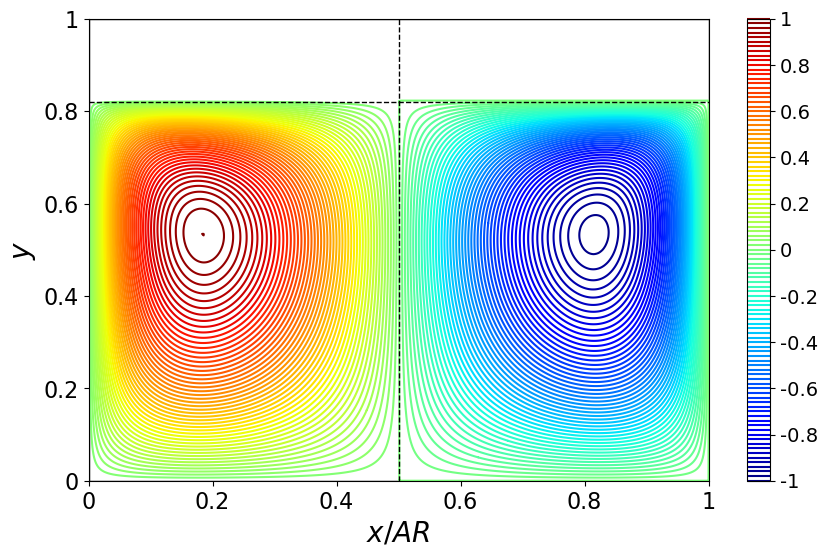}}
	\qquad
	\subfloat[$h=0.92, \big|b/\Ha\big|\le 0.012$]{\includegraphics[width=0.4\textwidth,clip]{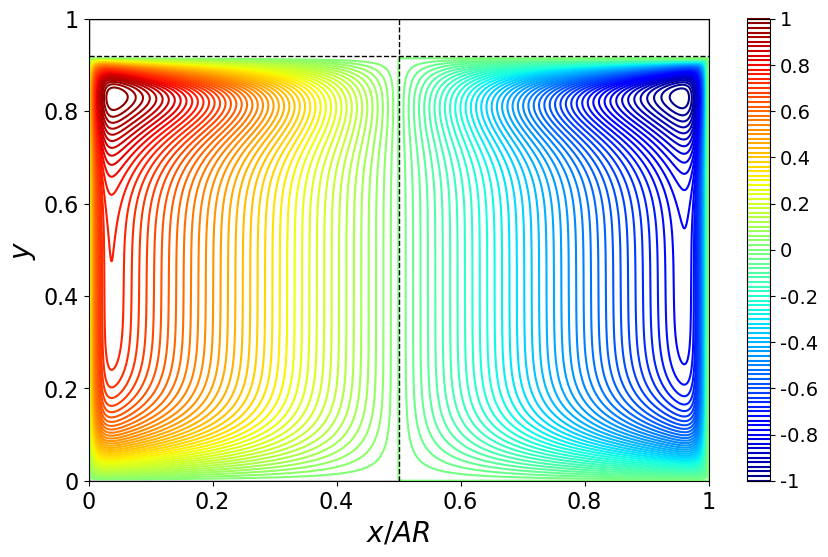}}
	\\
	\subfloat[$h=0.85, \big|b/\Ha\big|\le 0.38$]{\includegraphics[width=0.4\textwidth,clip]{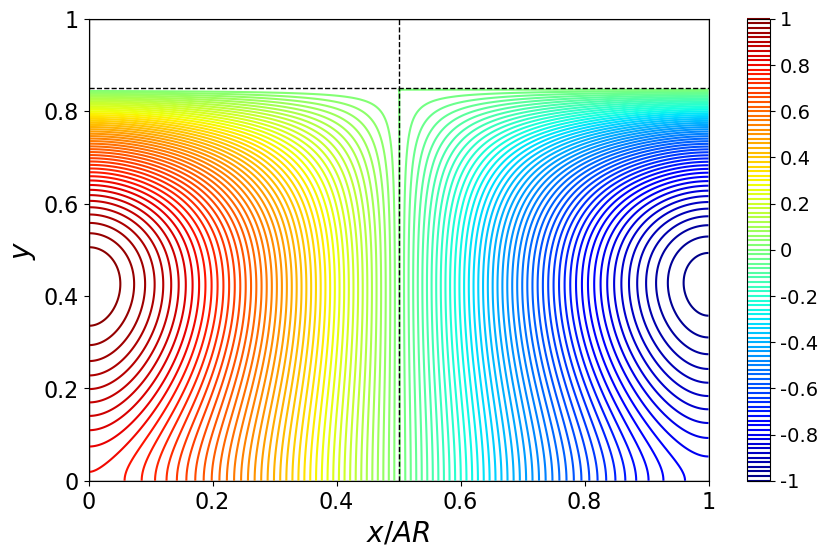}}
	\qquad
	\subfloat[$h=0.97, \big|b/\Ha\big|\le 0.174$]{\includegraphics[width=0.4\textwidth,clip]{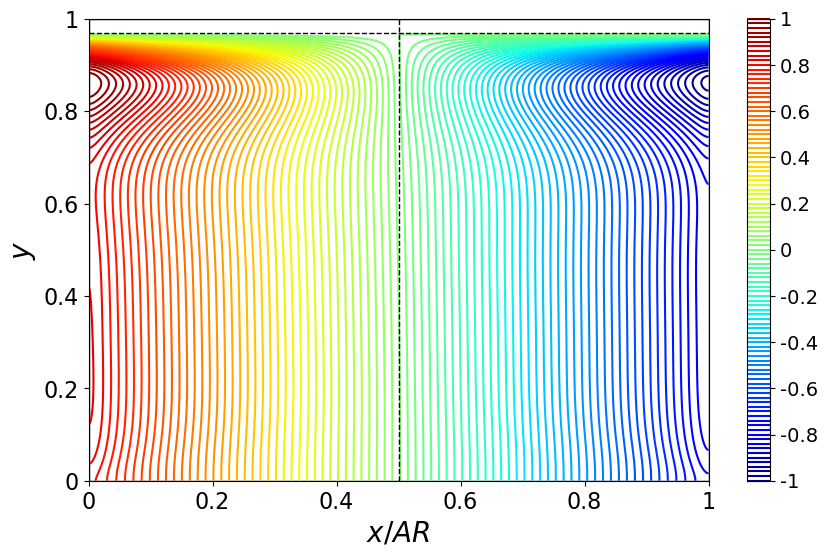}}
	\\
	\subfloat[$h=0.84, \big|b/\Ha\big|\le 0.318$]{\includegraphics[width=0.4\textwidth,clip]{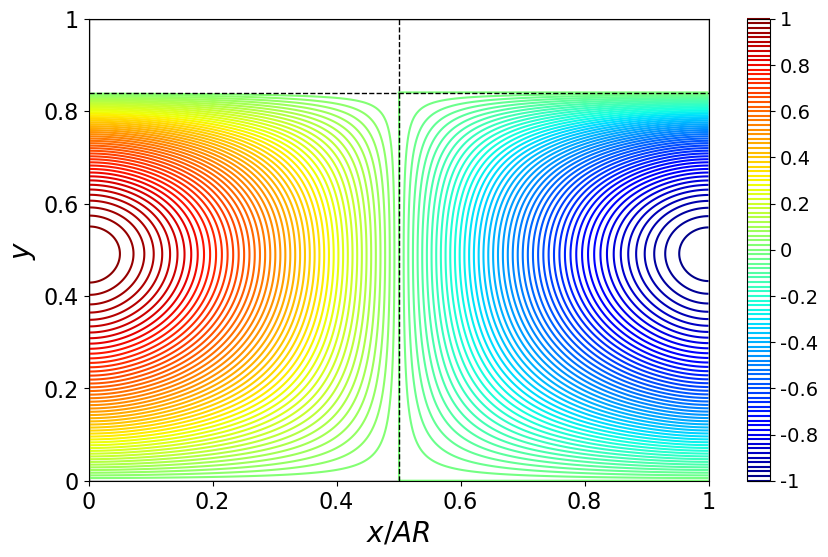}}
	\qquad
	\subfloat[$h=0.97, \big|b/\Ha\big|\le 0.134$]{\includegraphics[width=0.4\textwidth,clip]{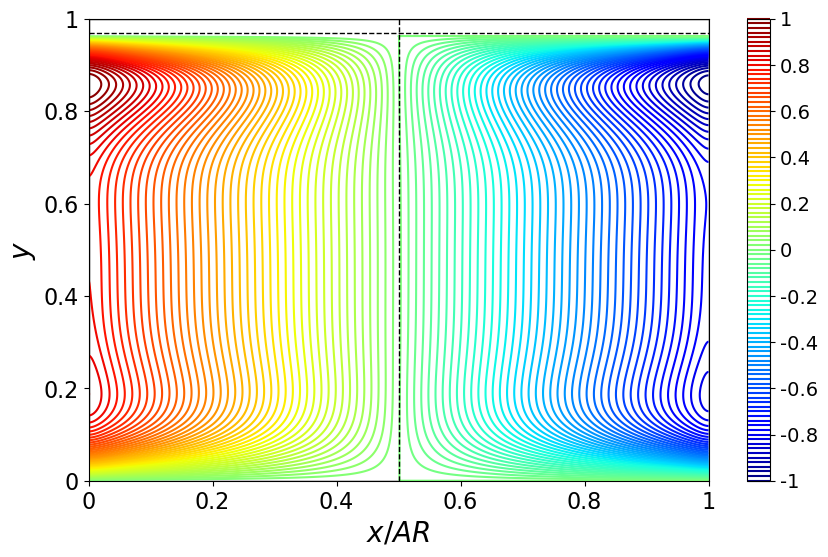}}
	\caption{\label{Fig: 4_23_b_contours_hor}Induced magnetic field contours in a square duct for $Q_{21}=1$. $\Ha=5.181$ and $\Ha=103.625$ in the l.h.s. and r.h.s. frames, respectively. (a, b) fully insulating duct, $I_bI_s$; (c, d) perfectly conducting duct, $C_bC_s$; (e, f) bottom insulating and conducting side walls, $I_bC_s$. }
\end{figure}

Examination of the velocity fields for different aspect ratios in the ducts with insulating side walls ($I_b I_s$ and $C_b I_s$ ducts) revealed that they are similar to those obtained in square ducts (Figs.\ \ref{Fig: 4_19_U_contours_B_hor}a, \ref{Fig: 4_20_Velocity_charact_B_hor}a, and \ref{Fig: 4_21_Centerline_Velocity_B_hor}a). However, with conducting side walls ($C_b C_s$ or $I_b C_s$), the effect of $AR$ on the velocity field is significant and is demonstrated in Figs.\ \ref{Fig: 4_20_Velocity_charact_B_hor}b,c, \ref{Fig: 4_21_Centerline_Velocity_B_hor}b,c, and \ref{Fig: 4_22_Umercury_contours_hor}. The centerline velocity profiles for the $I_b C_s$ ducts of different $AR$ are shown in Fig.\ \ref{Fig: 4_21_Centerline_Velocity_B_hor}c, while the cross-sectional contours are shown in Fig.\ \ref{Fig: 4_22_Umercury_contours_hor}. Similar to the case of $B_{0|y}$, under $B_{0|x}$ the central backflow region diminishes in narrow ducts, while the jetting velocities increase. The behavior of the velocity field observed in the figures can be explained by the lateral confinement effect, which restricts the axial recirculation, but somewhat enhances the maximal velocity at the interface and the jetting near the bottom wall. The backflow region disappears in narrow ducts with $AR<0.46$, in which the maximal mercury velocity decreases, contrary to the trend in wider ducts (see Fig.\ \ref{Fig: 4_20_Velocity_charact_B_hor}b). In wider ducts, e.g. $AR=2.8$ in Fig.\ \ref{Fig: 4_21_Centerline_Velocity_B_hor}c, the larger cross-section allows a thicker central zone of back flow, which extends across the entire gap between the two jets. Moreover, as shown in Fig.\ \ref{Fig: 4_20_Velocity_charact_B_hor}b, in rectangular $I_b C_s$ ducts of $AR>1$, the maximal velocity monotonically decreases with increasing $AR$, whereas the strongest backflow ($-0.408 U_{1S}$) is achieved in $AR\approx2.8$ ducts. The backflow region stretches in the spanwise direction and weakens in wider ducts (see Fig.\ \ref{Fig: 4_22_Umercury_contours_hor}c,d). It splits into two separate regions of weak backflow for $AR=5.25$ (Fig.\ \ref{Fig: 4_22_Umercury_contours_hor}d) and then disappears in ducts with $AR\gtrsim 5.35$. The jetting flow near the bottom wall also weakens with increasing the $AR$ (see Fig.\ \ref{Fig: 4_21_Centerline_Velocity_B_hor}c), disappears at $AR\approx10$ and is not found for wider ducts, in which the local (lower) maximum in the velocity profile is no longer obtained. 

In perfectly conducting ($C_b C_s$) narrow ducts ($AR<1$), the lateral confinement effects on the backflow and the maximal velocity are similar to those obtained in $I_b C_s$ ducts discussed above (see Figs.\ \ref{Fig: 4_20_Velocity_charact_B_hor}c and \ref{Fig: 4_21_Centerline_Velocity_B_hor}b). The backflow is the strongest ($U_\text{min}=-0.455 U_{1S}$) in a rectangular duct of $AR\approx2$. With further increase of $AR$, the backflow only slightly weakens, but pushed downward, resulting in reversed wall shear at the bottom wall, as seen in Fig.\ \ref{Fig: 4_20_Velocity_charact_B_hor}c for $AR=10$. In even wider ducts, the near-wall backflow region weakens with increasing $AR$. It disappears for $AR\gtrsim25$, out of the range shown in Fig.\ \ref{Fig: 4_20_Velocity_charact_B_hor}c.    
\begin{figure}[h!]
	\centering
	\subfloat[]{\includegraphics[width=0.33\textwidth,clip]{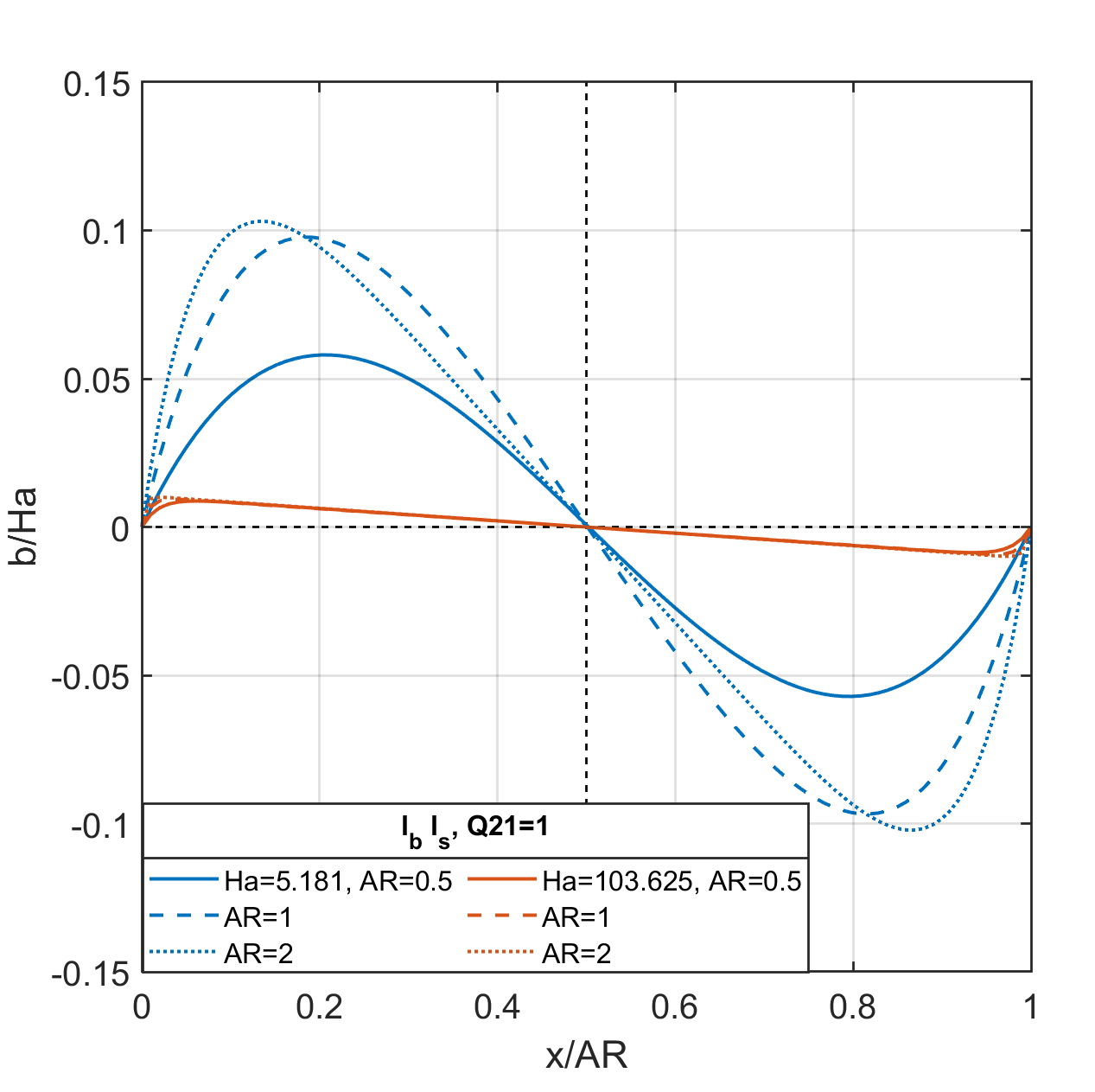}}
	\subfloat[]{\includegraphics[width=0.33\textwidth,clip]{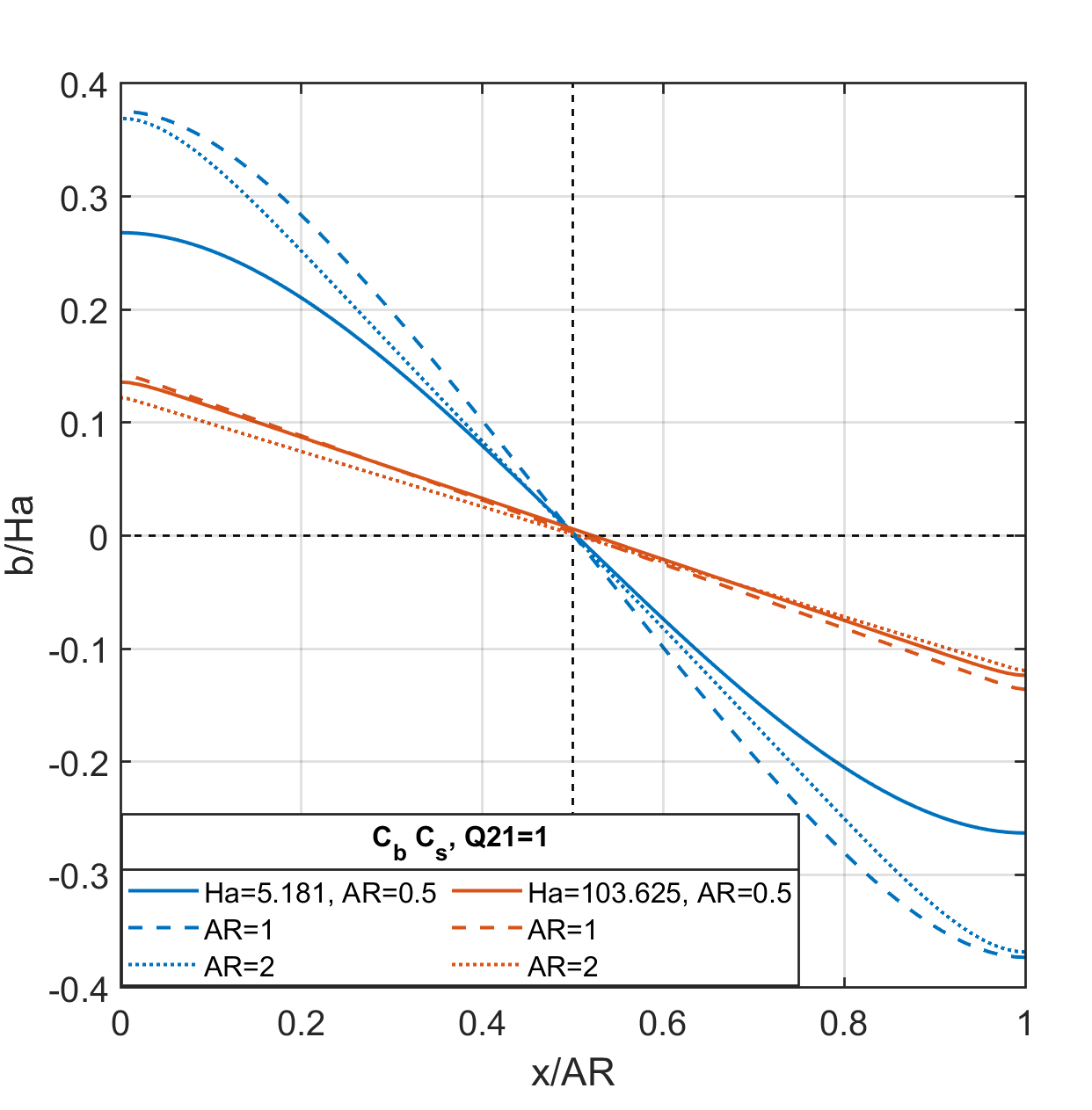}}
	\subfloat[]{\includegraphics[width=0.33\textwidth,clip]{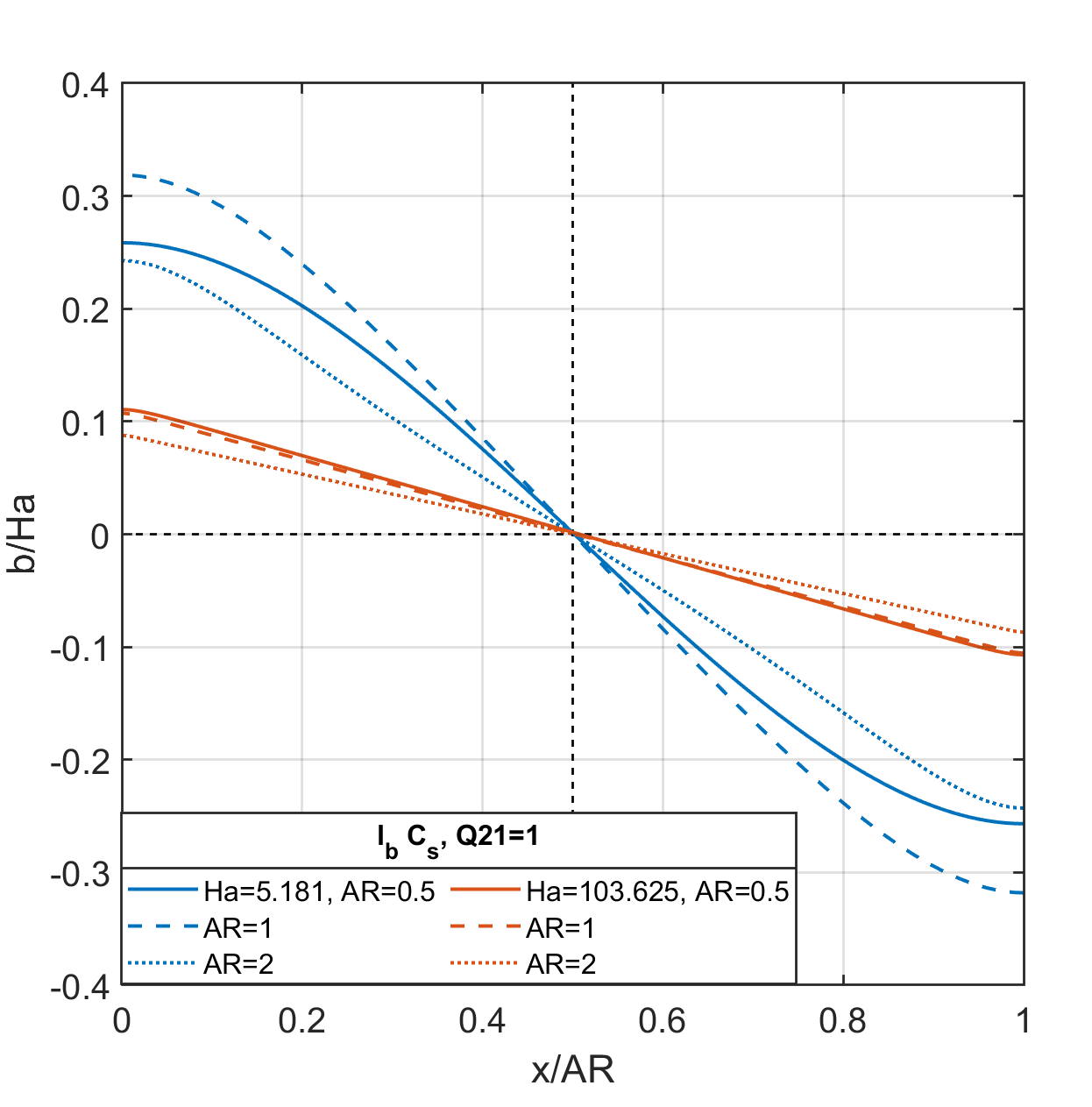}}
	\caption{\label{Fig: 4_24_b_linegraph_center_hor}Variation of the induced magnetic field profiles, $b/\Ha$ vs. $x/AR$ across the center of the conductive liquid for $Q_{21}=1$ and various $AR$: (a) $I_bI_s$, (b) $C_bC_s$, and (c) $I_bC_s$ ducts.}
\end{figure}

Figure\ \ref{Fig: 4_23_b_contours_hor} presents the contours of the induced magnetic field, $b/Ha$, in a square duct subjected to a horizontal magnetic field for low $\Ha$ ($=5.181$, l.h.s. frames) and high $\Ha$ ($=103.625$, r.h.s. frames). For all wall conductivity combinations, the contours are antisymmetric across the vertical centerline, $x/AR=0.5$. As seen in Fig.\ \ref{Fig: 4_23_b_contours_hor}, with increasing $\Ha$, the extrema of the induced magnetic field ($b_\text{max}/\Ha$ or $b_\text{min}/\Ha$) within the conductive liquid shifts toward the interface, and for $\Ha=103.625$ (r.h.s frames) it is located almost at the interface. The spanwise (horizontal) position of the extrema depends also on side wall conductivity. With insulating side walls ($I_b I_s$ or $C_b I_s$, for which the contours are similar so that for the latter they not shown), the extrema are located in the core region of the conductive liquid at low $\Ha$ (Fig.\ \ref{Fig: 4_23_b_contours_hor}a), but shift toward the side walls at high $\Ha$ (Fig.\ \ref{Fig: 4_23_b_contours_hor}b). When the side walls are conducting, $C_b C_s$ or $I_b C_s$, the extrema appear on the side walls regardless of $\Ha$ (Fig.\ \ref{Fig: 4_23_b_contours_hor}c-f), and the intensity of $b/Ha$ across the conductive liquid core varies almost linearly in the spanwise direction (see Fig.\ \ref{Fig: 4_24_b_linegraph_center_hor}b). In the $I_b C_s$ configuration (Fig.\ \ref{Fig: 4_23_b_contours_hor}f), corresponding to a jetting flow region near the bottom wall, a secondary pair of local extrema can be observed near the bottom corners of the duct. Overall, providing the same wall-conductivity configuration, the general pattern of the induced magnetic field contours are qualitatively similar in non-square ($AR\ne 1$) and square ducts when plotted in normalized coordinates $x/AR$ and $y/h$. This is demonstrated in Fig.\ \ref{Fig: 4_24_b_linegraph_center_hor} by showing the spanwise variation of $b/Ha$ across the conductive liquid core ($y/h=0.5$). 
\begin{figure}[h!]
	\centering
	\subfloat[]{\includegraphics[width=0.49\textwidth,clip]{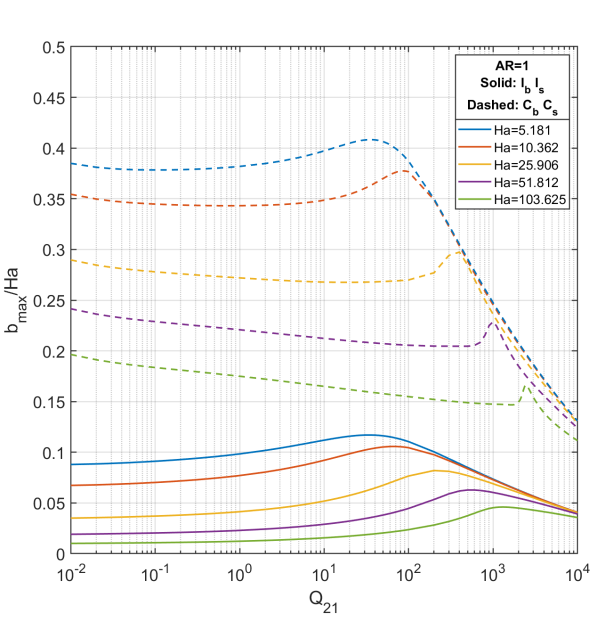}}
	\subfloat[]{\includegraphics[width=0.49\textwidth,clip]{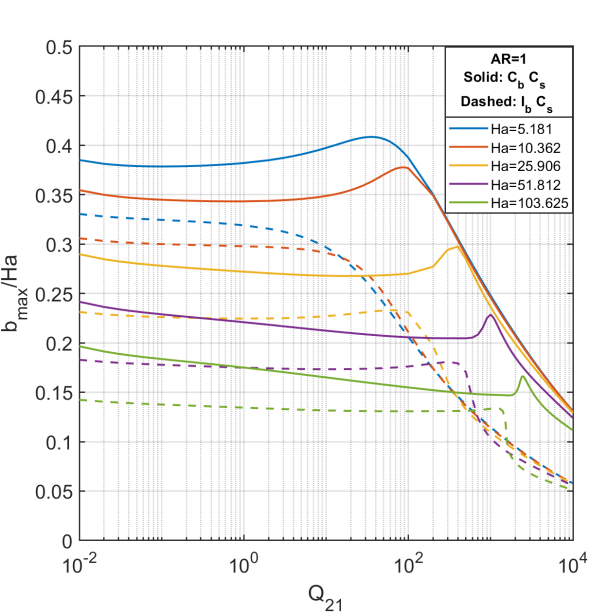}}
	\caption{\label{Fig: 4_25_b_effect_Q21_hor}Variation of $b_\text{max}/\Ha$ with $Q_{21}$ in a square duct: (a) Fully insulating ($I_bI_s$, solid lines) compared to perfectly conducting ducts ($C_bC_s$, dashed). (b) $C_bC_s$ (solid) ducts compared to ducts with bottom insulating and conducting side walls ($I_bC_s$, dashed).}
\end{figure}

Figure\ \ref{Fig: 4_25_b_effect_Q21_hor} presents the variation of $b_\text{max}/\Ha$ with $Q_{21}$ for different $\Ha$ in square ducts. Similar to the case of $B_{0|y}$ (Fig.\ \ref{Fig: 4_7_bmax_effect_Q21}), $b_\text{max}/\Ha$ decreases with increasing $\Ha$ and increases when replacing insulating with conducting walls. However, comparison of Fig.\ \ref{Fig: 4_25_b_effect_Q21_hor}a and Fig.\ \ref{Fig: 4_7_bmax_effect_Q21}a reveals that under $B_{0|x}$, the $b_\text{max}/\Ha$ values are considerably lower, e.g., approximately half those obtained with a vertical field in a conducting duct ($C_b C_s$). Replacing the bottom wall with an insulating one ($I_b C_s$) further reduces $b_\text{max}/\Ha$ (Fig.\ \ref{Fig: 4_25_b_effect_Q21_hor}b). Since in these cases $b/Ha$ varies almost linearly across the spanwise direction (see Fig.\ \ref{Fig: 4_24_b_linegraph_center_hor}b,c), smaller $b_\text{max}/\Ha$ corresponds to lower $\partial b/\partial x$, and consequently, a weaker Lorentz force. As a result, the holdup in these ducts is significantly lower than that obtained under a vertical field for the same $\Ha$ and $Q_{21}$, as discussed with reference to Figs.\ \ref{Fig: 4_17_h_effect_AR_hor} and \ref{Fig: 4_18_h_effect_Q21_hor}.   
\begin{figure}[h!]
	\centering
	\subfloat[]{\includegraphics[width=0.49\textwidth,clip]{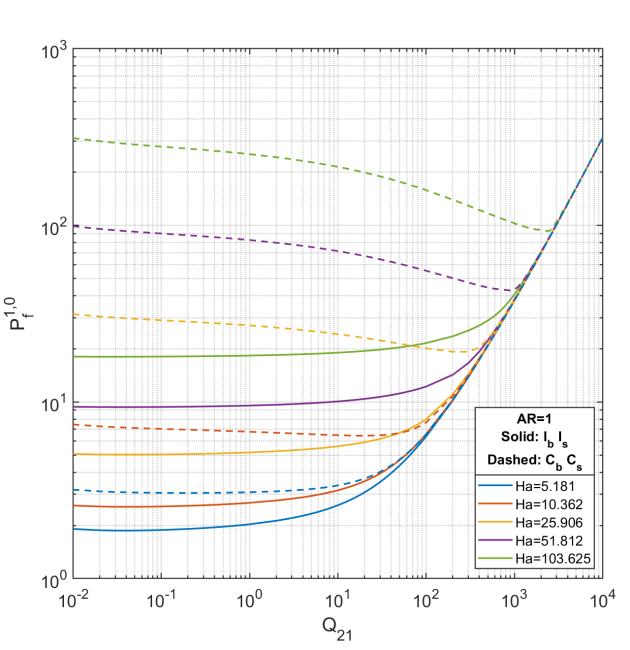}}
	\subfloat[]{\includegraphics[width=0.49\textwidth,clip]{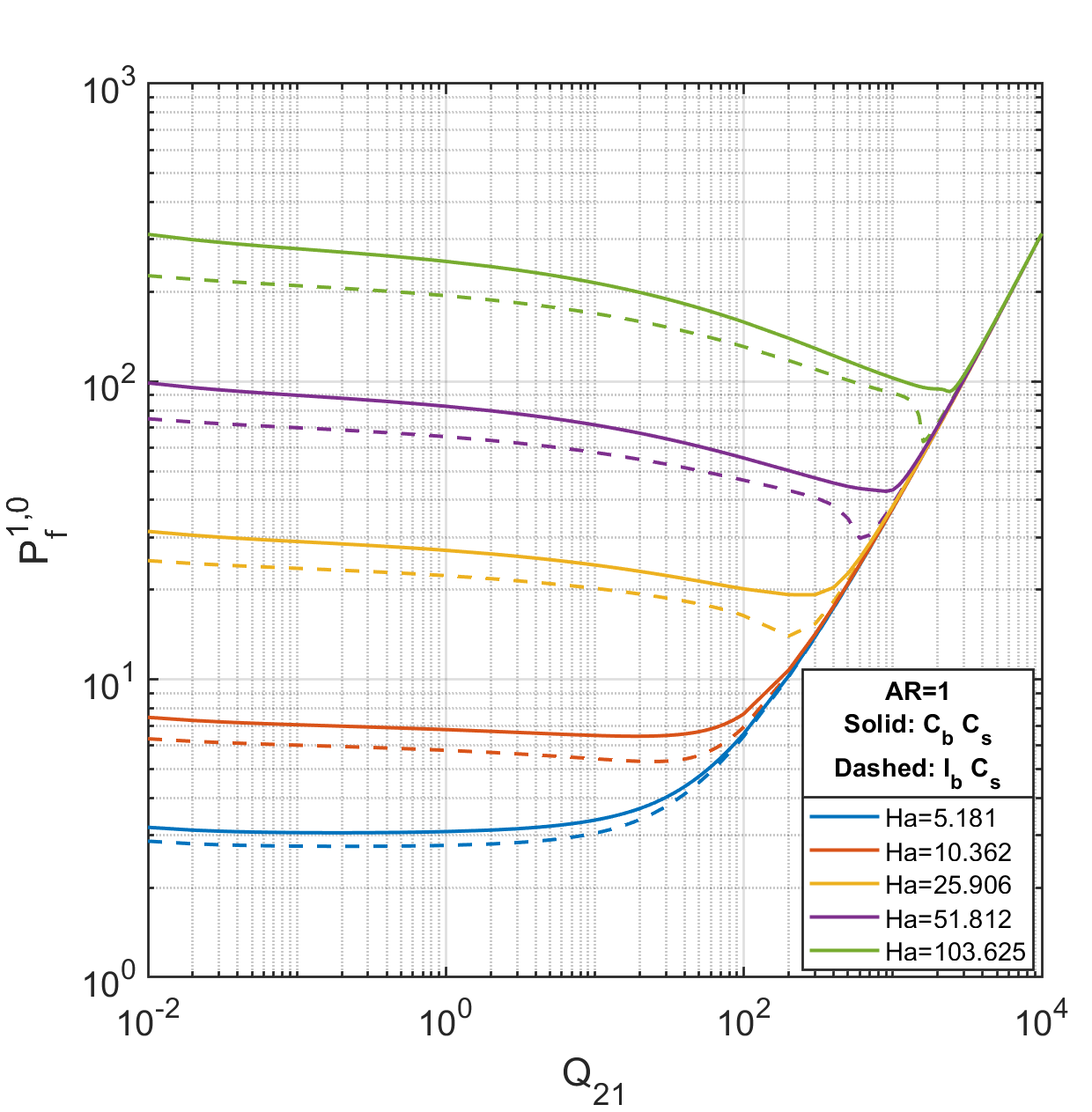}}
	\caption{\label{Fig: 4_26_P10_f_hor}Variation of the dimensionless pressure gradient $P_f^{1,0}$ (scaled by single phase flow of the conductive liquid between two-plates without magnetic field, $\Ha=0$) with $Q_{21}$ in a square ducts under vertical magnetic field for various $\Ha$: (a) fully insulating ($I_bI_s$, solid) compared to perfectly conducting ($C_bC_s$, dashed) ducts. (b) perfectly conducting ($C_bC_s$, solid) compared to insulating bottom and conducting side walls ($C_bI_s$,dashed) ducts.}
\end{figure}

It is further noted that unlike the vertical-field case, the monotonic increase of $b_\text{max}/\Ha$ with $Q_{21}$ is not observed under a horizontal field. Except at low $\Ha$, a global or local maximum (depending on whether it is higher or lower than the single-phase value, respectively) appears at relatively high gas flow rates ($Q_{21}>1$), followed by pronounced decline. The $Q_{21}$ value corresponding to this maximum increases with $\Ha$. Similar trends of  $b_\text{max}/\Ha$ are obtained in non-square ducts (not shown). However, increasing $AR$ (for wide ducts, $AR>1$) reduces the local peak value of $b_\text{max}/\Ha$ and shifts it toward lower $Q_{21}$, whereas in narrow ducts it increases and shifts it toward higher $Q_{21}$.  
\begin{figure}[h!]
	\centering
	\subfloat[]{\includegraphics[width=0.49\textwidth,clip]{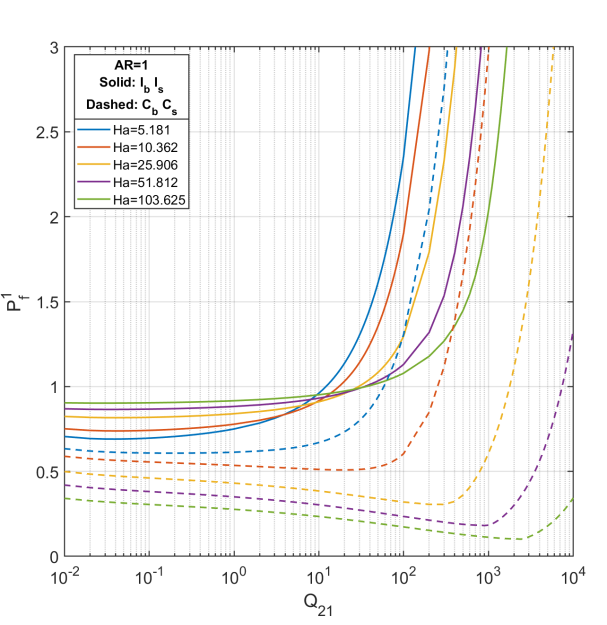}}
	\subfloat[]{\includegraphics[width=0.49\textwidth,clip]{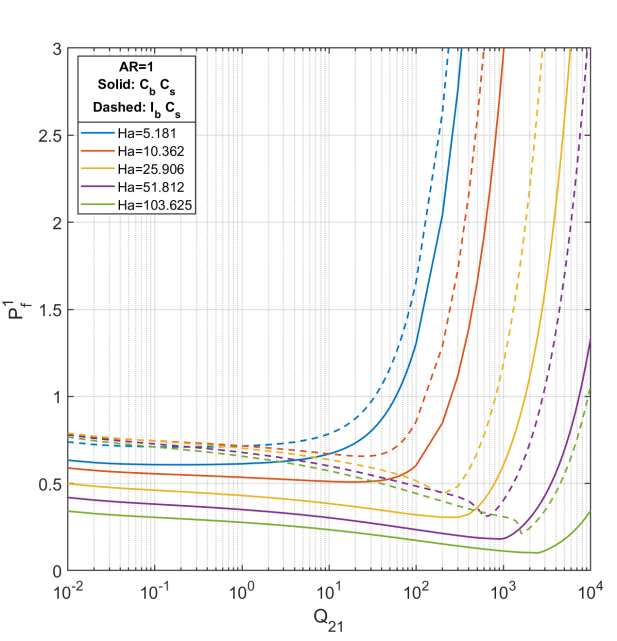}}
	\caption{\label{Fig: 4_27_P1_f_hor}The lubrication factor $P_f^1$ in square ducts (pressure gradient scaled by single phase flow in the same duct and for the same horizontal magnetic field strength): variation with $Q_{21}$ in a square ducts under vertical magnetic field for various $\Ha$: (a) fully insulating ($I_bI_s$, solid) compared to perfectly conducting ($C_bC_s$, dashed) ducts. (b) perfectly conducting ($C_bC_s$, solid) compared to insulating bottom and conducting side walls ($C_bI_s$,dashed) ducts. }
\end{figure}

Figure\ \ref{Fig: 4_26_P10_f_hor} illustrates the effects of gas flow rate and the Hartmann number on the dimensionless pressure gradient, $P_f^{1,0}$, in square ducts of various wall conductivity configuration under $B_{0|x}$. The overall trends observed in Fig.\ \ref{Fig: 4_26_P10_f_hor}a are similar to those observed under $B_{0|y}$: the pressure gradient values obtained in insulating ducts are lower than those obtained in conducting ducts for the same $\Ha$ and $Q_{21}$. Comparison of Fig.\ \ref{Fig: 4_26_P10_f_hor}a with Fig.\ \ref{Fig: 4_12_P_effect_Q21}a indicates that for $B_{0|x}$ the pressure gradient in fully insulating ducts is slightly higher (for the same $\Ha$ and $Q_{21}$), whereas in perfectly conducting ducts it is considerably lower. This is attributed to the lower Lorentz force and holdup of the conductive liquid. Moreover, under $B_{0|x}$, introducing gas flow reduces the pressure gradient over a wide range of $Q_{21}$, and this range broadens with increasing $\Ha$. Due to the larger effective flow area for the gas, the eventual increase in pressure gradient occurs only at much higher $Q_{21}$ values compared with the vertical field case. The minimum of $P_f^{1,0}$ is more pronounced in $I_b C_s$ ducts, where the pressure gradient remains below that of $C_b C_s$ ducts across the entire $Q_{21}$ range (see Fig.\ \ref{Fig: 4_26_P10_f_hor}b). These values are still higher than those obtained under $B_{0|y}$ for the same conditions, where the flow characteristics in $I_b C_s$ were found to be practically the same as those in $I_b I_s$ ducts. On the other hand, the pressure gradient in $C_b I_s$ ducts is practically the same as in insulating ducts under $B_{0|x}$ (presented in Fig.\ \ref{Fig: 4_26_P10_f_hor}a), and is much lower than that obtained under $B_{0|y}$ (Fig.\ \ref{Fig: 4_11_U_contours_CbIs_nonsquare}d) for the same $\Ha$ and $Q_{21}$.
\begin{figure}[h!]
	\centering
	\subfloat[]{\includegraphics[width=0.49\textwidth,clip]{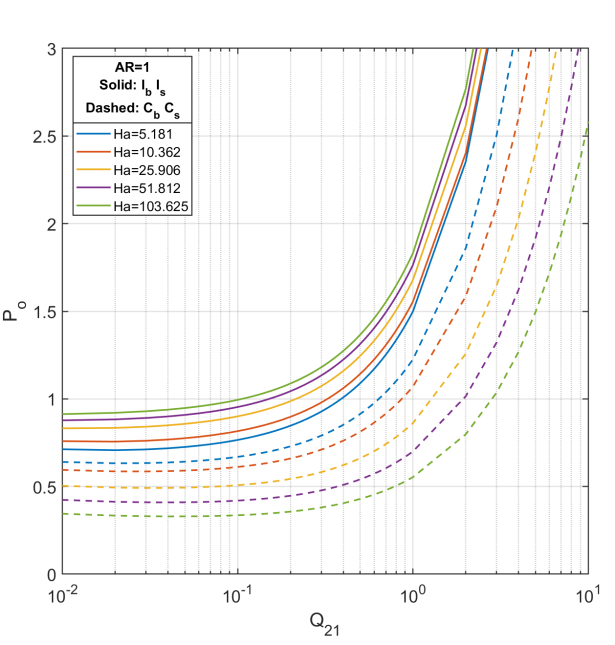}}
	\subfloat[]{\includegraphics[width=0.49\textwidth,clip]{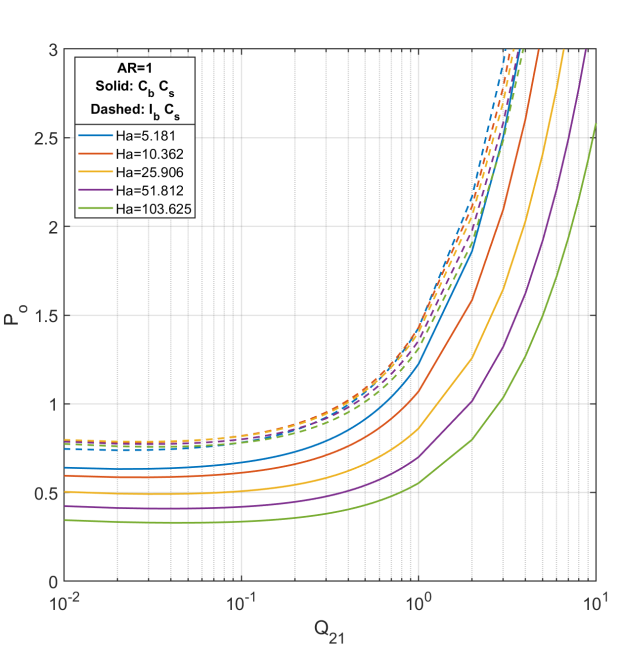}}
	\caption{\label{Fig: 4_28_Po_hor}Pumping power savings corresponding to Fig.\ \ref{Fig: 4_27_P1_f_hor}, $\Po= P_f^1(1+Q_{21})$. }
\end{figure}

To clarify the effect of increasing $Q_{21}$ on the pressure gradient in $C_b C_s$ and $I_b C_s$ ducts under $B_{0|x}$, the corresponding variations in the induced magnetic field ($b/\Ha$) and velocity field were examined (not shown). In $C_b C_s$ ducts, increasing $Q_{21}$ results in reduced $b_\text{max}/\Ha$ (see Fig.\ \ref{Fig: 4_25_b_effect_Q21_hor}b) while its location shifts along the side wall toward the bottom wall. Simultaneously, the backflow region approaches the bottom wall, resulting in a reversed wall shear. The combination of reduced Lorentz force and reversed wall shear causes the pressure gradient to decrease with $Q_{21}$, as observed in Fig.\ \ref{Fig: 4_26_P10_f_hor}b. The minimum pressure gradient occurs just before the backflow vanishes, e.g. at $Q_{21}\approx 900$ for $\Ha=51.812$. At this flow rate ratio the location of $b_\text{max}/\Ha$ reaches the lower duct corner, and it then attains its local peak value at $Q_{21}\approx 1000$ (see Fig.\ \ref{Fig: 4_25_b_effect_Q21_hor}b for $\Ha=51.812$). Beyond this point, further increase of $Q_{21}$ leads to a decreasing $b_\text{max}/\Ha$, but an increasing pressure gradient. The weak dependence of the pressure gradient on $\Ha$ at high $Q_{21}$ indicates that pressure losses are dominated by the flow resistance of the non-conductive gas phase.

In $I_b C_s$ ducts, the backflow region remains separated from the bottom wall; however, both the backflow intensity and the near-wall jetting weaken as $Q_{21}$ increases. The reduction in wall shear results in a decreasing pressure gradient. The local minimum in the pressure gradient at high $Q_{21}$ (see Fig.\ \ref{Fig: 4_26_P10_f_hor}b) corresponds to $Q_{21}$ (e.g., $\approx 600$ for $\Ha=51.812$), where a sharp decline of $b_\text{max}/\Ha$ (and thus of the Lorentz force) is observed in Fig.\ \ref{Fig: 4_25_b_effect_Q21_hor}. It occurs once the backflow is no longer present. As in $C_b C_s$ ducts, further increase in $Q_{21}$ leads to a pronounced rise in the pressure gradient, with values independent on $\Ha$. 
\begin{figure}[h!]
	\centering
	\subfloat[]{\includegraphics[width=0.49\textwidth,clip]{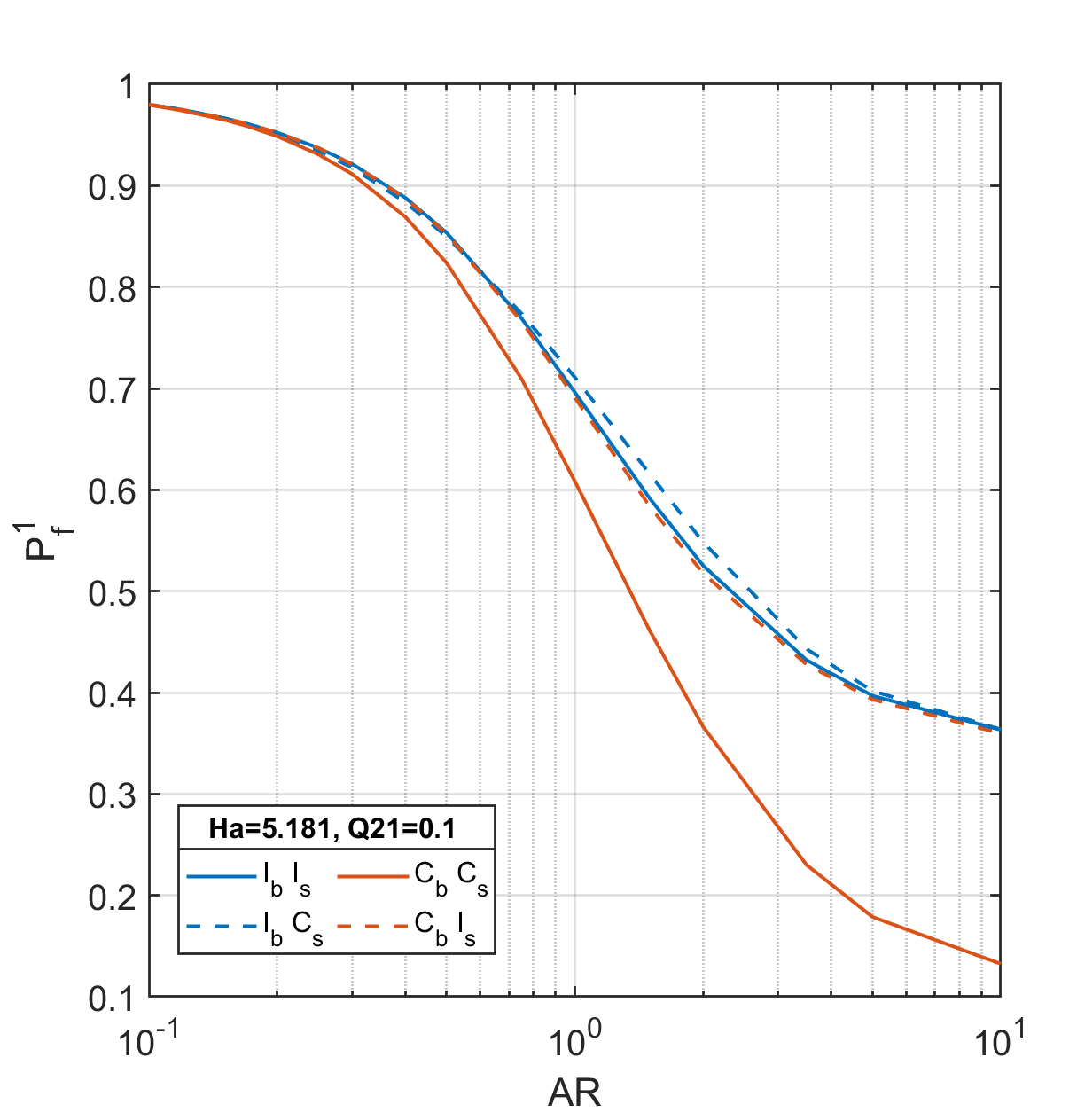}}
	\subfloat[]{\includegraphics[width=0.49\textwidth,clip]{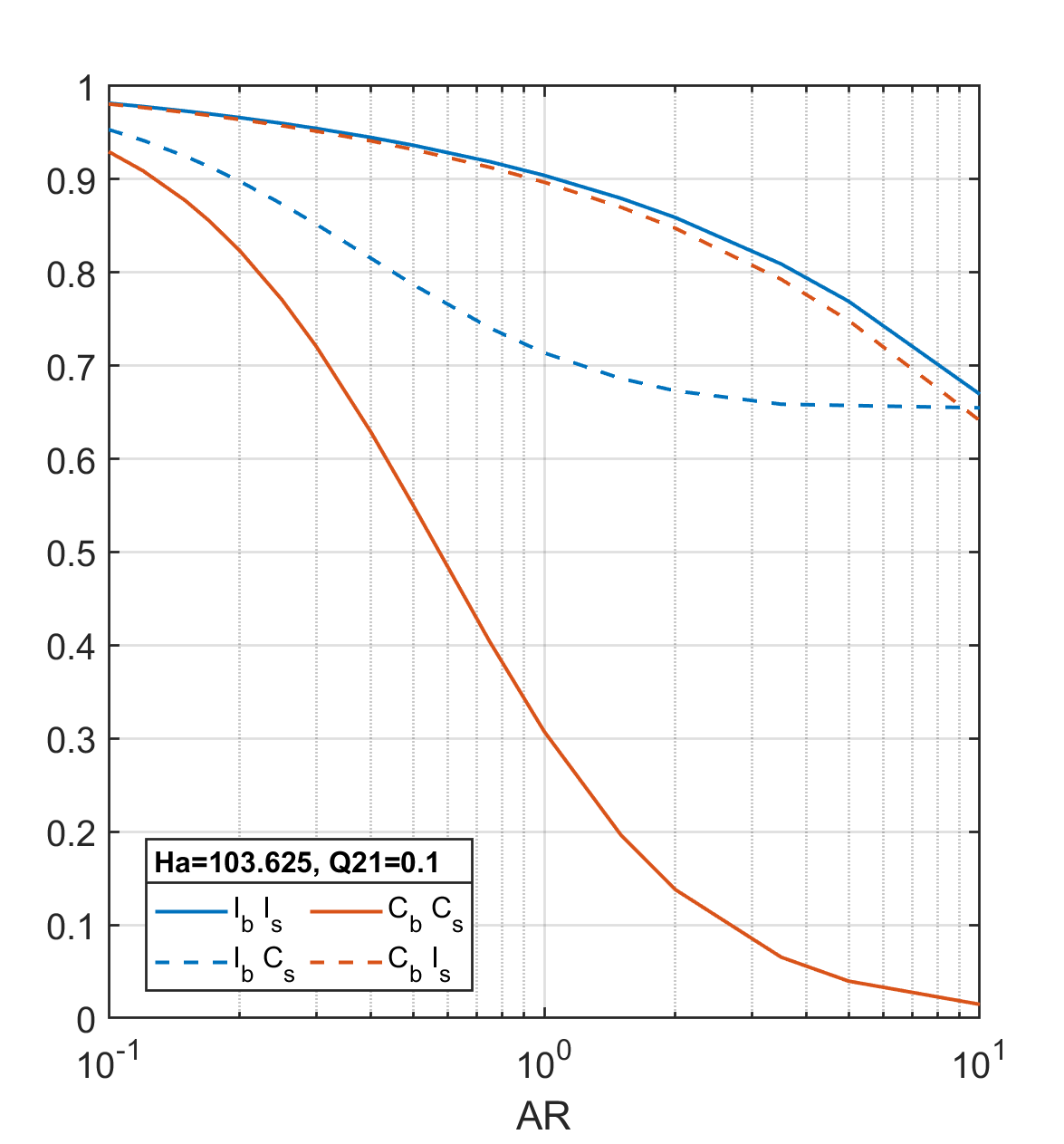}}
	\caption{\label{Fig: 4_29_P1_f_effect_AR_hor}Effect of $AR$ on the lubrication factor, $P_f^1$, in ducts with different wall conductivities configurations ($Q_{21}=0.1$): (a) $\Ha=5.181$ and (b) $\Ha=103.625$.}
\end{figure}

The lubrication effect induced by the gas flow in square ducts with different wall-conductivity combinations is quantified by the values of the $P_f^1$ factor shown in Fig.\ \ref{Fig: 4_27_P1_f_hor}. The corresponding power factor values, $\Po$, are presented in Fig.\ \ref{Fig: 4_28_Po_hor}. In contrast to the results obtained under $B_{0|y}$ (Fig.\ \ref{Fig: 4_12_P_effect_Q21}a), the lubrication effect ($P_f^1<1$) is stronger when the side walls are conducting, and is most pronounced in $C_b C_s$ ducts. In both configurations the lubrication effect intensifies and extends over a broader range of $Q_{21}$ with increasing $\Ha$, with $P_f^1$ reaching considerably lower values than those obtained under $B_{0|y}$. The pronounced gas lubrication effect in $C_b C_s$ and $I_bC_s$ ducts also leads to a substantially greater potential for pumping power savings under $B_{0|x}$ (compare Fig.\ \ref{Fig: 4_28_Po_hor}a with Fig.\ \ref{Fig: 4_15_P1f_effect_Q21}b). Conversely, in fully insulating ducts, the gas-induced lubrication effect weakens with increasing Ha. It is lower than that obtained under $B_{0|y}$, and the associated potential for power savings reduces accordingly.

The effect of $AR$ on the lubrication factor for various wall conductivity combinations under $B_{0|x}$ is demonstrated for $\Ha=5.181$ and $\Ha=103.625$ in Fig.\ \ref{Fig: 4_29_P1_f_effect_AR_hor}a and Fig.\ \ref{Fig: 4_29_P1_f_effect_AR_hor}b, respectively, with the gas flow rate set to $10\%$ of the conductive liquid flow rate ($Q_{21}=0.1$). The same parameters were used in Fig.\ \ref{Fig: 4_14_P_effect_AR} to examine the $AR$ effect for $B_{0|y}$. Comparison of these figures reveals distinct differences in the $AR$ influence even for low $\Ha$. Specifically, in $C_b I_s$ ducts, where $P_f^1$ varies non-monotonically with $AR$ under $B_{0|y}$ (due to the 'jetting' flow pattern), the variation becomes monotonic under $B_{0|x}$, with $P_f^1$ values nearly identical to those of fully insulating ducts. The latter show little sensitivity to the magnetic field orientation. However, in $C_b C_s$ ducts, the $P_f^1$ values are already significantly lower under $B_{0|x}$, particularly in rectangular ducts with $AR>1$. 

These differences in the $AR$ effects under horizontal and vertical magnetic fields become even more pronounced at high $\Ha$ (compare Fig.\ \ref{Fig: 4_29_P1_f_effect_AR_hor}b and Fig.\ \ref{Fig: 4_14_P_effect_AR}b). Under $B_{0|x}$, gas flow lubrication strengthens with increasing $AR$, reaching pressure drop reduction of about $86\%$ ($P_f^1\approx 0.14$) in $C_b C_s$ ducts with $AR=10$, whereas under $B_{0|y}$, gas lubrication is not achieved ($P_f^1\approx1$) over the same $AR$ range. The high lubrication achieved in wide $C_b C_s$ ducts under horizontal magnetic field results from the reduced or even reversed bottom-wall shear due to the near-wall backflow region (e.g., see the centerline velocity profile for $AR=10$ in Fig.\ \ref{Fig: 4_21_Centerline_Velocity_B_hor}c). 

It is worth mentioning that although the pressure gradient in $C_b C_s$ ducts is lower than in $I_b C_s$ ducts (see Fig.\ \ref{Fig: 4_26_P10_f_hor}b), the much lower $P_f^1$ values obtained in the latter are from the substantially higher single-phase pressure gradient of the conductive liquid in $C_b C_s$ ducts for the same $\Ha$ and $AR$. In both $I_b C_s$ and $I_b I_s$ ducts, $P_f^1$  decreases with increasing $AR$, however, the values remain higher under $B_{0|x}$, indicating a weaker lubrication effect. 

\section{Conclusions} \label{Sec: Conclusions} 
In this study, we investigate the two-phase flow characteristics of fully developed horizontal stratified MHD flow involving a conductive liquid and a non-conductive gas in rectangular ducts. We show that the solutions for the liquid holdup, dimensionless pressure gradient, and velocity field are governed by the Hartmann number ($\Ha$), the gas-to-liquid flow rate and viscosity ratios ($Q_{21}$ and $\eta_{1 2}$), the duct aspect ratio ($AR$), and the conductivities of the bottom and side walls. The latter strongly influences the induced magnetic field and thus the Lorentz force acting on the conductive phase. This parameter set also determines the dimensionless induced magnetic field scaled by the magnetic Reynolds number, $\Rey_m$, indicating that the results obtained are applicable for any $\Rey_m$.

The model equations are used to examine how the characteristics of stratified flow in ducts of various aspect ratios and wall-conductivity configurations respond to both the intensity and orientation of an external magnetic field. Although the formulation accommodates any oblique magnetic-field orientation, the results presented in this work were obtained for vertical ($B_{0|y}$) and horizontal ($B_{0|x}$) magnetic fields only. These two limiting cases were analyzed using mercury--air flow as a representative system. The presence of a non-conductive gas layer breaks the symmetry between the top and bottom duct walls existed in single-phase MHD flows. To investigate the effect of wall conductivity, we considered fully insulating ($I_bI_s$) and perfectly conducting ($C_bC_s$) ducts, as well as mixed-conductivity configurations ($I_bC_s$) and ($C_bI_s$). We showed that the solutions for MHD flows under a vertical magnetic field differ qualitatively from those for a flow under a horizontal magnetic field, especially for conducting or mixed-conductivity ducts. 

In all cases, the ratio of the maximum induced magnetic field to the external field remains below $\Rey_m$. Because conductive liquids  have typically a very low magnetic Prandtl number (e.g., for mercury  $Pr_m = 1.383\times10^{-7}$), $\Rey_m\ll 1$ even for laminar flows at relatively high Reynolds numbers, and the induced magnetic field is negligible compared to $B_0$. Nevertheless, it cannot be ignored since its variation with wall conductivity has a significant influence on the two-phase flow characteristics.

For a given gas-to-liquid flow-rate ratio ($Q_{21}$), Ha, and AR, the retarding Lorentz force is weakest when all duct walls are perfectly insulating ($I_bI_s$), for both vertical and horizontal magnetic fields. As a result, the bulk velocity of the conductive liquid is the highest and the liquid holdup is the lowest in this configuration. The reduced holdup and Lorentz force also lead to the lowest pressure gradient among all conductivity configurations and magnetic-field orientations. However, this does not necessarily imply that gas-phase lubrication is maximized in fully insulating ducts. The gas lubrication effect and pumping power savings are evaluated by comparing the pressure gradient and pumping power to those obtained in single phase flow of the conductive liquid in the same duct, i.e., identical AR and wall-conductivity configuration. Under $B_{0|y}$, the gas lubrication effect in fully insulating duct is more pronounced than that in perfectly conducting ducts. The strongest gas lubrication effect is achieved in a fully conducting duct subjected to a horizontal magnetic field.   
The influence of side-wall conductivity depends on both the bottom-wall conductivity and the orientation of the applied magnetic field. Under a vertical magnetic field, the side-wall conductivity has only a minor effect when the bottom wall is insulating, but it significantly influences the liquid holdup, pressure gradient, and other flow characteristics when the bottom wall is conducting. In contrast, under a horizontal magnetic field, the side-wall conductivity is negligible for an insulating bottom wall but becomes pronounced when the bottom wall is conducting.

In both cases involving mixed wall conductivities, the effect of the side walls manifests in the vicinity of insulating surfaces, where the reduced Lorentz force permits the formation of localized high-velocity (jet-like) regions of the conductive liquid, accompanied by a backflow region in the core. Accordingly, for $B_{0|y}$ with a conducting bottom wall, the high-velocity regions develop near the insulating side walls, whereas for $B_{0|x}$ with conducting side walls, they appear near the interface and the insulating bottom wall. In both configurations, the backflow intensifies with increasing magnetic-field strength. In the absence of magnetic field, backflow of the liquid phase in gas-liquid stratified flows can be encountered only in upward inclined systems and is known to introduce substantial disturbance to the stratified flow configuration at the duct entrance, which affect flow pattern transition downstream (e.g., to slug flow) and increased pressure drop. Therefore, cases with significant backflow require careful design of the inlet device to prevent undesired downstream disruption of the stratified flow configuration. 

Future work will involve a rigorous stability analysis of the base-flow solutions obtained in this study to evaluate the effects of design parameters and operating conditions on the stability of MHD stratified flow.


\bmhead{Acknowledgments}

This research was supported by Israel Science Foundation (ISF) grant No 1363/23.

\begin{appendices}
	
	\section{Analytical solutions for vertical external magnetic field} \label{Sec: appendix_analyt_vertical}
	\numberwithin{equation}{section}
	\setcounter{equation}{0}
	
	When the magnetic field is applied in the $y$-direction only, i.e., $B_0=B_{0|y}$, the governing equations \ref{Eq: Induction_b}-\ref{Eq: Momentum_2_dim} become:
	\begin{subequations} \label{Eq: Governing_B0y}
		\begin{align} 
		&\frac{\partial^2 b}{\partial x^2} 
		+ \frac{\partial^2 b}{\partial y^2}
		+ \Ha
		\frac{\partial U_1}{\partial y}
		= 0, 
		\\  
		&\frac{\partial^2 U_1}{\partial x^2} 
		+ \frac{\partial^2 U_1}{\partial y^2}
		+ \Ha 		
		\frac{\partial b}{\partial y}  		
		= G,
		\\
		&\frac{\partial^2 U_2}{\partial x^2} 
		+ \frac{\partial^2 U_2}{\partial y^2}
		= \eta_{1 2} G
		\end{align}
	\end{subequations}
	with the boundary conditions at the interface and on the duct walls (Eqs.\ \ref{Eq: BC_U_rectangular}-\ref{Eq: BC_induced_B_interface}). Note that in the analytical solution the $y$ coordinate is chosen differently from the numerical solution (main text): $y=0$ at the interface and $y=-h$ and $y=1-h$ at the bottom and top walls, respectively. In the following, $h_1=h$ and $h_2=1-h$.
	
	\subsection{Insulating side walls} \label{Sec: appendix_vertical_ins}
	If the side walls are insulating, $b$, as well as $U_{1,2}$, vanishes at $x=0,AR$, so they can be represented by the following series:
	\begin{subequations}
		\begin{align}
		&U_j = \sum_k U_{j,k}(y) \sin(\lambda_k x),
		\\
		&b = \sum_k b_k(y) \sin(\lambda_k x),
		\end{align}
	\end{subequations}
	where $\lambda_k=\pi k/AR$, $j=1,2$, and $k=1,2,...$. In the following, differentiation with respect to $y$ will be denoted by the prime mark (e.g., $U'_{1,k}$). $U_{j,k}$ and $b_k$ should satisfy Eqs.\ \ref{Eq: Governing_B0y}a-c:
	\begin{subequations}
		\begin{align}
		&b''_k - \lambda_k^2 b_k + \Ha U'_{1,k} = 0,
		&U''_{1,k} - \lambda_k^2 U_{1,k} + \Ha b'_k = -f_k,
		&U''_{2,k} - \lambda_k^2 U_{2,k} = -\eta_{1 2} f_k
		\end{align}
	\end{subequations}
	subject to boundary conditions 4-5 at the interface and on the bottom and top walls:
	\begin{subequations}
		\begin{align}
			&U_{1,k}(0) = U_{2,k}(0), 
			\quad
			\eta_{1 2} U'_{1,k}(0) = U'_{2,k}(0), 
			\quad
			b_k(0) = 0;
			\\
			&U_{1,k}(-h_1) = 0,
			\quad
			U_{2,k}(h_2) = 0,
			\quad
			b_k(-h_1) - c_{bw} b'_k \bigr\rvert_{y=-h_1} = 0,
		\end{align}
	\end{subequations}
	where $f_k = -\gamma_k G$
	\begin{equation} \label{Eq: press_grad_aux}
	1\approx \sum_{k} \gamma_k \sin(\lambda_k x),
	\gamma_k = \dfrac{\int_{0}^{AR} \sin(\lambda_k x) dx}{\int_{0}^{AR} \sin^2(\lambda_k x) dx} = \frac{2}{\pi k}\bigl[1-(-1)^k\bigr].
	\end{equation}
	Following the approach of \cite{Shercliff53}, we replace $U_{1,k}$ and $b_k$ with:
	\begin{subequations}
		\begin{align}
		&u_k = U_{2,k}
		\\
		&v_k = U_{1,k} + b_k,
		\\
		&w_k = U_{1,k} - b_k
		\end{align}
	\end{subequations}
	to render equations decoupled for these variables for each $k$ that will be omitted for simplicity in the following:
	\begin{subequations} \label{Eq: Governing_B0y_ins_wo_k}
		\begin{align}
		&u'' - \lambda_k^2 u = -\eta_{1 2} f
		\\
		&v'' - \lambda_k^2 v + \Ha v' = -f,
		\\
		&w'' - \lambda_k^2 w - \Ha w' = -f,			
		\\
		&u(0) = v(0) = w(0), 
		\eta_{1 2} v'(0) + \eta_{1 2} w'(0) = 2 u'(0);
		\\
		&u(h_2) = 0, v(-h_1) + w(-h_1) = 0, 
		v(-h_1) + w(-h_1) = 0,
		\\
		&(1 - c_{bw} \frac{d}{dy})(v - w)_{y=-h_1} = 0;
		\end{align}
	\end{subequations}
	By denoting $M=\Ha/2$ and $N=\sqrt{\lambda^2 + M^2}$, Eqs.\ \ref{Eq: Governing_B0y_ins_wo_k}a-c can be written as
	\begin{subequations}
		\begin{align}
		&\bigl(\frac{d}{dy} - \lambda\bigr)
		\bigl(\frac{d}{dy} + \lambda\bigr)
		u = -\eta_{1 2} f,
		\\
		&\bigl(\frac{d}{dy} + M - N\bigr)
		\bigl(\frac{d}{dy} + M + N\bigr)
		v = -f
		\\
		&\bigl(\frac{d}{dy} - M - N\bigr)
		\bigl(\frac{d}{dy} - M + N\bigr)
		w = -f
		\end{align}
	\end{subequations}
	
	\subsubsection{Insulating bottom wall} \label{Sec: appendix_vertical_Ib_Is}
	
	In case of fully insulating bottom wall $c_{bw}=0$, so that Eq.\ \ref{Eq: Governing_B0y_ins_wo_k}f becomes:
	\begin{equation} \label{Eq: BC_bottom_ins}
	v(-h_1) = w(-h_1) = 0
	\end{equation}
	Let denote $y_1 = y + h_1$ and $y_2 = y - h_2$ (and also $\eta_{2 1}=1 / \eta_{1 2}$) and present the solution of Eqs.\ \ref{Eq: Governing_B0y_ins_wo_k}a-\ref{Eq: Governing_B0y_ins_wo_k}c that satisfies Eqs.\ \ref{Eq: Governing_B0y_ins_wo_k}d,e and Eq.\ \ref{Eq: BC_bottom_ins} :
	\begin{subequations} \label{Eq: u_v_w_Ib_Is}
		\begin{align}
		&u = \frac{f}{\lambda^2 \eta_{21}} 
		\bigg(1 - \cosh(\lambda y_2) - C^u \sinh(\lambda y_2)\bigg),
		\\
		&v = \frac{f}{\lambda^2} 
		\bigg(1 - e^{-M y_1}\cosh(N y_1) + C^v e^{-M y_1} \sinh(N y_1)\bigg),
		\\
		&w= \frac{f}{\lambda^2} 
		\bigg(1 - e^{M y_1}\cosh(N y_1) + C^W e^{M y_1} \sinh(N y_1)\bigg).
		\end{align}
	\end{subequations}
	Rewriting $u(0) = v(0) = w(0) = Af/\lambda^2$ gives
	\begin{subequations} \label{Eq: u0_v0_w0}
		\begin{align}
		&\eta_{21} \lambda^2 u(0) / f 
		= 1 - \cosh(\lambda h_2) + C^u \sinh(\lambda h_2) 
		= \eta_{21} A
		\\
		&\lambda^2 v(0) / f 
		= 1 - e^{-M h_1}\cosh(N h_1) + C^v e^{-M h_1} \sinh(N h_1)
		= A
		\\
		&\lambda^2 w(0) / f 
		= 1 - e^{M h_1}\cosh(N h_1) + C^w e^{M h_1} \sinh(N h_1)
		= A
		\end{align}
	\end{subequations}
	Then the balance of shear stresses, Eq.\ \ref{Eq: Governing_B0y_ins_wo_k}d, gives:
	\begin{align} \label{Eq: Balance_shear_ins}
	\begin{aligned}
	&N \bigg[-\cosh(M h_1) \sinh(N h_1) 
	+ \dfrac{C^v e^{-M h_1} + C^w e^{M h_1}}{2} \cosh(N h_1)\bigg]
	\\
	&= \lambda \big[\sinh(\lambda h_2) - C^u \cosh(\lambda h_2)\big],
	\end{aligned}
	\end{align}
	where from Eqs.\ \ref{Eq: u0_v0_w0}a-\ref{Eq: u0_v0_w0}c:
	\begin{subequations}
		\begin{align}
		&C^u = \dfrac{\eta_{21} A - 1}{\sinh(\lambda h_2)}
		+ \dfrac{\cosh(\lambda h_2)}{\sinh(\lambda h_2)},
		\\
		&\frac{1}{2} \big(C^v e^{-M h_1} + C^w e^{M h_1}\big)
		= \dfrac{A-1}{\sinh(N h_1)} 
		+ \dfrac{\cosh(N h_1)}{\sinh(N h_1)} \cosh(M h_1)
		\end{align}
	\end{subequations}
	Hence Eq.\ref{Eq: Balance_shear_ins} can be written as:
	\begin{align}
	&\begin{aligned}
	&N \bigg[\dfrac{A-1}{\sinh(N h_1)} 
	+ \dfrac{\cosh(N h_1)}{\sinh(N h_1)} \cosh(M h_1)\bigg] 
	\cosh(N h_1)
	- N \cosh(M h_1) \sinh(N h_1)
	\\
	&= \lambda \sinh(\lambda h_2)
	- \lambda \bigg[\frac{\eta_{21} A - 1}{\sinh(\lambda h_2)} 
	+ \frac{\cosh(\lambda h_2)}{\sinh(\lambda h_2)}\bigg] \cosh(\lambda h_2),
	\end{aligned}
	\end{align}
	where $A$ is given by:
	\begin{equation}
	A = \dfrac{N \big[\cosh(N h_1) - \cosh(M h_1)\big] \sinh(\lambda h_2)
		+ \lambda \big[\cosh(\lambda h_2) - 1\big] \sinh(N h_1)}{N \cosh(N h_1) \sinh(\lambda h_2) + \eta_{21} \lambda \cosh(\lambda h_2) \sinh(N h_1)}.
	\end{equation}
	The coefficients $C^u, C^v, C^w$ are found from Eqs.\ \ref{Eq: u0_v0_w0}a-c:
	\begin{subequations}
		\begin{align}
		&C^u = \dfrac{\eta_{21} A - 1}{\sinh(\lambda h_2)}
		+ \coth(\lambda h_2),
		\\
		&C^v = \dfrac{A - 1}{\sinh(N h_1)} e^{M h_1}
		+ \coth(N h_1),
		\\
		&C^w = \dfrac{A - 1}{\sinh(N h_1)} e^{-M h_1}
		+ \coth(N h_1).
		\end{align}
	\end{subequations}	
	Finally, $u,v,w$ can be obtained from Eqs.\ \ref{Eq: u_v_w_Ib_Is}a-c and then $U_2=u$, $U_1 = (v+w)/2$, $b=(v-w)/2$.
	
	\subsubsection{Conducting bottom wall} \label{Sec: appendix_vertical_Cb_Is}
	
	In case of perfectly conducting bottom wall $c_{bw}\to\infty$, so that Eq.\ \ref{Eq: Governing_B0y_ins_wo_k}f becomes:
	\begin{equation} \label{Eq: BC_bottom_cond}
	v'(-h_1) = w'(-h_1) = 0
	\end{equation}
	Similarly to Eqs.\ \ref{Eq: u_v_w_Ib_Is}, the following solutions can be found:
	\begin{subequations} \label{Eq: u_v_w_Cb_Is}
		\begin{align}
		&u = \frac{f}{\lambda^2 \eta_{21}} 
		\bigg(1 - \cosh(\lambda y_2) - C^u \sinh(\lambda y_2)\bigg),
		\\
		&v = \frac{f}{\lambda^2} 
		\bigg(1 - (1-D) e^{-M y_1}\cosh(N y_1) + C^v e^{-M y_1} \sinh(N y_1)\bigg),
		\\
		&w= \frac{f}{\lambda^2} 
		\bigg(1 - (1+D) e^{M y_1}\cosh(N y_1) + C^w e^{M y_1} \sinh(N y_1)\bigg).
		\end{align}
	\end{subequations}
	From Eq.\ \ref{Eq: BC_bottom_cond}, one can get:
	\begin{equation}
	M(1-D) + N C^v = - M (1+D) + N C^w
	\end{equation}
	Hence, the coefficient $C^v$ and $C^w$ can be represented as:
	\begin{equation}
	C^v = (E - M) / N, C^w = (E + M) / N.
	\end{equation}
	Rewriting $u(0) = v(0) = w(0) = Af/\lambda^2$ gives
	\begin{subequations} \label{Eq: u0_v0_w0_Cb}
		\begin{align}
		&\eta_{21} \lambda^2 u(0) / f 
		= 1 - \cosh(\lambda h_2) + C^u \sinh(\lambda h_2) 
		= \eta_{21} A
		\\
		&\begin{aligned}
			\lambda^2 v(0) / f 
				= 1 - (1-D) e^{-M h_1}\cosh(N h_1) 
				+ (E - M) / N e^{-M h_1} \sinh(N h_1)
				= A
		\end{aligned}
		\\
		&\begin{aligned}
			\lambda^2 w(0) / f 
				= 1 - (1+D) e^{M h_1}\cosh(N h_1) 
				+ (E + M) / N e^{M h_1} \sinh(N h_1)
				= A
		\end{aligned}		
		\end{align}
	\end{subequations}
	The coefficient $C^u$ is then obtained from Eq.\ \ref{Eq: u0_v0_w0_Cb} and substituted in Eq.\ \ref{Eq: u_v_w_Cb_Is}:
	\begin{subequations}
		\begin{align}
		&C^u = \dfrac{\eta_{21} A - 1}{\sinh(\lambda h_2)}
		+ \coth(\lambda h_2),
		\\
		&u = \frac{f}{\lambda^2 \eta_{21}} 
		\bigg(1 - \cosh(\lambda y_2) - (\eta_{21} A - 1) \dfrac{\sinh(\lambda y_2)}{\sinh(\lambda h_2)} - \sinh(\lambda y_2) \coth(\lambda h_2)\bigg)
		\end{align}
	\end{subequations}
	The only unknowns left are $A,D,E$, which can be found as:
	\begin{subequations}
		\begin{align}
		&\begin{aligned}
		A =& \Bigg[N \coth(N h_1) \bigg(1 + \dfrac{\sinh^2(M h_1)}{\cosh^2(N h_1)}\bigg) + \frac{\lambda}{\eta_{1 2}} \coth(\lambda h_1)\Bigg]^{-1} \times
		\\
		&\Bigg[N \dfrac{\cosh(N h_1) - \cosh(M h_1)}{\sinh(N h_1)}
		+ \bigg(N \dfrac{\sinh(M h_1)}{\sinh(N h_1)} - M\bigg)
		\dfrac{\sinh(M h_1)}{\cosh(N h_1)} 
		\\
		&+ \lambda
		\dfrac{\cosh(\lambda h_1)-1}{\sinh(\lambda h_1)}\Bigg],
		\\
		D =& (A - 1) \dfrac{\sinh(M h_1)}{\cosh(N h_1)}
		+ \frac{M}{N} \tanh(N h_1),
		\\
		E =& N (A - 1) \dfrac{\cosh(M h_1)}{\sinh(N h_1)}.
		+ N \coth(N h_1)
		\end{aligned}
		\end{align}
	\end{subequations}
	
	Then $u,v,w$ can be obtained from Eqs.\ \ref{Eq: u_v_w_Cb_Is}a-c and then $U_2=u$, $U_1 = (v+w)/2$, $b=(v-w)/2$.
	
	\subsection{MHD flow profiles and flow rates} \label{Sec: appendix_profiles}
	
	Once $u_k$, $v_k$, and $w_k$ are found using the solutions presented in Sec.\ \ref{Sec: appendix_vertical_Ib_Is} and Sec.\ \ref{Sec: appendix_vertical_Cb_Is}, the velocity and induced magnetic field profiles can be expressed as:
	\begin{subequations}
		\begin{align}
		&U_1(x,y) = \sum_{k} U_{1,k}(y) \sin(\lambda_k x)
		= \sum_{k} \frac{v_k + w_k}{2} \sin(\lambda_k x),
		\\
		&U_2(x,y) = \sum_{k} U_{2,k}(y) \sin(\lambda_k x)
		= \sum_{k} u_k \sin(\lambda_k x),
		\\
		&b(x,y) = \sum_{k} b_k(y) \sin(\lambda_k x)
		= \sum_{k} \frac{v_k - w_k}{2} \sin(\lambda_k x)
		\end{align}
	\end{subequations}
	
	
	The flow rates of each phase are then obtained by integration of the velocity profile over the corresponding cross section occupied by the phase:
	\begin{subequations}
		\begin{align}
		&Q_1 = \sum_{k}\int_{-h_1}^{0} \int_{0}^{AR} U_{1,k} \sin{\lambda_k x} dx dy,
		\\
		&Q_2 = \sum_{k}\int_{0}^{h_2} \int_{0}^{AR} U_{2,k} \sin{\lambda_k x} dx dy
		\end{align}
	\end{subequations}
	The final expressions for the flow rates are:
	\begin{subequations} \label{Eq: Q1_Q2}
		\begin{align}
		&\begin{aligned}
		Q_1 &= \sum_{k} \dfrac{-\gamma_k G}{\lambda_k^3}
		\Bigg[h_1 + (C^v + C^w + 2D)
		\dfrac{\cosh\big((N_k - M)h_1\big) - 1}{4(N_k - M)}
		\\
		&- (C^w - C^v + 2) \dfrac{\sinh\big((N_k - M)h_1\big)}{4(N_k - M)}
		\\
		&+ (C^v + C^w - 2D) \dfrac{\cosh\big((N_k + M)h_1\big)}{4(N_k + M)}
		\\
		&- (C^w - C^v + 2) \dfrac{\sinh\big((N_k + M)h_1\big)}{4(N_k + M)}\Bigg]
		\Big[1 - (-1)^k\Big]
		\end{aligned}
		\\
		&\begin{aligned}
		Q_2 &= \eta_{1 2} \sum_{k} \dfrac{-\gamma_k G}{\lambda_k^3}
		\Bigg[h_2 - \frac{1}{\lambda_k} \sinh(\lambda_k h_2)
		\\
		&+ \frac{C^u}{\lambda_k} \big(\cosh(\lambda_k h_2) - 1\big)\Bigg]
		\Big[1 - (-1)^k\Big]
		\end{aligned}
		\end{align}
	\end{subequations}
	
	The flow rate ratio, $Q_{21}=Q_2/Q_1$, is independent of the pressure gradient. For prescribed $Q_{21}$, the holdup can be found. Then the (dimensionless) pressure gradient, $G$, can be obtained from either Eq.\ \ref{Eq: Q1_Q2}a or Eq.\ \ref{Eq: Q1_Q2}b.
	\newpage
	\section{Analytical solutions for horizontal external magnetic field} \label{Sec: appendix_analyt_horizontal}
	\numberwithin{equation}{section}
	
	When the magnetic field is applied in the $x$-direction only, i.e., $B_0=B_{0|x}$, the governing equations \ref{Eq: Induction_b}-\ref{Eq: Momentum_2_dim} become:
	\begin{subequations} \label{Eq: Governing_B0x}
		\begin{align} 
		&\frac{\partial^2 b}{\partial x^2} 
		+ \frac{\partial^2 b}{\partial y^2}
		+ \Ha
		\frac{\partial U_1}{\partial x}
		= 0, 
		\\  
		&\frac{\partial^2 U_1}{\partial x^2} 
		+ \frac{\partial^2 U_1}{\partial y^2}
		+ \Ha 		
		\frac{\partial b}{\partial x}  		
		= G,
		\\
		&\frac{\partial^2 U_2}{\partial x^2} 
		+ \frac{\partial^2 U_2}{\partial y^2}
		= \eta_{1 2} G
		\end{align}
	\end{subequations}
	and the boundary conditions at the interface and on the duct walls (Eqs.\ \ref{Eq: BC_U_rectangular}-\ref{Eq: BC_induced_B_interface}). Similarly to Appendix\ \ref{Sec: appendix_analyt_vertical}, the $y$-coordinate adopted here is such that $y=0$ at the interface, and $y=-h$ and $y=1-h$ at the bottom and top walls, respectively. In the following, $h_1=h$ and $h_2=1-h$.
	
	\subsection{Conducting side walls} \label{Sec: appendix_horizontal_cond}
	If the side walls are conducting, $\partial b/\partial x$ as well as $U_{1,2}$, vanishes at $x=0,AR$, so they can be represented by the following series:
	\begin{subequations}
		\begin{align}
		&U_j = \sum_k U_{j,k}(y) \sin(\lambda_k x),
		\\
		&b = b_0(y) + \sum_k b_k(y) \cos(\lambda_k x),
		\end{align}
	\end{subequations}
	where $\lambda_k=\pi k/AR$, $j=1,2$, and $k=1,2,...$. Since $b''_0=0$, $b_0(y)$ can be written as $b_0(y) = C_{0b} + C_{1b}y$. Utilizing boundary conditions at the interface, $y=0$, $b_0 = C_{0b} = 0$. Meanwhile, for the bottom wall, $y=-h_1$, $\displaystyle C_{1b}(1 + h / c_{bw})$, so that $C_{1b}=0$ for both perfectly conducting and fully insulating walls.
	
	Following the annotations of vertical magnetic field, $U_{j,k}$ and $b_k$ should satisfy Eqs.\ \ref{Eq: Governing_B0x}a-c:
	\begin{subequations}
		\begin{align}
		&b''_k - \lambda_k^2 b_k + \lambda_k \Ha U_{1,k} = 0,
		&U''_{1,k} - \lambda_k^2 U_{1,k} - \lambda_k \Ha b_k = -f_k,
		&U''_{2,k} - \lambda_k^2 U_{2,k} = -\eta_{1 2} f_k
		\end{align}
	\end{subequations}
	subject to boundary conditions 4-5 at the interface and on the bottom and top walls:
	\begin{subequations}
		\begin{align}
		&U_{1,k}(0) = U_{2,k}(0), 
		\eta_{1 2} U'_{1,k}(0) = U'_{2,k}(0), 
		b_k(0) = 0;
		\\
		&U_{1,k}(-h_1) = 0,
		U_{2,k}(h_2) = 0,
		b_k(-h_1) - c_{bw} b'_k \bigr\rvert_{y=-h_1} = 0.
		\end{align}
	\end{subequations}
	For convenience we replace $U_{j,k}$ with:
	\begin{subequations}
		\begin{align}
		&u_k = U_{2,k}
		\\
		&v_k = U_{1,k},
		\end{align}
	\end{subequations}
	to render equations decoupled for these variables for each $k$ that will be omitted for simplicity in the following:
	\begin{subequations} \label{Eq: Governing_B0x_cond_wo_k}
		\begin{align}
		&u'' - \lambda_k^2 u = -\eta_{1 2} f
		\\
		&v'' - \lambda_k^2 v - \lambda \Ha b = -f,
		\\
		&b'' - \lambda_k^2 b + \lambda \Ha v = 0,			
		\\
		&u(0) = v(0), b(0)=0, 
		\eta_{1 2} v'(0) = u'(0);
		\\
		&u(h_2) = 0, v(-h_1) = 0,(1 - c_{bw} \frac{d}{dy})b_{y=-h_1} = 0;
		\end{align}
	\end{subequations}
	Let us apply $(\frac{d^2}{dy^2}-\lambda^2)$ to Eq.\ \ref{Eq: Governing_B0x_cond_wo_k}(b) and substitute  Eq.\ \ref{Eq: Governing_B0x_cond_wo_k}(c) into the result:
	\begin{equation}
	(\frac{d^2}{dy^2}-\lambda^2-i\lambda \Ha)(\frac{d^2}{dy^2}-\lambda^2+i\lambda \Ha)v= \lambda^2 f
	\end{equation}
	By denoting $y_1=y+h_1$,  $y_2=y-h_2$, $\alpha^2=\frac{\lambda}{2} (\lambda+\sqrt{\Ha^2+\lambda^2})$, $\beta=\lambda \Ha / 2\alpha$ so that $(\alpha \pm i\beta)^2=\lambda^2 \pm i\lambda \Ha$, and also considering $u(y=h_2)=0$, $v(y=-h_1)=0$ and Eq.\ \ref{Eq: Governing_B0x_cond_wo_k}(a)-(b), we can then represent $u$, $v$ and $b$ as:
	\begin{subequations} \label{v_hori}
		\begin{align}
		&\begin{aligned}
		u &=
		\frac{f}{\lambda^2\eta_{21}}  \Big( 1-\cosh(\lambda y_2 )-C^u  \sinh(\lambda y_2)  \Big)
		\end{aligned}
		\\
		&\begin{aligned}
		v&=
		\frac{f}{\lambda^2 + \Ha^2} \Big(1+\sinh(\alpha y_1) [C_1  \sin(\beta y_1) + C_2 \cos(\beta y_1)] 
		\\
		&+ \cosh(\alpha y_1)[ C_3 \sin(\beta y_1)-\cos(\beta y_1) ] \Big)
		\end{aligned}
		\\
		&\begin{aligned}
		b&=
		\frac{f}{\lambda^2 + \Ha^2} \Big(\frac{\Ha}{\lambda} + \cosh(\alpha y_1) [C_1 \cos(\beta y_1)- C_2 \sin(\beta y_1)] 
		\\
		&+\sinh(\alpha y_1)[C_3 \cos(\beta y_1) + \sin(\beta y_1) ] \Big)
		\end{aligned}
		\end{align}
	\end{subequations}
	The unknown constants $C_1$,$C_2$, $C_3$ and $C^u$ must be found from the conditions at the interface and the condition for $b$ on the bottom wall. We consider only two limiting cases: a perfectly insulating bottom wall and a perfectly conducting bottom wall.
	
	\subsubsection{Insulating bottom wall} \label{Sec: appendix_horizontal_Ib_Cs}
	
	In case of fully insulating bottom wall $b(y=-h_1)=0$, therefore
	\begin{equation} 
	\frac{\Ha}{\lambda}+C_1 = 0.
	\end{equation}
	When considering conditions at the interface, for brevity, let us introduce
	\begin{subequations} \label{Eq: u_v_w_Ib_Cs}
		\begin{align}
		&t_{cc} = \cosh(\alpha h_1)\cos(\beta h_1), \hspace{1cm} t_{cs} = \cosh(\alpha h_1)\sin(\beta h_1),
		\\
		&t_{sc} = \sinh(\alpha h_1)\cos(\beta h_1), \hspace{1cm} t_{ss} = \sinh(\alpha h_1)\sin(\beta h_1).
		\end{align}
	\end{subequations}
	The continuity of velocity $u=v$ gives
	\begin{equation}  \label{continuity_vel}
	\frac{1}{\lambda^2 \eta_{21}} \Big(1-\cosh(\lambda h_2) - C^u \sinh(\lambda h_2) \Big) = \frac{1}{\lambda^2 + \Ha^2} \Big( 1+C_1 t_{ss} +C_2 t_{sc} +C_3 t_{cs} - t_{cc} \Big)
	\end{equation}
	The continuity of stress $u'=\eta_{12}v'$ gives
	\begin{subequations} \label{continuity_stress}
		\begin{align} \nonumber
		\frac{\lambda \sinh(\lambda h_2)-\lambda C^u \cosh(\lambda h_2)}{\lambda^2}= &\frac{\alpha(C_1 t_{cs} +C_2 t_{cc} +C_3 t_{ss} - t_{sc})}{\lambda^2 + \Ha^2}
		\\
		&+ \frac{\beta (C_1 t_{sc} -C_2 t_{ss} +C_3 t_{cc} + t_{cs})}{\lambda^2 + \Ha^2}
		\end{align}
	\end{subequations}
	And since the upper phase is insulating, we have $b=0$, which gives
	\begin{equation} \label{interface_b}
	\frac{\Ha}{\lambda}+C_1 t_{cc} -C_2 t_{cs} +C_3 t_{sc} + t_{ss}=0. 
	\end{equation}
	Let, $u(o)=v(0)=\frac{Af}{\lambda^2}$, so that:
	\begin{equation} \label{cu}
	C^u=\frac{A}{\eta_{12}\sinh(\lambda h_2)} + \frac{\cosh(\lambda h_2)-1}{\sinh(\lambda h_2)}.
	\end{equation}
	Now we substitute $C_1=-\frac{\Ha}{\lambda}$ into Eqs. \ref{continuity_vel} and \ref{interface_b}:
	\begin{subequations}
		\begin{align}
		&C_3 t_{cs} + C_2 t_{sc}= (t_{cc}-1)+\frac{\Ha}{\lambda}t_{ss}+A \Big( 1+\frac{\Ha^2}{\lambda^2} \Big),
		\\
		&C_3 t_{sc}-C_2 t_{cs}=\frac{\Ha}{\lambda}(t_{cc}-1)-t_{ss};
		\end{align}
	\end{subequations}
	so we can express $C_2$ and $C_3$ as
	\begin{equation} \label{c2c3}
	C_2 = K_2A + F_2, \hspace{1cm} C_3= K_3A + F_3,
	\end{equation}
	where,
	\begin{subequations}
		\begin{align}
			&\begin{aligned}
				&K_2=\frac{t_{sc}}{D}\Big( 1+ \frac{\Ha^2}{\lambda^2} \Big), 
				\\
				&F_2=\frac{1}{D} \Big[ (t_{cc}-1)\Big(t_{sc}-\frac{\Ha}{\lambda}t_{cs} \Big) +t_{ss}\Big(\frac{\Ha}{\lambda}t_{sc} + t_{cs} \Big) \Big],
			\end{aligned}
			\\
			&\begin{aligned}
				&K_3=\frac{t_{cs}}{D}\Big( 1+ \frac{\Ha^2}{\lambda^2} \Big), 
				\\
				&F_3=\frac{1}{D} \Big[ (t_{cc}-1)\Big(t_{cs}+\frac{\Ha}{\lambda}t_{sc} \Big) +t_{ss}\Big(\frac{\Ha}{\lambda}t_{cs} - t_{sc} \Big) \Big],
			\end{aligned}
			\\
			&D= t_{cs}^2 + t_{sc}^2.
		\end{align}
	\end{subequations}
	Now we substitute these into Eq.\ \ref{continuity_stress}:
	\begin{equation}
	A=\frac{F_u-\alpha(F_2t_{cc} +  F_3t_{ss}-t_{sc}-\frac{\Ha}{\lambda}t_{cs}) - \beta(F_3t_{cc}-F_2t_{ss}+t_{cs}-\frac{\Ha}{\lambda}t_{sc})}{K_u+\alpha(K_2t_{cc}+K_3t_{ss}) + \beta(K_3t_{cc}-K_2t_{ss})},
	\end{equation}
	where
	\begin{subequations} \label{kufu}
		\begin{align}
		&K_u=\frac{\lambda}{\eta_{12}}.\frac{\cosh(\lambda h_2)}{\sinh(\lambda h_2)}.\Big( 1+\frac{\Ha^2}{\lambda^2} \Big),
		\\
		&F_u=\lambda \Big( 1+\frac{\Ha^2}{\lambda^2} \Big) \Big[ \sinh(\lambda h_2) - \frac{\cosh(\lambda h_2)-1}{\sinh(\lambda h_2)}.\cosh(\lambda h_2) \Big].
		\end{align}
	\end{subequations}
	Then $C_2$, $C_3$ and $C^u$ can be found using Eqs. \ref{c2c3} and \ref{cu}.
	
	\subsubsection{Conducting bottom wall} \label{Sec: appendix_horizontal_Cb_Cs}
	
	In case of perfectly conducting bottom wall $b'(y=-h_1)=0$, therefore
	\begin{equation} \label{cv}
	\alpha C_3-\beta C_2 = 0,\hspace{1cm} \implies C^v=\alpha C_3=\beta C_2.
	\end{equation}
	Using Eq.\ \ref{cv}, we rewrite Eqs.\ \ref{continuity_stress} and \ref{interface_b} as:
	\begin{subequations}
		\begin{align}
		&C_1 t_{ss} + C^v \Big(\frac{t_{sc}}{\beta} + \frac{t_{cs}}{\alpha} \Big) = (t_{cc}-1)+A \Big( 1+\frac{\Ha^2}{\lambda^2} \Big),
		\\
		&C_1 t_{cc} + C^v \Big(\frac{t_{sc}}{\alpha} - \frac{t_{cs}}{\beta} \Big) = -t_{ss}-\frac{\Ha}{\lambda}. 
		\end{align}
	\end{subequations}
	We can express $C_1$ and $C^v$ as
	\begin{equation} \label{c1cv}
	C_1 = K_1A + F_1, \hspace{1cm} C^v= K_vA + F_v,
	\end{equation}
	where,
	\begin{subequations}
		\begin{align}
			&\begin{aligned}
				&K_1=\frac{1}{D}\Big( 1+ \frac{\Ha^2}{\lambda^2} \Big)\Big(\frac{t_{sc}}{\alpha} - \frac{t_{cs}}{\beta} \Big), 
				\\
				&F_1=\frac{1}{D} \Big[ (t_{cc}-1)\Big(\frac{t_{sc}}{\alpha} - \frac{t_{cs}}{\beta} \Big) +\Big(t_{ss}+\frac{\Ha}{\lambda}\Big) \Big(\frac{t_{sc}}{\beta} + \frac{t_{cs}}{\alpha} \Big)\Big],
			\end{aligned}
		\\
		&K_v=-\frac{t_{cc}}{D}\Big( 1+ \frac{\Ha^2}{\lambda^2} \Big), \hspace{1cm} F_v=-\frac{1}{D} \Big[ t_{cc}(t_{cc}-1) +t_{ss}\Big(t_{ss}+\frac{\Ha}{\lambda} \Big) \Big],
		\\
		&D= t_{ss}\Big(\frac{t_{sc}}{\alpha} - \frac{t_{cs}}{\beta} \Big) - t_{cc}\Big(\frac{t_{sc}}{\beta} + \frac{t_{cs}}{\alpha} \Big).
		\end{align}
	\end{subequations}
	Now we substitute these into Eq.\ \ref{continuity_stress}:
	\begin{equation}
	A=\frac{F_u-\Big[(\alpha t_{cs} +  \beta t_{sc})F_1+\Big(\frac{\alpha}{\beta} +\frac{\beta}{\alpha}\Big)t_{cc}F_v\Big] + (\alpha t_{sc}-\beta t_{cs})}{K_u+(\alpha t_{cs}+\beta t_{sc})K_1 + \Big(\frac{\alpha}{\beta} +\frac{\beta}{\alpha}\Big)t_{cc}K_v},
	\end{equation}
	where $k_u$ and $F_u$ are given by Eqs. \ref{kufu}. Then $C_1$, $C_2$, $C_3$, $C^u$ and $C^v$ can be found using Eqs. \ref{c1cv}, \ref{cu} and \ref{cv}.

	\subsection{MHD flow profiles and flow rates} \label{Sec: appendixB_profiles}
	
	Once $u_k$, $v_k$, and $b_k$ are found using the solutions presented in Sec.\ \ref{Sec: appendix_horizontal_Ib_Cs} and Sec.\ \ref{Sec: appendix_horizontal_Cb_Cs}, the velocity and induced magnetic field profiles can be expressed as:
	\begin{subequations}
		\begin{align}
		&U_1(x,y) = \sum_{k} U_{1,k}(y) \sin(\lambda_k x)
		= \sum_{k} v_k  \sin(\lambda_k x),
		\\
		&U_2(x,y) = \sum_{k} U_{2,k}(y) \sin(\lambda_k x)
		= \sum_{k} u_k \sin(\lambda_k x),
		\\
		&b(x,y) = \sum_{k} b_k(y) \sin(\lambda_k x)
		= \sum_{k} b_k  \sin(\lambda_k x)
		\end{align}
	\end{subequations}
	
	The flow rates of each phase are then obtained by integration of the velocity profile over the corresponding cross section occupied by the phase:
	\begin{subequations}
		\begin{align}
		&Q_1 = \sum_{k}\int_{-h_1}^{0} \int_{0}^{AR} U_{1,k} \sin{\lambda_k x} dx dy,
		\\
		&Q_2 = \sum_{k}\int_{0}^{h_2} \int_{0}^{AR} U_{2,k} \sin{\lambda_k x} dx dy
		\end{align}
	\end{subequations}
	The final expressions for the flow rates are:
	\begin{subequations} \label{Eq: Q1_Q2_hori}
		\begin{align}
		&\begin{aligned}
		Q_1 &= \sum_{k} \dfrac{-\gamma_k G}{\lambda_k^3 + \lambda_k \Ha^2}
		\Bigg[h_1 + \frac{1}{\alpha_k^2 + \beta_k^2}
		\Big\{ \sinh (\alpha_k h_1)[(\alpha_k C_3 + \beta_k C_2)\sinh(\beta_k h_1)
		\\
		&- (\alpha_k +\beta_k C_1)\cos(\beta_k h_1)]+\cosh(\alpha_k h_1)[(\alpha_k C_2- \beta_k C_3)\cos(\beta_k h_1)] 
		\\
		&+\beta_k C_3-\alpha_k C_2 \Big\}\Bigg]	\Big[1 - (-1)^k\Big]
		\end{aligned}
		\\
		&\begin{aligned}
		Q_2 &= \eta_{1 2} \sum_{k} \dfrac{-\gamma_k G}{\lambda_k^3}
		\Bigg[h_2 - \frac{1}{\lambda_k} \sinh(\lambda_k h_2)
		\\
		&+ \frac{C^u}{\lambda_k} \big(\cosh(\lambda_k h_2) - 1\big)\Bigg]
		\Big[1 - (-1)^k\Big]
		\end{aligned}
		\end{align}
	\end{subequations}
	
	The flow rate ratio, $Q_{21}=Q_2/Q_1$, similar to the relations obtained in Appendix \ref{Sec: appendix_analyt_vertical}, is independent of the pressure gradient, and can be used to obtain the holdup for specified $\Ha$ and $Q_{21}$. Then the (dimensionless) pressure gradient, $G$, can be obtained from either Eq.\ \ref{Eq: Q1_Q2_hori}a or Eq.\ \ref{Eq: Q1_Q2_hori}b.
	\newpage
	
	\section{Pressure gradients for single phase MHD flow } \label{Sec: appendix_single_phase}
	\numberwithin{equation}{section}
	{\renewcommand{\arraystretch}{1.5}
		\begin{table}[h!]
			\caption{\label{Tab: singlephase_pgrad} Dimensionless single-phase pressure gradient, $P_{fs}^1 = G/12$, for the problem with a vertical external magnetic field ($B_0 = B_{0|y}$).}
			\begin{tabular*}{\textwidth}{@{\extracolsep{\fill}}c c c c c@{\extracolsep{\fill}}}
				\hline%
				\hline
				\multirow{2}{*}{$AR$} 
				& $I_bI_s$ & $I_bC_s$ & $C_bC_s$ & $C_b I_s$
				\\
				\cmidrule{2-5}
				\multicolumn{5}{c}{$\Ha=5.181$}
				\\
				\hline
				0.1  & 106.747 & 106.794 & 109.027 & 106.796
				\\
				0.5       & 6.021 & 6.113 & 8.363 & 6.457
				\\
				1       & 2.707 & 2.744 & 5.028 & 3.877
				\\
				2   & 1.851 & 1.861 & 4.177 & 3.677
				\\
				5  & 1.537 & 1.540 & 3.822 & 3.641 	
				\\
				\hline
				\multicolumn{5}{c}{$\Ha=103.625$}
				\\
				\hline
				0.1  & 113.796 & 120.715 & 1009.869 & 126.516
				\\
				0.5       & 23.110 & 23.922 & 915.678 & 176.798
				\\
				1       & 19.988 & 20.286 & 910.567 & 295.239
				\\
				2   & 18.723 & 18.852 & 908.022 & 445.291
				\\
				5  & 18.037 & 18.085 & 906.502 & 640.643 	
				\\
				\hline
				\hline
			\end{tabular*} \label{A3:Ha103}         
		\end{table}
	}
\end{appendices}

\section*{Declarations}

\subsubsection*{Ethical approval}
Not applicable.

\subsubsection*{Competing interests}
None.

\subsubsection*{Authors' contributions}

\subsubsection*{Funding} 
Israel Science Foundation (ISF) grant No 1363/23.

\subsubsection*{Availability of data and materials}
The datasets generated and analyzed during the current study are available from the corresponding author on reasonable request.

\bibliography{MHD_rectangular_2phase}

\end{document}

%% file: figures/Fig1a.pdf_tex
\begingroup%
  \makeatletter%
  \providecommand\color[2][]{%
    \errmessage{(Inkscape) Color is used for the text in Inkscape, but the package 'color.sty' is not loaded}%
    \renewcommand\color[2][]{}%
  }%
  \providecommand\transparent[1]{%
    \errmessage{(Inkscape) Transparency is used (non-zero) for the text in Inkscape, but the package 'transparent.sty' is not loaded}%
    \renewcommand\transparent[1]{}%
  }%
  \providecommand\rotatebox[2]{#2}%
  \newcommand*\fsize{\dimexpr\f@size pt\relax}%
  \newcommand*\lineheight[1]{\fontsize{\fsize}{#1\fsize}\selectfont}%
  \ifx\svgwidth\undefined%
    \setlength{\unitlength}{333.77252053bp}%
    \ifx\svgscale\undefined%
      \relax%
    \else%
      \setlength{\unitlength}{\unitlength * \real{\svgscale}}%
    \fi%
  \else%
    \setlength{\unitlength}{\svgwidth}%
  \fi%
  \global\let\svgwidth\undefined%
  \global\let\svgscale\undefined%
  \makeatother%
  \begin{picture}(1,0.41496348)%
    \lineheight{1}%
    \setlength\tabcolsep{0pt}%
    \put(0,0){\includegraphics[width=\unitlength,page=1]{Fig1a.pdf}}%
    \put(0.34508359,0.19927496){\color[rgb]{0,0,0}\makebox(0,0)[lt]{\lineheight{1.25}\smash{\begin{tabular}[t]{l}$Q_2$\end{tabular}}}}%
    \put(0.34508359,0.06801798){\color[rgb]{0,0,0}\makebox(0,0)[lt]{\lineheight{1.25}\smash{\begin{tabular}[t]{l}$Q_1$\end{tabular}}}}%
    \put(0,0){\includegraphics[width=\unitlength,page=2]{Fig1a.pdf}}%
    \put(0.23296678,0.06804636){\color[rgb]{0,0,0}\makebox(0,0)[lt]{\lineheight{1.25}\smash{\begin{tabular}[t]{l}$h_1$\end{tabular}}}}%
    \put(0.23296678,0.17857493){\color[rgb]{0,0,0}\makebox(0,0)[lt]{\lineheight{1.25}\smash{\begin{tabular}[t]{l}$h_2$\end{tabular}}}}%
    \put(0.8534038,0.01958498){\color[rgb]{0,0,0}\makebox(0,0)[lt]{\lineheight{1.25}\smash{\begin{tabular}[t]{l}$z$\end{tabular}}}}%
    \put(0.92528594,0.157472){\color[rgb]{0,0,0}\makebox(0,0)[lt]{\lineheight{1.25}\smash{\begin{tabular}[t]{l}$x$\end{tabular}}}}%
    \put(0.75527406,0.38309533){\color[rgb]{0,0,0}\makebox(0,0)[lt]{\lineheight{1.25}\smash{\begin{tabular}[t]{l}$y$\end{tabular}}}}%
    \put(0.54105171,0.02942559){\color[rgb]{0,0,0}\makebox(0,0)[lt]{\lineheight{1.25}\smash{\begin{tabular}[t]{l}$\rho_1, \eta_1, \sigma_{e1}$\end{tabular}}}}%
    \put(0.54105171,0.31221998){\color[rgb]{0,0,0}\makebox(0,0)[lt]{\lineheight{1.25}\smash{\begin{tabular}[t]{l}$\rho_2, \eta_2$\end{tabular}}}}%
    \put(0.01476216,0.14013415){\color[rgb]{0,0,0}\makebox(0,0)[lt]{\lineheight{1.25}\smash{\begin{tabular}[t]{l}$H$\end{tabular}}}}%
    \put(0,0){\includegraphics[width=\unitlength,page=3]{Fig1a.pdf}}%
    \put(0.08144722,0.3183301){\color[rgb]{0,0,0}\makebox(0,0)[lt]{\lineheight{1.25}\smash{\begin{tabular}[t]{l}$W$\end{tabular}}}}%
    \put(0,0){\includegraphics[width=\unitlength,page=4]{Fig1a.pdf}}%
    \put(0.91582905,0.29777767){\color[rgb]{0,0,0}\makebox(0,0)[lt]{\lineheight{1.25}\smash{\begin{tabular}[t]{l}$\textbf{g}$\end{tabular}}}}%
    \put(0,0){\includegraphics[width=\unitlength,page=5]{Fig1a.pdf}}%
    \put(0.37459244,0.27567094){\color[rgb]{0,0,0}\makebox(0,0)[lt]{\lineheight{1.25}\smash{\begin{tabular}[t]{l}$\textbf{B}_{0|x}$\end{tabular}}}}%
    \put(0,0){\includegraphics[width=\unitlength,page=6]{Fig1a.pdf}}%
    \put(0.39474393,0.32510981){\color[rgb]{0,0,0}\makebox(0,0)[lt]{\lineheight{1.25}\smash{\begin{tabular}[t]{l}$\textbf{B}_{0|y}$\end{tabular}}}}%
    \put(0.30017581,0.31814777){\color[rgb]{0,0,0}\makebox(0,0)[lt]{\lineheight{1.25}\smash{\begin{tabular}[t]{l}$\textbf{B}_0$\end{tabular}}}}%
  \end{picture}%
\endgroup%

%% file: figures/Fig1b.pdf_tex
\begingroup%
  \makeatletter%
  \providecommand\color[2][]{%
    \errmessage{(Inkscape) Color is used for the text in Inkscape, but the package 'color.sty' is not loaded}%
    \renewcommand\color[2][]{}%
  }%
  \providecommand\transparent[1]{%
    \errmessage{(Inkscape) Transparency is used (non-zero) for the text in Inkscape, but the package 'transparent.sty' is not loaded}%
    \renewcommand\transparent[1]{}%
  }%
  \providecommand\rotatebox[2]{#2}%
  \newcommand*\fsize{\dimexpr\f@size pt\relax}%
  \newcommand*\lineheight[1]{\fontsize{\fsize}{#1\fsize}\selectfont}%
  \ifx\svgwidth\undefined%
    \setlength{\unitlength}{373.74649168bp}%
    \ifx\svgscale\undefined%
      \relax%
    \else%
      \setlength{\unitlength}{\unitlength * \real{\svgscale}}%
    \fi%
  \else%
    \setlength{\unitlength}{\svgwidth}%
  \fi%
  \global\let\svgwidth\undefined%
  \global\let\svgscale\undefined%
  \makeatother%
  \begin{picture}(1,0.56272563)%
    \lineheight{1}%
    \setlength\tabcolsep{0pt}%
    \put(0,0){\includegraphics[width=\unitlength,page=1]{Fig1b.pdf}}%
    \put(0.71478337,0.13856591){\color[rgb]{0,0,0}\makebox(0,0)[lt]{\lineheight{1.25}\smash{\begin{tabular}[t]{l}$W$\end{tabular}}}}%
    \put(0.82318383,0.14383042){\color[rgb]{0,0,0}\makebox(0,0)[lt]{\lineheight{1.25}\smash{\begin{tabular}[t]{l}$x$\end{tabular}}}}%
    \put(0.1421826,0.53390014){\color[rgb]{0,0,0}\makebox(0,0)[lt]{\lineheight{1.25}\smash{\begin{tabular}[t]{l}$y$\end{tabular}}}}%
    \put(0.14256948,0.46236373){\color[rgb]{0,0,0}\makebox(0,0)[lt]{\lineheight{1.25}\smash{\begin{tabular}[t]{l}$H$\end{tabular}}}}%
    \put(0.14256948,0.3240028){\color[rgb]{0,0,0}\makebox(0,0)[lt]{\lineheight{1.25}\smash{\begin{tabular}[t]{l}$h_1$\end{tabular}}}}%
    \put(0.34031205,0.21861129){\color[rgb]{0,0,0}\makebox(0,0)[lt]{\lineheight{1.25}\smash{\begin{tabular}[t]{l}$\rho_1, \eta_1, \sigma_{e1}$\end{tabular}}}}%
    \put(0.34031205,0.38405798){\color[rgb]{0,0,0}\makebox(0,0)[lt]{\lineheight{1.25}\smash{\begin{tabular}[t]{l}$\rho_2, \eta_2$\end{tabular}}}}%
    \put(0,0){\includegraphics[width=\unitlength,page=2]{Fig1b.pdf}}%
    \put(0.00254194,0.27807533){\color[rgb]{0,0,0}\makebox(0,0)[lt]{\lineheight{1.25}\smash{\begin{tabular}[t]{l}$\textbf{B}_{0|x}$\end{tabular}}}}%
    \put(0,0){\includegraphics[width=\unitlength,page=3]{Fig1b.pdf}}%
    \put(0.75644623,0.39959667){\color[rgb]{0,0,0}\makebox(0,0)[lt]{\lineheight{1.25}\smash{\begin{tabular}[t]{l}$\textbf{g}$\end{tabular}}}}%
    \put(0,0){\includegraphics[width=\unitlength,page=4]{Fig1b.pdf}}%
    \put(0.38177504,0.04075337){\color[rgb]{0,0,0}\makebox(0,0)[lt]{\lineheight{1.25}\smash{\begin{tabular}[t]{l}$\textbf{B}_{0|y}$\end{tabular}}}}%
    \put(0,0){\includegraphics[width=\unitlength,page=5]{Fig1b.pdf}}%
    \put(0.17776876,0.09260431){\color[rgb]{0,0,0}\makebox(0,0)[lt]{\lineheight{1.25}\smash{\begin{tabular}[t]{l}$c_{bw}$\end{tabular}}}}%
    \put(0.05134608,0.18777348){\color[rgb]{0,0,0}\makebox(0,0)[lt]{\lineheight{1.25}\smash{\begin{tabular}[t]{l}$c_{sw}$\end{tabular}}}}%
    \put(0.71355966,0.26985083){\color[rgb]{0,0,0}\makebox(0,0)[lt]{\lineheight{1.25}\smash{\begin{tabular}[t]{l}$c_{sw}$\end{tabular}}}}%
  \end{picture}%
\endgroup%

%% file: MHD_rectangular_2phase.bib
@Book{Muller01,
  author    = {M{\"u}ller, Ulrich and B{\"u}hler, Leo},
  publisher = {Springer Science \& Business Media},
  title     = {Magnetofluiddynamics in channels and containers},
  year      = {2001},
  address   = {Berlin, Germany},
}

@Book{Davidson01,
  author    = {Davidson, P. A},
  publisher = {Cambridge University Press},
  title     = {An introduction to magnetohydrodynamics},
  year      = {2001},
  address   = {Cambridge, UK},
}

@Book{Branover78,
  author    = {Branover, H.},
  publisher = {John Wiley},
  title     = {Magnetohydrodynamic flow in ducts},
  year      = {1978},
  address   = {New {Y}ork, USA},
}

@Article{Hartman37a,
  author  = {Hartmann, J.},
  journal = {K. Dan. Vidensk. Selsk. Mat. Fys. Medd.},
  title   = {Theory of the laminar flow of an electrically conductive liquid in a homogeneous magnetic field},
  year    = {1937},
  number  = {6},
  pages   = {1--28},
  volume  = {15},
}

@Article{Hunt65,
  author  = {Hunt, J. C. R.},
  journal = {J. Fluid Mech.},
  title   = {Magnetohydrodynamic flow in rectangular ducts},
  year    = {1965},
  number  = {4},
  pages   = {577--590},
  volume  = {21},
}

@Article{Hussameddine08,
  author  = {Hussameddine, S. K. and Martin, J. M. and Sang, W. J.},
  journal = {J. Fluids Eng.},
  title   = {Analytical prediction of flow field in magnetohydrodynamic-based microfluidic devices},
  year    = {2008},
  pages   = {091204},
  volume  = {130},
}

@Article{Lohrasbi88,
  author  = {Lohrasbi, J. and Sahai, V.},
  journal = {Appli. Sci. Res.},
  title   = {Magnetohydrodynamic heat transfer in two-phase flow between parallel plates},
  year    = {1988},
  pages   = {53--66},
  volume  = {45},
}

@Article{Malashetty97,
  author  = {Malashetty, M. S. and Umavathi, J. C.},
  journal = {Int. J. Multiphase Flow},
  title   = {Two-phase magnetohydrodynamic flow and heat transfer in an inclined channel},
  year    = {1997},
  pages   = {545--560},
  volume  = {23},
}

@Book{Molokov07,
  author    = {Molokov, S. and Moreau, R. and Moffatt, K.},
  publisher = {Springer},
  title     = {Magnetohydrodynamics: Historical Evolution and Trends},
  year      = {2007},
  address   = {Dordrecht, Netherlands},
  series    = {Fluid Mechanics and Its Applications (FMIA)},
  volume    = {80},
}

@Article{Parfenov24,
  author  = {Parfenov, A. and Gelfgat, A. and Ullmann, A. and Brauner, N.},
  journal = {Fluids},
  title   = {Hartmann flow of two-layered fluids in horizontal and inclined channels},
  year    = {2024},
  number  = {6},
  pages   = {1--29},
  volume  = {9},
}

@Article{Shail73,
  author  = {Shail, R.},
  journal = {lnt. J. Eng. Sci.},
  title   = {On laminar two-phase flow in magnetohydrodynamics},
  year    = {1973},
  pages   = {1103--1108},
  volume  = {11},
}

@Article{Shercliff53,
  author  = {Shercliff, J. A.},
  journal = {Math. Proc. Camb. Philos. Soc.},
  title   = {Steady motion of conducting fluids in pipes under transverse magnetic fields},
  year    = {1953},
  number  = {1},
  pages   = {136--144},
  volume  = {49},
}

@Article{Wang22,
  author  = {Wang, Y. and Huang, H. and Lu, P. and Liu, Z.},
  journal = {Ind. Eng. Chem. Res.},
  title   = {Numerical investigation of gas–liquid metal two-phase glow in a multiple-entrance magnetohydrodynamic generator},
  year    = {2022},
  pages   = {4980--4995},
  volume  = {61},
}

@Article{Weston10,
  author  = {Weston, M. C. and Gerner, M. D. and Fritsch, I.},
  journal = {Anal. Chem.},
  title   = {Magnetic fields for fluid motion},
  year    = {2010},
  pages   = {3411},
  volume  = {82},
}

@Article{Yi02,
  author  = {Yi, M. and Qian, S. and Bau, H.},
  journal = {J. Fluid Mech.},
  title   = {A magnetohydrodynamic chaotic stirrer},
  year    = {2002},
  pages   = {153--177},
  volume  = {468},
}

@Article{Krasnov10,
  author  = {Krasnov, D. and Zikanov, O. and Rossi, M. and Boeck, T.},
  journal = {J. Fluid Mech.},
  title   = {Optimal linear growth in magnetohydrodynamic duct flow},
  year    = {2010},
  pages   = {273--299},
  volume  = {653},
}

@Article{Boeck14,
  author  = {Boeck, T. and Krasnov, D.},
  journal = {Phys. Fluids},
  title   = {A mixing-length model for side layers of magnetohydrodynamic channel and duct flows with insulating walls},
  year    = {2014},
  number  = {2},
  pages   = {025106},
  volume  = {26},
}

@Article{Shah22,
  author  = {Shah, N. A. and Alrabaiah, H. and Vieru, D. and Yook, S.-J.},
  journal = {Sci. Rep.},
  title   = {Induced magnetic field and viscous dissipation on flows of two immiscible fluids in a rectangular channel},
  year    = {2022},
  pages   = {39},
  volume  = {12},
}

@Book{Moreau90,
  title     = {Magnetohydrodynamics},
  publisher = {Kluwer Academic Publishers},
  year      = {1990},
  author    = {Moreau, R.J.},
  address   = {Dordrecht, The Netherlands},
}

@Article{Kirillov95,
  author  = {Igor R. Kirillov and Claude B. Reed and Leopold Barleon and Keiji Miyazaki},
  journal = {Fusion Eng. Des.},
  title   = {Present understanding of MHD and heat transfer phenomena for liquid metal blankets},
  year    = {1995},
  pages   = {553--569},
  volume  = {27},
}

@Article{Shanmugadas25,
  author  = {Shanmugadas, Sindu E. and Bau, Haim H.},
  journal = {Phys. Fluids},
  title   = {Onsager coefficients for liquid metal flow in a conduit under a magnetic field},
  year    = {2025},
  number  = {7},
  pages   = {073614},
  volume  = {37},
}

@Article{Buhler20,
  author  = {L. Bühler and C. Mistrangelo and H.-J. Brinkmann},
  journal = {Fusion Eng. Des.},
  title   = {Experimental investigation of liquid metal MHD flow entering a flow channel insert},
  year    = {2020},
  pages   = {111484},
  volume  = {154},
}

@Article{Nishio25,
  author  = {Ryunosuke Nishio and Masatoshi Kondo and Teruya Tanaka and Naoko Oono-Hori},
  journal = {Nucl. Mater. Energy},
  title   = {Experimental and analytical investigations to reduce MHD pressure drop for liquid LiPb fusion blanket systems: use of ODS-FeCrAl alloys with electrically insulating $\alpha$-Al2O3 layer in optimal flow channel geometry},
  year    = {2025},
  pages   = {101965},
  volume  = {44},
}

@Article{Chang61,
  author  = {Chang, Chieh C. and Lundgren, Thomas S.},
  journal = {Zeitschrift für angewandte Mathematik und Physik ZAMP},
  title   = {Duct flow in magnetohydrodynamics},
  year    = {1961},
  pages   = {100--114},
  volume  = {12},
}

@Article{Uflyand61,
  author  = {Uflyand, Ya S.},
  title   = {Flow stability of a conducting fluid in a rectangular channel in a transverse magnetic field},
  journal = {Soviet Physics: Technical Physics},
  year    = {1961},
  volume  = {5},
  pages   = {1191--1193},
}

@Article{Sloan66,
  author  = {Sloan, D. M. and Smith, P.},
  journal = {Z. Angew. Math. Mech. (J. Appl. Math. Mech.)},
  title   = {Magnetohydrodynamic Flow in a Rectangular Pipe between Conducting Plates},
  year    = {1966},
  pages   = {439--443},
  volume  = {46},
}

@Article{Barmak25,
  author        = {Ilya Barmak and Subham Pal and Alexander Gelfgat and Neima Brauner},
  journal       = {Theor. Comput. Fluid Dyn.},
  title         = {Two-phase stratified MHD flows in wide rectangular ducts: analytical and numerical solutions},
  year          = {in press, 2025},
  archiveprefix = {arXiv},
  eprint        = {2506.13357},
  url           = {https://arxiv.org/abs/2506.13357},
}

@Article{Butler69,
  author  = {Butler, G. F.},
  journal = {Math. Proc. Cambridge Philos. Soc.},
  title   = {A note on magnetohydrodynamic duct flow},
  year    = {1969},
  number  = {3},
  pages   = {655--662},
  volume  = {66},
}

@Article{Tao15,
  author  = {Tao, Zhen and Ni, MingJiu},
  journal = {Sci. China Physics Mech. Astronomy},
  title   = {Analytical solutions for MHD flow at a rectangular duct with unsymmetrical walls of arbitrary conductivity},
  year    = {2015},
  number  = {2},
  pages   = {1--18},
  volume  = {58},
}

@Article{Knaepen08,
  author  = {Knaepen, Bernard and Moreau, Ren\'e},
  journal = {Annu. Rev. Fluid Mech.},
  title   = {Magnetohydrodynamic Turbulence at Low Magnetic {R}eynolds Number},
  year    = {2008},
  pages   = {25--45},
  volume  = {40},
}

@Article{Bau03,
  author  = {Haim H. Bau and Jianzhong Zhu and Shizhi Qian and Yu Xiang},
  journal = {Sens. Actuators, B},
  title   = {A magneto-hydrodynamically controlled fluidic network},
  year    = {2003},
  number  = {2},
  pages   = {205--216},
  volume  = {88},
}
